\documentclass[aps,prd,preprint,groupedaddress,nofootinbib,byrevtex]{revtex4-1}
\usepackage{amsmath,amssymb}
\usepackage[dvips]{graphicx}
\usepackage{mathrsfs}

\begin{document}
\title{ Transition Probability for  the  Neutrino Wave in Muon Decay and Oscillation 
Experiments 
}
\author{Kenzo Ishikawa$^{(1)}$, Tasuku Nozaki$^{(1)}$, Masashi Sentoku$^{(1)}$, and Yutaka Tobita$^{(2)}$}

\affiliation{(1) Department of Physics, 
Faculty of Science, Hokkaido
University, Sapporo 060-0810,Japan} 
\affiliation{(2) Department of Mathematics and Physics, Faculty of
Science, Hirosaki 
University, Hirosaki 036-8561, Japan}
\date{\today}


\begin{abstract}  
This paper elucidates the anomalous  decay of the muon ascribed to extended waves. 
Due to  a large overlap of  the parent and daughters,  the transition amplitude 
and probability for the neutrinos are modified from  the standard formula.     A rigorous probability  
 from the von Neumann's fundamental principle of the quantum mechanics 
 at a large time interval $T$ is, $P=T\Gamma + P^{(d)}$,
 where  $\Gamma$ is derived from the Fermi's golden rule, and
 $P^{(d)}$ is a correction term.  A new term $P^{(d)}$ has origins in the overlapping waves, and
 influences the  determinations of physical parameters.  
 By including  $P^{(d)}$,  short-baseline neutrino experiments with 
LSND, KARMEN, and MiniBooNE 
become consistent each other and with  the solar, long-baseline, and 
reactor  experiments within the three neutrinos. 
Byproduct is that the absolute neutrino mass of the neutrinos  
can be measured.    

\end{abstract}
\maketitle

\section{Introduction}
The muon decays  
\begin{eqnarray}
& &\mu^- \to e^- + \bar{\nu}_e + \nu_\mu,\label{mu-decay}\\
& &\mu^+ \to e^+ + \nu_e + \bar{\nu}_\mu,\label{mu+decay} \nonumber
\end{eqnarray}
are fundamental processes  described by the interaction ${\mathcal L}_{\text int}=\frac{G_F}{\sqrt 2} J_{\rho}(\mu, x) {J^{\rho}(e,x)}^{\dagger}$, where $G_F$ is the Fermi coupling constant and $J_{\rho}(l,x)$ is the $V-A$ current of the charged lepton and the neutrino. The observed electron spectrum is in accord with the theoretical spectrum computed with the Fermi's golden rule precisely. It is expected that   the neutrino spectrum would be consistent with 
 the  golden rule   \cite{Dirac,
von Neumann, Wigner-Weisskopf}.

The decay is described  by a many-body wavefunction,   $|\Psi(t) \rangle $,  which is a solution of the time-dependent Schr\"{o}dinger equation with the  Hamiltonian $H_0+H_\text{int}$, where $H_0$ is the free part and $H_\text{int}=-\int d{\vec x} {\mathcal L}_\text{int}$.  That is initially at a time $t=0$  a one-muon state and develops a three-lepton component at later times.  The state  at  a large $t$ normally satisfies      
  $ \langle \Psi(t)|H_\text{int}| \Psi(t) \rangle=0$, which represents   free  independent 
particles; we call this
 region a particle zone. A left figure in Fig.1 shows the waves in the particle zone. Using  the transition probability $P(T)$ at $t=T$, which is defined with a square of the inner product of the states  from
 the von Neumann's fundamental principle of the quantum mechanics (FQM) for normalized states, the average probability  per unit of time $\Gamma=\frac{P(T)-P(T_0)}{T-T_0}$ between  a large $T$ and a small $T_0$ is determined by the Fermi's golden rule\cite{Dirac,
von Neumann, Wigner-Weisskopf} for $P(T) \ll 1$. Now,  states    of 
  $ \langle \Psi(t)|H_{\text {int}}| \Psi(t) \rangle \neq 0$  shown in a right
 figure of Fig.1 also exist and give an independent  contribution  to the probability at $T_0$.  
These  represent interacting waves; we call this region a wave zone. Thus 
   \begin{eqnarray}
   P(T)= P(T_0)+\Gamma (T-T_0) \label{probability:0}
   \end{eqnarray}
    \cite{Ishikawa-Tobita-PTEP,Ishikawa-Tobita-ANA,Ishikawa-Tajima-Tobita, Ishikawa-Tobita-II}. 
 The neutrino in the muon decay is one of the cleanest systems that reveal  intriguing properties of the wave zone,  which is studied  in the present paper.
\begin{figure}[t]
\includegraphics[scale=.5]{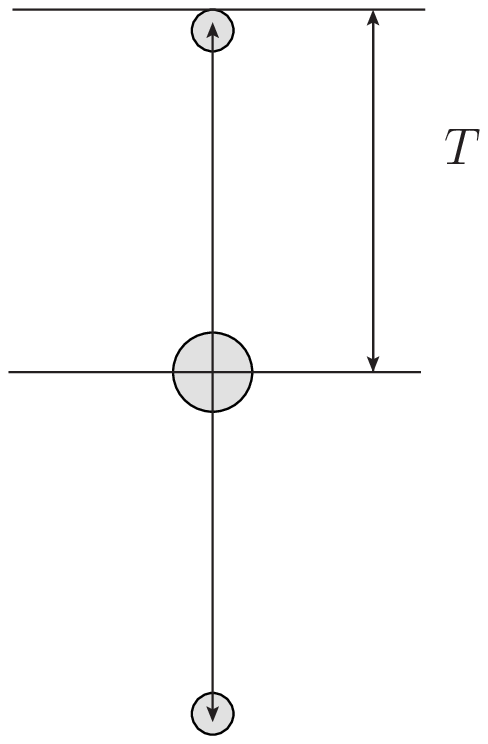}
\includegraphics[scale=.6]{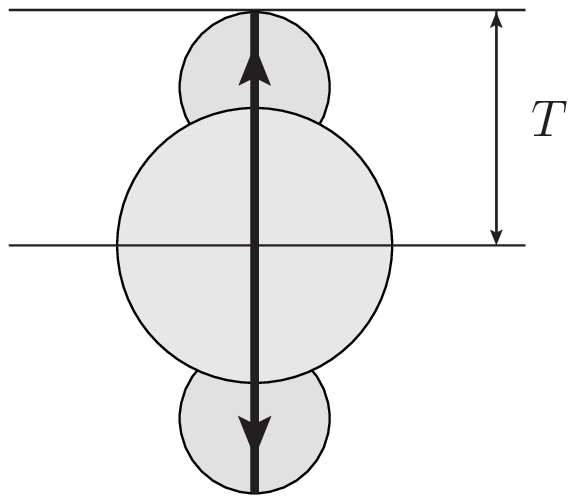}
\caption{Wavefunction of a two-body decay of a particle at rest in a configuration space 
at $T$ in the particle zone  (left) and in the wave zone (right).  The shadow areas show the regions of the waves. In the particle zone,  the 
 states do not overlap and behave like particles.  In the wave zone,  they  overlap
 and behave like interacting waves.  }
\label{fig:zone}
\end{figure}

Neutrino oscillation experiments   disclosed  the neutrino's masses and mixing angles.  The values obtained with single-particle formula  in long-distance region are consistent with 
each other \cite{pdg}, and the  mass-squared differences of the neutrinos  $\Delta m_{ij}^2=m_{\nu_i}^2-m_{\nu_j}^2 $ are $(7.53\pm 0.18)\times 10^{-5}\
\text{eV}^2/c^4$ and $ (2.44\pm 0.06)\times 10^{-3}\ \text{eV}^2/c^4$ (normal hierarchy)
 or $(2.51\pm 0.06)\times 10^{-3}\ \text{eV}^2/c^4$ (inverted).
The  formula gives negligible effect in short distance regions.  Nevertheless, signals have been found in experiments of  LSND and 
 MiniBooNE and fitted with much higher mass, but not in KARMEN. 
 The discrepancy 
 may be a signal  beyond the standard model.  Sterile neutrino is one of interesting  possibilities and projects for  its search are under ways. 
 In these analyses, the neutrino in the particle zone were  assumed.    If   neutrinos 
  in the wave zone participate in the short-distance region,  the oscillation formula  will be modified.  That could  be related with the 
   existing discrepancy, and deserves of intensive  study.

  In the transitions at the particle zone, the final waves separate quickly and the state is  described by a stationary state of  {$\displaystyle{\lim_{t\to \pm\infty}} \langle \Psi(t)|H_\text{int}|\Psi(t)\rangle=0$}   \cite{Schwinger,Tomonaga,Feynmann}. They are described by a standard method of field theory,   an  S-matrix, $S[\infty]$  with  an interaction Hamiltonian $e^{- \epsilon |t|} H_\text{int}$, in which the limit $\epsilon \rightarrow 0$ is taken at the end of the calculation. The interaction switches off adiabatically (ASI) and {$\displaystyle{\lim_{t\to \pm\infty}} \langle \Psi_o(t)|e^{-\epsilon|t|}H_\text{int}|\Psi_i(t)\rangle=0$}  
holds for arbitrary states $|\Psi_i(t) \rangle$ and $|\Psi_o(t) \rangle$.  The probability proportional to the time interval $T$, $\Gamma T$, is derived. The neutrino spectrum is given by  the single-particle formula.

 The  states of   $\langle \Psi(t)| H_\text{int}|\Psi (t)\rangle \neq  0$ at large $t$ are not described 
by ASI   \cite{Goldberger-Watson-paper}, and the 
transition probability  from these states is not included in that of  the golden rule or of $S[\infty]$.  The probability   computed  with  FQM  \cite{Ishikawa-Tobita-PTEP,Ishikawa-Tobita-ANA,Ishikawa-Tajima-Tobita, Ishikawa-Tobita-II} includes both, and is expressed   in   the two-body decays 
as 
\begin{align}
 P(T) = T\Gamma + P^{(d)}. \label{probability}
\end{align}
 Here   $\Gamma $ agrees with that of the golden rule.  From  Eq.$(\ref{probability:0})$, 
 $P^{(d)}$ given by $P^{(d)}=P(T_0)-\Gamma T_0$  shows a rapid transition  of the extended waves at a small time.  This  is characterized by the angular velocity of the non-stationary neutrino at a position traveling  with the speed of  light $\omega=\frac{E_{\nu}({\vec p})-cp}{\hbar} $, where $\hbar=\frac{h}{2\pi}$ and $h$ is the Planck constant,
${\vec p}$, $E_\nu({\vec p}\,)=\sqrt{p^2c^2 + m_\nu^2c^4}$, $m$, and $c$ are the
momentum, energy, mass, and the speed of  light, and $p=|{\vec p}\,|$.  
This 
 $\omega $   corresponds to a  length $L_c=\frac{c}{\omega}$ in the configuration space. $L_c$   is  twice of 
 the product of the
 neutrino's Compton wavelength with   the boost factor $\frac{E_\nu(\vec{p}\,)}{m_{\nu}c^2}$, and  much 
longer than the typical size of the stationary wave, de Broglie wavelength $\frac{\hbar}{p}$.  Accordingly $P^{(d)}$ is large compared with  $\Gamma T_0$ in
 the neutrino, but is small  in the electron.   The flavor-changing probability  is modified from the single-particle formula.

The   time-dependent Schr\"{o}dinger equation    which describes the muon decay is
\begin{align}
 i\hbar\frac{\partial}{\partial t}|\Psi(t)\rangle &= H|\Psi(t)\rangle=\left(H_0 + H_\text{int}\right)|\Psi(t)\rangle,\label{shrodinger-eq}\\
H_0 &= \int d\vec{x}\sum_{l=e,\mu} \left( \bar{l}(x)\left(\vec{\alpha}\cdot\vec{\nabla} + \beta m_l\right)l(x) + \bar{\nu}_l
(x)\left(
\vec{\alpha}\cdot\vec{\nabla} + \beta m_{\nu_l}\right)\nu_l(x)\right),\nonumber\\
H_\text{int} &=\frac{G_F}{\sqrt{2}}\int d\vec{x}\left(
\bar{\mu}(x)\gamma_\rho\left(1 - \gamma_5\right)
 \nu_{\mu}(x)\right)\gamma^\rho\left(\bar e(x)\left(1 - \gamma_5\right)
 \nu_{e}(x) \right)^{\dagger},
 \nonumber
\end{align}
for a case of no-mixing, and  solved easily in the lowest order of $G_F$. 
  For a muon of the lifetime $\tau_{\mu}$ with average momentum ${\vec p}_{\mu}$ and the range in space covered by the wavefunction $\sigma_{\mu}$ in the initial state,
   a normalized solution $|\Psi(t),{\vec p}_{\mu}\rangle$  is 
\begin{align}
& |\Psi(t),\vec{p}_{\mu}\,\rangle = a_0(t)|\vec{p}_\mu; \sigma_{\mu}\rangle+\int
 d\vec{p}_e d\vec{p}_{\nu_e}d\vec{p}_{\nu_\mu}
a_1(t,\vec{p}_e,\vec{p}_{\nu_e},\vec{p}_{\nu_\mu})|\vec{p}_e,\vec{p}_{\nu_e},\vec{p}_{\nu_\mu}\rangle+
O(G_F^2),\label{weight}\\ 
& a_0(t) = \left({1 \over 1+\zeta(t)}\right)^{1/2}e^{-i\frac{E_\mu}{\hbar}t -
 \frac{t}{2 \tau_\mu} }, \nonumber \\ 
& a_1(t,\vec{p}_e,\vec{p}_{\nu_e},\vec{p}_{\nu_\mu}) = 
\left({1 \over 1+\zeta(t)}\right)^{1/2} e^{-i\frac{E_\mu}{\hbar}t}\frac{e^{-i\frac{\Delta E }{\hbar}t} 
- e^{-\frac{t}{2 \tau_\mu}}}{\Delta E  + i\frac{\hbar}{2 \tau_\mu}}\langle \vec{p}_e,\vec{p}_{\nu_e},
\vec{p}_{\nu_\mu}|H_\text{int}|\vec{p}_\mu; \sigma_{\mu} \rangle, 
 \nonumber 
\end{align}
 where
$\Delta E = E_e + E_{\nu_e} +E_{\nu_\mu} - E_\mu$. The coefficients satisfy $a_0(0)=1,a_1(0,\vec{p}_e,\vec{p}_{\nu_e},\vec{p}_{\nu_\mu}) = 0$.  Using the matrix element given later,
\begin{eqnarray}
& &(1-e^{-\frac{t}{\tau_\mu}})= \int_{\Delta E \approx 0}
 d\vec{p}_e d\vec{p}_{\nu_e}d\vec{p}_{\nu_\mu} 
\left|\frac{e^{-i\frac{\Delta E }{\hbar}t} 
- e^{-\frac{t}{2\tau_\mu}}}{\Delta E  + i\frac{\hbar}{2\tau_\mu}}\langle \vec{p}_e,\vec{p}_{\nu_e},
\vec{p}_{\nu_\mu}|H_\text{int}|\vec{p}_\mu, \sigma_{\mu} \rangle\right|^2, \label{tau} \\ 
& &\zeta(t) =\int_{\Delta E  \neq 0}
 d\vec{p}_e d\vec{p}_{\nu_e}d\vec{p}_{\nu_\mu} \left|\frac{e^{-i\frac{\Delta E }{\hbar}t} 
- e^{-\frac{t}{2\tau_\mu}}  }{\Delta E  + i\frac{\hbar}{2\tau_\mu}}\langle \vec{p}_e,\vec{p}_{\nu_e},
\vec{p}_{\nu_\mu}|H_\text{int}|\vec{p}_\mu,\sigma_{\mu} \rangle\right|^2, \label{zeta}
\end{eqnarray}
where Eq.$(\ref{zeta})$ is evaluated at a  small $\frac{t}{\tau_{\mu}}$. 
 $\tau_{\mu}$ in Eq.$(\ref{tau})$ and $\zeta(t)$ in Eq.$(\ref{zeta})$ signify the asymptotic behavior ascribed to the wavefunction in the particle zone and in the wave zone  respectively. The  right-hand side of Eq.$(\ref{tau})$ varies slowly with $t$, and 
 the right-hand side of Eq.$(\ref{zeta})$ increases rapidly  at a small $t$ and becomes constant at later times.  
Ignoring the rapid change,    
\begin{align}
 \zeta(0)=0,\ \zeta(t) =\zeta <1 (t >0),
 \end{align}
where $\zeta$ is a constant.  Norm  of  $ |
\Psi(t),{\vec p}_{\mu}\rangle $ is the sum of  the  integrals over the momenta in  
 $\Delta E  \approx 0$ and $\Delta E  \neq 0$,
 \begin{eqnarray}
 \langle \Psi(t),{\vec p}_{\mu} | \Psi(t),{\vec p}_{\mu} \rangle  =\frac{1}{1+\zeta(t)}e^{-\frac{t}{\tau_{\mu}}}+\frac{1}{1+\zeta(t)}(1-e^{-\frac{t}{\tau_{\mu}}})+\frac{\zeta(t)}{1+\zeta(t)},
\end{eqnarray}
where the first term in the right-hand side shows the norm of the muon, the second term shows that of the three leptons of  $\Delta E \approx 0$, and the third one shows these of $\Delta E  \neq 0$.    
A component of  $\Delta E \approx 0$ results to  $\Gamma=\frac{1}{\tau_{\mu}}$, and that  of  $\Delta E  \neq 0$ results 
to $\zeta(t)$. 
 The  muon number decreases rapidly at a short time from  $\zeta(t)$.     
  
For $t \gg \tau_{\mu} $, $|a_0(t)|=0$, 
\begin{align}
& a_1(t) =  \left({1 \over 1+\zeta}\right)^{1/2}  e^{-i\frac{E_\mu}{\hbar}t}\frac{e^{-i\frac{\Delta E }{\hbar}t} 
}{\Delta E + i\frac{\hbar}{2\tau_\mu}}\langle \vec{p}_e,\vec{p}_{\nu_e},
\vec{p}_{\nu_\mu}|H_\text{int}|\vec{p}_\mu, \sigma_{\mu}\rangle,  
\end{align}
and          
\begin{align}
& \langle \Psi(t),\vec{p}_{\mu}\ |\Psi(t),\vec{p}_{\mu}\,\rangle ={1 \over 1+\zeta}
 (1+\zeta)=1\\
& \delta E^2=\langle \Psi(t)|(i\hbar\frac{\partial}{\partial t} -
 E_\mu)^2|\Psi(t)\rangle = \langle
 \Psi(t)|H_\text{int}^2|\Psi(t)\rangle \\
& ~~~~~ \geq  \langle \Psi(t)|H_\text{int}|\Psi(t) \rangle \langle \Psi(t) |
H_\text{int}| \Psi(t)\rangle\label{ratio of energy} 
 = \zeta^2. \nonumber
\end{align}
$\zeta \neq 0$ leads  $ \delta E^2 \neq 0$. Thus $|\Psi(\infty) \rangle$ includes the states of the wave zone.

Substituting  $\frac{t}{\tau_{\mu}} \ll 1$ to Eqs.$( \ref{tau})$  and $(  \ref{zeta})$,  $\Gamma=\frac{1}{\tau_{\mu}}$ and  $\zeta=P^{(d)}$ are expressed with the integrals in the right-hand sides. The former is  equivalent to the Weisskopf-Wigner formula.  
  In Eq.$(\ref{zeta})$, the upper limit  is  $\Delta E = \infty$.    The states of $\Delta E \rightarrow \infty$  are characterized  by Lorentz invariant spectrum, and
  lead a universal form to $P^{(d)} $.  That  is independent of the particle's spin and evaluated for a scalar neutrino first.   Writing the interaction Hamiltonian  in the form $\int d^3 x J(x) \nu(x)$ with the neutrino $\nu(x)$ and a product of the 
parent $|\text P \rangle $ and other daughters $|\text D \rangle$ $J(x)$,  
the amplitude that the neutrino of the momentum $p_{\nu}$ is detected  is proportional to $ \langle p_{\nu},\text D |\int d^4 x J(x) \nu(x) | \text P \rangle $.  The probability is proportional to
$\int d^4x d^4y \Delta_J(x-y)\Delta_{\nu}(x-y)$, where  $\Delta_J(x-y)=\sum_{\text D} \langle \text P| J^{\dagger}(y) |\text D \rangle \langle \text D |
J(x)|\text P \rangle $ is a source correlation function in
the state of the parent   $|\text P\rangle$ summed over   $|\text D \rangle $, and  $ \Delta_{\nu}(x-y;p_{\nu})=\langle p_{\nu}|\nu(x)
\nu^{\dagger}(y)|p_{\nu} \rangle $ is the neutrino propagator. $\Delta_J(x-y)$ has a singular and long-range component and a short-range one.
The former one comes from the states of  $E_D \rightarrow \infty$  and is of the form   $\Delta_J(x-y) \approx \delta((x-y)^2)$. 
Substituting this term,  
\begin{eqnarray}
\Delta_J(x-y)\Delta_{\nu}(x-y)
=e^{i(E({\vec p_{\nu}}\,)-c|{\vec p_{\nu}}\,| \cos \theta)(x^0-y^0)/{\hbar}}\delta((x-y)^2), \label{wave-light-cone}
\end{eqnarray}
where $ \theta$ is the angle between ${\vec p_{\nu}}$ and ${\vec
 x-\vec{y}}$, and 
 $E({\vec{p}_{\nu}}\,)=\sqrt{\vec{p}_{\nu}^{\,2}c^2+m_{\nu}^2c^4}$, which becomes at
 $\theta =0$,
\begin{eqnarray}
e^{i{\omega} (x^0-y^0)} \delta((x-y)^2),\ \omega=\frac{E({\vec p}_{\nu})-c p_{\nu}}{\hbar}.
\end{eqnarray}
The phase varies extremely slowly over the region 
$l \leq L_c$,
\begin{eqnarray}
 L_c =\frac{c}{\omega}=\frac{\hbar}{m_{\nu}c}\times\frac{2E(\vec{p}_{\nu})}{m_{\nu}c^2}.
 \label{coherence-length1}
\end{eqnarray}
   Thus in the extreme forward direction,  $\theta =0$, the
wavefunction has an extremely long correlation, and 
 the integrand Eq.$(\ref{wave-light-cone})$ varies with $x^0-y^0$ extremely slowly. The integral over $x$ and $y$  gets
an extra   contribution from large $|x^0-y^0|$ region.   A calculation showed that the probability $P^{(d)}$ is proportional to the  
range in space covered by the wavefunction $\sigma_{\nu}$,  
\begin{eqnarray}
P^{(d)} = \sigma_{\nu} \frac{\hbar E(\vec{p}_{\nu})}{  m_{\nu}^2c^3} \times (\text {numerical~factor}). \label{diffraction-form}
\end{eqnarray}

The length $L_c$ and $\sqrt \sigma_{\nu}$ are  much longer than the  de Broglie wavelength. 
 $L_c$ is $2 \times 10^2$ meter for $E=10$ MeV and $m_{\nu}c^2 =10^{-1}$ eV, and its ratio over the de Broglie wavelength is roughly $(\frac{E}{m_{\nu}c^2})^2 =10^{16}$. 
 Accordingly,  a frequency of the events determined by  Eq.$(\ref{diffraction-form})$  is high.  These are characteristic features of the interacting quantum waves in the wave zone. 
$L_c$ is slightly smaller than the muon mean free length $c \tau_{\mu}$, and a delicate competition between this  diffraction and the decay gives a sensitive signal to the absolute neutrino mass.     
 The long-range correlation similar to the EPR correlation  \cite{EPR,Jaguar} appears and gives $P^{(d)}$. 

Rigorous calculation is made later using the   S-matrix $S[T]$ defined with the normalized functions of satisfying the boundary conditions at  $T$ based on FQM. The transition 
probability at $T$, $P(T)$, is computed  without facing   difficulty  mentioned in \cite{Goldberger-Watson-paper}.    
 $S[T]$  is formulated with the
M{\o}ller operator, and wave packets localized in space,  \cite{Ishikawa-Shimomura,
Ishikawa-Tobita-PTEP,Ishikawa-Tobita-ANA, Ishikawa-Tajima-Tobita,Ishikawa-Tobita-II}. LSZ formula
\begin{eqnarray}
& &\lim_{t \rightarrow -  T/2}\langle\alpha| \phi^f (t)|\beta
 \rangle=  \langle \alpha| \phi_\text{in}^f |\beta \rangle,\label{boundary-condition}\\
& &\lim_{t \rightarrow +  T/2}\langle \alpha|\phi^f(t) |\beta \rangle=  \langle
 \alpha| \phi_\text{out}^f |\beta \rangle, \nonumber
\end{eqnarray}
are applied,
where $\phi_\text{in}(x)$ and $\phi_\text{out}(x)$ satisfy the free wave equation,
and $\phi^f(t),\ \phi_\text{in}^f(t)$, and $\phi_\text{out}^f(t)$ are the expansion
coefficients of  $\phi(x),\ \phi_\text{in}(x)$, and $\phi_\text{out}(x)$, with the
normalized wave functions $f(x)$ of the form
\begin{eqnarray}
\phi^f(t)=i\int d^3 x f^{*}({\vec x},t) \overleftrightarrow{\partial_0}
 \phi({\vec x},t) \label{field-expansion}.
\end{eqnarray}
It is noticed that a complete set of the normalized  functions includes those  that are specified by   their centers in  momentum space and in  configuration space
\cite{Ishikawa-Shimomura,Ishikawa-tobita-ptp}.

   $P^{(d)}$  has the origin in the long-range component of the  wavefunction that depends on the initial and final states, and causes   the rapid transition in short times.  
    That shows  unique  properties.  Because  $P^{(d)}$  depends on the absolute mass, that of the neutrino is  much larger  than that of the electron, and    the neutrino flavor
 change  distinct from the standard oscillation may arise in the short-distance regions.
 $P^{(d)}$ may be  directly connected with the inconsistency among
 previous experiments within three neutrino flavor.

The absolute neutrino masses  are   unknown  now.  
  One of existing
 methods is to study tritium
beta decay, but the existing upper bound for the effective electron-neutrino mass-squared difference is approximately 2 eV${}^2$/$c^4$ 
\cite{Troitsk}. 
From cosmology, the bounds for a sum of masses are  $0.44$ eV/$c^2$ \cite{Wmap7,Wmap9} and $ 0.23$ eV/$c^2$ \cite{Planck}.  They are important in structure formations in 
cosmology. They may be determined by precision experiments in short-distance regions.

The present paper analyzes the neutrino properties in the wave zone   
in detail including the mixing effects, and  is organized in the following manner. In Sec. 2, the transition amplitudes  and probability of the muon decay processes including the overlap effects 
are derived.  In Sec. 3, its
implications in the neutrino experiments  are presented.  Summary is given in Sec. 4.  
\section{The transition amplitude and probability }
The normalized  functions in configuration space  in   Eqs. $(\ref{boundary-condition})$ and
$ (\ref{field-expansion})$   for outgoing states represent microscopic states that
 an outgoing state
interactes with \cite{Ishikawa-Tobita-PTEP}.
For    the processes that  the neutrinos are detected, they represent nucleus or
atom bound in solid, which have almost uniform density and   decrease  smoothly 
toward the edge. They are approximated
well with  Gaussian wave packets  
\cite{Ishikawa-Tobita-ANA}.  For incoming states, they  express the
 states in the beam. The muon  in the 
beam  has a finite range in space  determined by the  
mean free path in medium.  From here the natural unit is used in most places.
\subsection{S-matrix at a finite-time interval $S[T]$ }
\begin{figure}[t]
 \includegraphics[scale=.4]{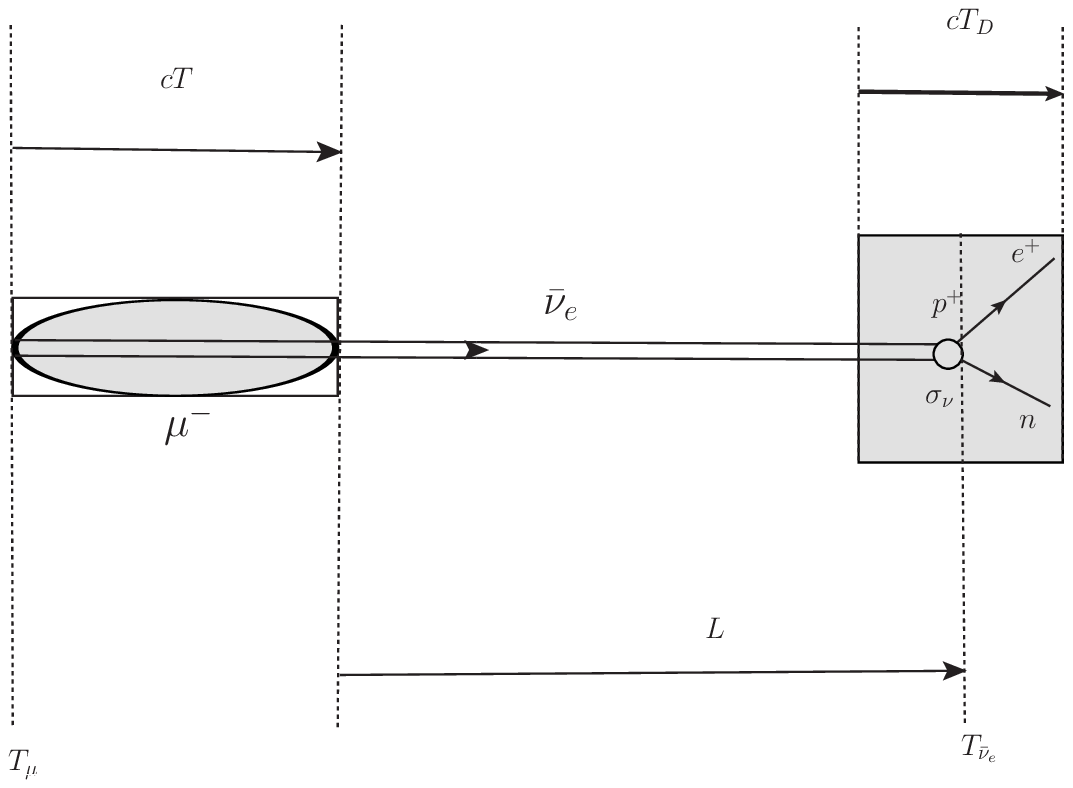}
\caption{The geometry of the $\mu^-$ decay at rest is shown. In real experiments, a detector with a length of $cT_D$ is located away from the decay region. However, in Sec. 3 we study a case where a detector is located in front of the decay region, i.e., $L=0$, for simplicity.}
\label{fig:mu-decay-at-rest}
\end{figure}

The $S[T]$ defined from the wavefunction at $T$ satisfies the  unitarity  $
S[T] S^{\dagger}[T]=1$. Inserting the initial and final states of the muon decay,   
\begin{eqnarray}
\langle \mu |S[T]| \mu \rangle \langle \mu|S^{\dagger}[T]|\mu \rangle +\langle \mu |S[T]|e,\nu,\bar \nu \rangle
 \langle e,\nu,\bar \nu|S^{\dagger}[T] |\mu \rangle =1.
\end{eqnarray}
Thus the survivable probability, $P(\mu^- \to \mu^-)$, is written with the decay probability, $P(\mu^- \to e^- + \nu_\mu + \bar{\nu}_e)$, as
\begin{align}
P(\mu^- \to \mu^-)=1 - P(\mu^- \to e^- + \nu_\mu + \bar{\nu}_e).
\end{align}
Later $P( \mu^- \rightarrow  e^-+\nu_{\mu}+\bar \nu_e)=\Gamma T+P^{(d)}$ is computed. Depending on the magnitude of $P(\mu^- \to e^- + \nu_\mu + \bar{\nu}_e)$, that is written as
\begin{align}
P(\mu^- \to \mu^-;T)=
\begin{cases}
1-(\Gamma T+P^{(d)});\ \text{small}~~ P(\mu^-\to e^-+\nu_\mu + \bar{\nu}_e)\\
\frac{1}{1 + P^{(d)}}e^{-\Gamma T}; \text{ other}.
\end{cases}
\end{align}
The normalization of the muon is affected by the transition from Eq.$(\ref{weight})$.
 \subsubsection{$\mu \rightarrow e+\nu+\bar \nu$}
In the case  $m_{\nu} \approx 0.08 $ eV,  $T_c$ in
Eq. $(\ref{coherence-length1})$ satisfies $T_c \gg \tau_{\mu}$. The case $P \ll 1$
is studied first. 

The detection of $\bar \nu_e$  through the inverse $\beta$ process  is
shown in Fig. \ref{fig:mu-decay-at-rest}; $cT$ is the muon decay region
and $D$ is the detector.
The length between decay region and detector denoted
$L$ is set to zero in this section, but that is included in the latter section.
We study first the single neutrino case, using the same notation for  the particle and  anti-particle: $\mu$ for $\mu^{+}$ and $\mu^{-}$, etc.
 
The decay  amplitude of a muon of a wave packet denoted as  $|\mu\rangle$ to a final state of  plane waves for  $e$ and $\nu_\mu$ and a wave packet for $\nu_e$ $|e,\nu_\mu,\nu_e\rangle$ \cite{Ishikawa-tobita-wavepacket-letter} 
\begin{align}
& |\mu\rangle = |\vec{p}_\mu,\vec{X}_\mu,T_\mu\rangle,
\ |e,\nu_\mu,\nu_e\rangle = |\vec{p}_e,\vec{p}_{\nu_\mu},\vec{p}_{\nu_e},\vec{X}_{\nu_e},T_{\nu_e}\rangle
\end{align}
is 
\begin{align}
& \mathcal{M}=\frac{G_F}{\sqrt{2}}
 8N_{\mu}^{-1}N_{\nu_e}^{-1}\varrho_\mu\varrho_e\varrho_{\nu_\mu}\varrho_{\nu_e}f I(\delta p),\ \delta p = p_\mu-p_e-p_{\nu_\mu}-p_{\nu_e}\label{amp-3},\\
& I(\delta p)=\int_{T_\mu}^{T_{\nu_e}} dt\int d\vec{x} e^{-i\delta\phi(x)}
w(x,X_\mu;\sigma_\mu)w(x,X_{\nu_e};\sigma_{\nu_e}),\nonumber \\
&f = \bar{u}(\vec{p}_{\nu_\mu})\gamma_\rho(1-\gamma_5)u(\vec{p}_\mu)\bar{u}(\vec{p}_e)\gamma^\rho
(1-\gamma_5)u(\vec{p}_{\nu_e}),\nonumber \\
&\delta\phi(x) = \phi_\mu(x,\vec{p}_\mu)-\phi_e(x,\vec{p}_e) 
-\phi_{\nu_\mu}(x,\vec{p}_{\nu_\mu})-\phi_{\nu_e}(x,\vec{p}_{\nu_e}), \nonumber
\\
& N_\mu=\left(\frac{\sigma_\mu}{\pi}\right)^\frac{3}{4},
\ N_{\nu_e} = \left(\frac{\sigma_{\nu_e}}{\pi}\right)^\frac{3}{4},
\ \varrho_\alpha =
 \left(\frac{m_\alpha}{(2\pi)^3E_\alpha}\right)^\frac{1}{2}, \\
& \phi_\alpha(x,\vec{k}_\alpha)
= E(\vec{k}_\alpha)(t-T_\alpha) - \vec{k}_\alpha\cdot(\vec{x}-\vec{X}_\alpha),\ (\alpha=\mu,\nu_e)\label{phase-mu-nue}, \\
&\phi_{\beta}(x,\vec{p}_\beta)= E(\vec{p}_\beta) t -
 \vec{p}_\beta\cdot\vec{x},\ (\beta=e,\nu_\mu)\label{phase-e-numu}, 
\end{align}
where the wavefunctions are
\begin{eqnarray}
& &w(x,X_{\mu};\sigma_{\mu};\tau_{\mu})=  e^{-{1 \over
 2\sigma_{\mu}}({\vec x}-{\vec X}_{\mu}-{\vec v}_{\mu}(t-T_{\mu}))^2 -{t-T_{\mu}
 \over 2 \tau_{\mu}}};\ t -T_{\mu} \geq 0 ,\\
& &w(x,X_{\nu};\sigma_{\nu})= e^{-{1 \over
 2\sigma_{\nu}}({\vec x}-{\vec X}_{\nu}-{\vec v}_{\nu}(t-T_{\nu}))^2}. \nonumber
\end{eqnarray} 
Now  the spreading effect of the wave packet and a causality at
large $|t-T_{\nu_e}|$ that  the wave packet vanishes
at $(t-T_{\nu_e})^2-({\vec x}-{\vec X}_{\nu_e})\leq 0$
\cite{Ishikawa-Shimomura}, are negligible, and
are not expressed explicitly
in most places.

\subsection{Transition probability}
The amplitude $\mathcal M$ from Eq.$(\ref{amp-3})$, is substituted into the transition probability  at $T$
 \begin{eqnarray}
  P=\int d{\vec p}_e d{\vec p}_{\nu_{\mu}} d{\vec p}_{\nu_e} \frac{d {\vec X}_{\nu_e}}{(2\pi)^3} |\mathcal M|^2.
  \end{eqnarray}
  Averaging  over the initial spins and summing  over
the final spins,
\begin{align}
& P = \left(\frac{\pi^2}{\sigma_\mu\sigma_{\nu_e}}\right)^\frac{3}{2}
\frac{2^8G_F^2}{E_\mu(2\pi)^3}
\int \frac{d\vec{p}_e d\vec{p}_{\nu_\mu}d\vec{X}_{\nu_e}d\vec{p}_{\nu_e}}
{E_e E_{\nu_\mu}E_{\nu_e}(2\pi)^{12}}
(p_\mu\cdot p_{\nu_e})(p_e\cdot p_{\nu_\mu})\left|I(\delta p)\right|^2.\label{total-probability}
\end{align}
\subsubsection{ $\mu \rightarrow e \nu \bar \nu $: transition probability: normal term }
$|I(\delta p)|^2$ in Eq.$(\ref{total-probability})$ is expressed by an integrals over times. The integrand   in small $t_1-t_2$ region  is characterized by the de Broglie wavelength, and much shorter than the $cT$. Hence  the integral from this region determines $\Gamma T$ from Appendix C, and is evaluated by the integral  over the region  $-\infty <t_1-t_2< \infty$.   Then  the effect of the muon's life-time is neglected, and $T$ is  replaced with  $\infty$. It follows 
\begin{align}
  I^\text{normal}(\delta p)
=\int_{-\infty}^{\infty}dt\int
 d\vec{x}e^{-i\delta\phi(x)}w(x,X_\mu;\sigma_\mu)w(x,X_{\nu_e};\sigma_{\nu_e}).
 \label{I-normal}
\end{align}
and
\begin{align}
 \left|I^\text{normal}(\delta p)\right|^2 
= &\left(\frac{2\pi(\sigma_\mu+\sigma_{\nu_e})}{(\vec{v}_\mu-\vec{v}_{\nu_e})^2}
\right)\left(\frac{2\pi\sigma_\mu\sigma_{\nu_e}}{\sigma_\mu+\sigma_{\nu_e}}\right)^{3}
\exp\left[{-\frac{\sigma_\mu+\sigma_{\nu_e}}{(\vec{v}_\mu-\vec{v}_{\nu_e})^2}\left(
\delta p^0 - \vec{v}_0\cdot\delta \vec{p}
\right)^2}\right]\nonumber\label{sqr-modu}\\
&\times\exp\left[-\frac{\sigma_\mu\sigma_{\nu_e}}{\sigma_\mu+\sigma_{\nu_e}}\delta \vec{p}^{\,2}
-\frac{\left(\tilde{\vec{X}}_\mu-\tilde{\vec{X}}_{\nu_e}\right)_T^2}
{\sigma_\mu+\sigma_{\nu_e}}\right], 
\end{align}
where
\begin{align}
\vec{v}_0 = &\frac{\sigma_\mu\vec{v}_{\nu_e}+\sigma_{\nu_e}\vec{v}_\mu}{\sigma_\mu+\sigma_{\nu_e}},\ 
\tilde{\vec{X}}_\mu = \vec{X}_\mu-\vec{v}_\mu T_\mu,\ \tilde{\vec{X}}_{\nu_e} = \vec{X}_{\nu_e}-\vec{v}_{\nu_e}T_{\nu_e}, \\
(\tilde{\vec{X}}_\mu-\tilde{\vec{X}}_{\nu_e})_T = &\tilde{\vec{X}}_\mu-\tilde{\vec{X}}_{\nu_e} - \vec{n}
\left(\vec{n}\cdot(\tilde{\vec{X}}_\mu-\tilde{\vec{X}}_{\nu_e})\right),
\ \vec{n} =
 \frac{\vec{v}_\mu-\vec{v}_{\nu_e}}{|\vec{v}_\mu-\vec{v}_{\nu_e}|}. \nonumber
\end{align}
Integrating over ${\vec X}_{\nu_e}$, the probability
 is written as
 \begin{align}
P^0= \Gamma T  =  &T\frac{(\sigma_\mu\sigma_{\nu_e})^\frac{3}{2}}{|\vec{v}_\mu-\vec{v}_{\nu_e}|(\sigma_\mu+\sigma_{\nu_e})}
\frac{2^4G_F^2}{E_\mu(2\pi)^7}
\int \frac{d\vec{p}_e d\vec{p}_{\nu_\mu}d\vec{p}_{\nu_e}}{E_e E_{\nu_\mu}E_{\nu_e}}
(p_\mu\cdot p_{\nu_e})(p_e\cdot p_{\nu_\mu})\nonumber \label{normal-term}\\
&\times\exp\left[-\frac{\sigma_\mu\sigma_{\nu_e}}{\sigma_\mu+\sigma_{\nu_e}}\delta \vec{p}^{\,2}\right]
\exp\left[{-\frac{\sigma_\mu+\sigma_{\nu_e}}{(\vec{v}_\mu-\vec{v}_{\nu_e})^2}\left( 
\delta p^0 - {\vec{v}_0\cdot\delta \vec{p}}
\right)^2}\right]
\end{align}
\cite{Ishikawa-Tobita-PTEP},  where $T=T_{\nu_e}-T_\mu$.
The condition  $T\ll \tau_\mu$ is satisfied in all  experiments and assumed 
in this paper. Other region $T\approx \tau_\mu$  is given in Appendix \ref{App-normal}.
\subsubsection{Spectral representation : finite-size corrections}
 The probability  from the large $t_1-t_2$ region in Eq.$(\ref{total-probability})$  is found from the expression   
\begin{align}
P=&\frac{2^5G_F^2}{({\sigma_\mu\sigma_{\nu_e}})^\frac{3}{2}E_\mu}
\int\frac{d\vec{X}_{\nu_e}d\vec{p}_{\nu_e}}{E_{\nu_e}(2\pi)^6}
\int d^4x_1d^4x_2\Delta_{e,\nu_\mu}(\delta x)e^{ip_{\nu_e}\cdot\delta x},
\nonumber\\
&\times \prod_{i}w(x_i,X_\mu;\sigma_\mu;\tau_{\mu})w(x_i,X_{\nu_e};\sigma_{\nu_e}), \delta x=x_1-x_2,
\label{total-probability1}\\
& \Delta_{e,\nu_\mu}(\delta x)=\frac{1}{(2\pi)^6}\int \frac{d\vec{p}_ed\vec{p}_{\nu_\mu}}{E_e
E_{\nu_\mu}}(p_\mu\cdot p_{\nu_e})
(p_e\cdot p_{\nu_\mu})e^{i(p_e+p_{\nu_\mu}-p_\mu)\cdot \delta x},\label{correlation-1}
\end{align}
where $\Delta_{e,\nu_\mu}(\delta x)$ has a short-range component and a long-range one.   The long-range component is extracted by  writing it with  an integral
 representation  of Jost, Lehmann, and Dyson \cite{JDL-Rep1,JDL-Rep2}, given in Appendix B, 
\begin{align}
& \Delta_{e,\nu_\mu}(\delta x) = \frac{p_\mu\cdot p_{\nu_e}}{2(2\pi)^2}\int_{m_e^2}dm^2\rho(m^2)iD^+(\delta t,\delta\vec{x};p_\mu,m), \\
&\rho(m^2)= m^2-2m_e^2+\frac{m_e^4}{m^2},\nonumber
\end{align}
Using the momentum $Q^{\rho}=p_e^{\rho}+p_{\nu_{\mu}}^{\rho}-p_{\mu}^{\rho}$, $iD^+(\delta t,\delta\vec{x};p_\mu,m)$ is expressed  as 
\begin{align}
iD^+(\delta t,\delta\vec{x};p_\mu,m)=\frac{1}{(2\pi)^3}\int d^4 Q\delta(Q^2-m^2)\theta(Q^0)e^{iQ\cdot \delta x},
\end{align}
and 
\begin{align}
 m\leq m_\mu& &\nonumber\\
&iD^+(\delta t,\delta\vec{x};p_\mu,m)=iD^+(\delta t,\delta\vec{x};p_\mu,m)^{(1)} + iD^+(\delta t,\delta\vec{x})^{(2)}+iD^+(\delta t,\delta\vec{x};p_\mu,m)^{(3)} \\
&iD^+(\delta t,\delta\vec{x};p_\mu,m)^{(1)} =\frac{1}{(2\pi)^3}\int_{Q^0=0}^{Q^0=p_\mu^0}
\frac{d\vec{Q}}{2Q^0}e^{iQ\cdot\delta x} \\
& iD^+(\delta t,\delta\vec{x})^{(2)} = \frac{i}{4\pi}\delta(\lambda)\epsilon(\delta t)&\\
&iD^+(\delta t,\delta\vec{x};p_\mu,m)^{(3)} = \sum_{n=0}^{\infty}\frac{1}{n!}\left(\frac{\partial}{\partial \tilde{m}^2}\right)^n\left(-ip_\mu\cdot\frac{\partial}{\partial \delta x}
\right)^ni\tilde{D}^+(\delta t,\delta\vec{x};i\tilde{m})
,&\label{correlation}\\
m_\mu < m& &\nonumber\\
&iD^+(\delta t,\delta\vec{x};p_\mu,m)=0,
\end{align}
where $\tilde{D}^+(\delta t,\delta\vec{x};i\tilde{m})$ is the sum of the
Bessel functions. The short-range components of these functions, which   
  give the probability  $\Gamma T$ from Appendix A and C,  were  computed
 in the previous sub-section.   
 
Now,   $P^{(d)}$ is computed from   $ iD^+(\delta t,\delta\vec{x})^{(2)}$. This is derived from   the states of $|{\vec Q}| \rightarrow \infty$, and  has the long-range component, when the series in $ iD^+(\delta t,\delta\vec{x};p_\mu,m)^{(3)}$ converges. This condition is reduced to  that of the momenta
\begin{align}
 2p_\mu\cdot p_{\nu_e}\leq m_\mu^2-m^2\label{conv-cond}, 
\end{align}
which agrees also with a causality condition that the light-cone singularity is in the physical region.   Due to the singular nature, the computation is made   in the
 configuration space. That is  similar to that  of  \cite{Ishikawa-Tobita-ANA} and the
 details are given in Appendix \ref{app-gtilde}.
For a muon of large $\sigma_{\mu}$, $\frac{(\vec{v}_\mu -
 \vec{v}_{\nu_e} )^2T^2}{4\sigma_{\mu}} \ll 1$,   the relevant term $ \tilde {g}(\omega_{\nu_e},T;\tau_\mu))$  in  Eq.$(\ref{lifetime-g})$ is written as   
$\tau_\mu({g}(\omega_{\nu_e},T;\tau_\mu)-{g}(\omega_{\nu_e},\infty;\tau_\mu))$. Then  $\tau_\mu\tilde{g}(\omega_{\nu_e},T;\tau_\mu)$ gives the slowly varying component, whereas ${g}(\omega_{\nu_e},\infty;\tau_\mu)$ is combined  with other short-range terms. 
Substituting $ \tilde {g}(\omega_{\nu_e},T;\tau_\mu))$, it follows     
\begin{align}
P^{(d)}=\frac{2G_F^2}{E_\mu}\int\frac{d\vec{p}_{\nu_e}p_\mu\cdot p_{\nu_e}}{E_{\nu_e}(2\pi)^5}\int dm^2\rho(m^2)
\left[
\sigma_{\nu_e} \mathcal{C}\tilde{g}(\omega_{\nu_e},T;\tau_\mu) \right],
\end{align}
where the integral over the invariant mass 
\begin{align}
 &\int_{m_e^2}^{m_\mu^2-2p_\mu\cdot p_{\nu_e}}dm^2\rho(m^2)
\simeq \frac{\left(m_\mu^2-2p_\mu\cdot p_{\nu_e}\right)^2}{2}\theta(m_\mu^2-2p_\mu\cdot p_{\nu_e}),
\end{align}
is substituted and $\mathcal{C}$ is constant in $p_{\nu_e}$ and given in Appendix
\ref{app-gtilde} for various cases. 

Finally, 
\begin{eqnarray}
& &P = T\Gamma + P^{(d)},\\
& &P^{(d)} = \mathcal{C}\frac{G_F^2}{E_\mu}\int\frac{d\vec{p}_{\nu_e}}{E_{\nu_e}(2\pi)^5}
(p_\mu\cdot p_{\nu_e})(m_\mu^2 - 2p_\mu\cdot p_{\nu_e})^2\theta(m_\mu^2-2p_\mu\cdot p_{\nu_e})
\sigma_{\nu_e}\tilde{g}(\omega_{\nu_e},T;\tau_\mu),\label{diff-rate}  \\
& &\Gamma = 
\int dE_{\nu_e}\frac{d\Gamma }{dE_{\nu_e}} = \int dE_{\nu_e}\frac{G_F^2}{2\pi^3}m_\mu^2E_{\nu_e}^2\left(1-\frac{2E_{\nu_e}}{m_\mu}\right). \label{asy-rate}
\label{asy-spec}
\end{eqnarray}
$\Gamma$ is in agreement with  the known rate.  $P^{(d)}$ is expressed, after
the tedious calculations, as
\begin{align}
 P^{(d)} &= \frac{G_F^2}{(2\pi)^5E_\mu}\left(\tilde{J}_1(p_\mu) + \tilde{J}_2(p_\mu)\right)\label{Prob-diff}\\
&= \frac{G_F^2\sigma_{\nu_e}\mathcal{C}}{(2\pi)^4E_\mu|\vec{p}_\mu|}\left[
\int_0^{E_{\text{min}}} dE_{\nu_e}F_1(E_{\nu_e})\tilde{g}(\omega_{\nu_e},T;\tau_\mu)+
\int_{E_{\text{min}}}^{E_{\text{max}}} dE_{\nu_e}F_2(E_{\nu_e})\tilde{g}(\omega_{\nu_e},T;\tau_\mu)
\right]\nonumber \\
F_1(E_{\nu_e})
&=E_{\nu_e}^2
\left[
E_{\nu_e}^2\left({p_\mu^+}^4 - {p_\mu^-}^4\right)
-\frac{4}{3}E_{\nu_e}m_\mu^2\left({p_\mu^+}^3 - {p_\mu^-}^3\right)
+\frac{m_\mu^4}{2}\left({p_\mu^+}^2 - {p_\mu^-}^2\right)
\right] \nonumber\\
F_2(E_{\nu_e}) &=E_{\nu_e}^2\Biggl[
E_{\nu_e}^2\left\{{p_\mu^-}^4 - \left(\frac{m_\mu^2}{2E_{\nu_e}}\right)^4\right\}
-\frac{4}{3}E_{\nu_e}m_\mu^2\left\{{p_\mu^-}^3 - \left(\frac{m_\mu^2}{2E_{\nu_e}}\right)^3\right\}\nonumber\\
&+\frac{m_\mu^4}{2}\left\{{p_\mu^-}^2 - \left(\frac{m_\mu^2}{2E_{\nu_e}}\right)^2\right\}
\Biggr],p_{\mu}^{\pm}=E_{\mu} \pm |{\vec p}_{\mu}|.\nonumber
\end{align}
In the above equations, the integral over the angle $\theta$ between the
momenta of $\mu$ and $\nu_{e}$ is made following the condition Eq. \eqref{conv-cond},
\begin{align}
 \cos\theta \leq \cos\theta_c=\frac{E_\mu}{|\vec{p}_\mu|} - \frac{m_\mu^2}{2|\vec{p}_\mu||\vec{p}_{\nu_e}|}.
\end{align}
The $P^{(d)}$ depends on  the neutrino absolute
mass through $\tilde{g}(\omega_{\nu_e},T;\tau_\mu) $. 
\subsection{Energy spectrum}
\subsubsection{Diffraction component}
The energy spectrum of the diffraction component is given from Eq.$( \ref{Prob-diff})$ by
\begin{align}
\frac{dP^{(d)}}{dE_{\nu_e}}
=&
\frac{G_F^2\sigma_{\nu_e}\mathcal{C}}{(2\pi)^4E_\mu|\vec{p}_\mu|}
[
F_1(E_{\nu_e})\tilde{g}(\omega_{\nu_e},T;\tau_\mu)\theta(E_\text{min}-E_{\nu_e}) \nonumber\\
&+ F_2(E_{\nu_e})
\tilde{g}(\omega_{\nu_e},T;\tau_\mu)\theta(E_\text{max}-E_{\nu_e})
\theta(E_{\nu_e}-E_\text{min})
]. \label{nue-spec-low}
\end{align}
For the low-energy muon $|\vec{p}_\mu|\ll E_\mu\sim m_\mu$,
\begin{align}
& F_1(E_{\nu_e}) = |\vec{p}_\mu|\frac{m_\mu^7}{2}x^2(1-x)^2,\ F_2(E_{\nu_e})=0,\ x = \frac{2E_{\nu_e}}{m_\mu}.
\end{align}

\begin{figure}[t]
\includegraphics[angle=-90,scale=.32]{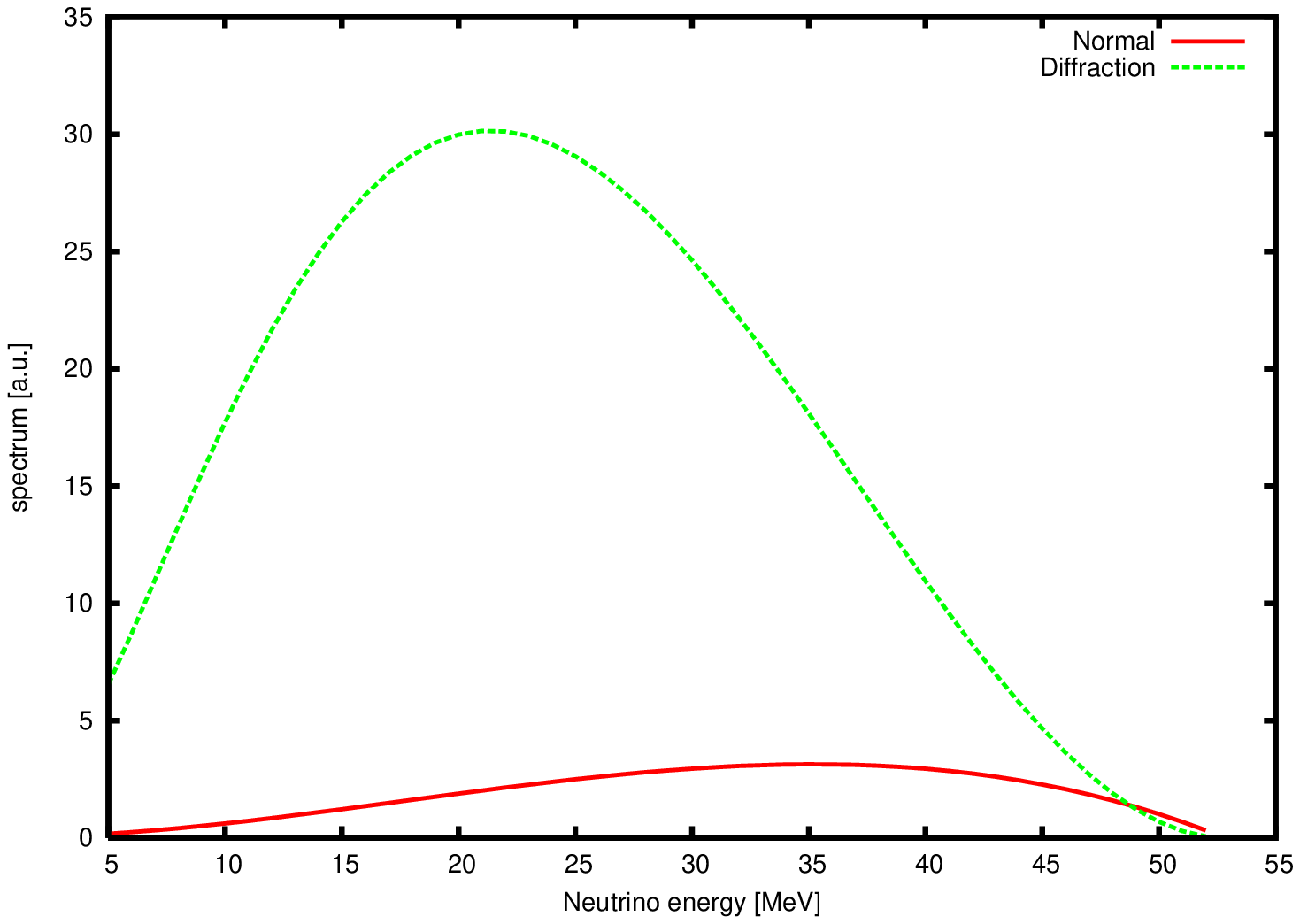}
\includegraphics[angle=-90,scale=.32]{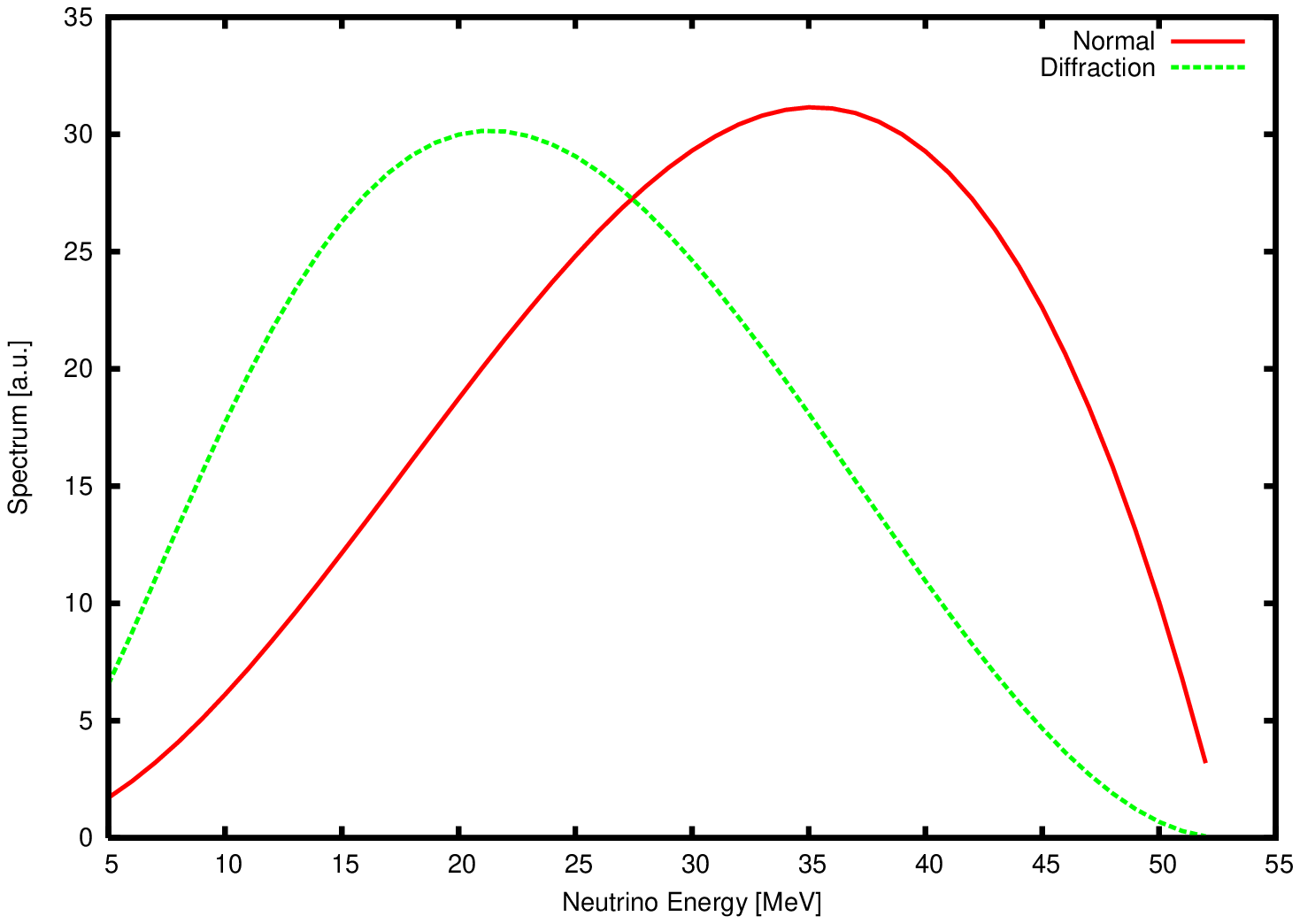}
\caption{(Color online) $\nu_e$ spectrum in $\mu$DAR.
 The red curve shows the normal component given in  Eq. \eqref{asy-spec-p}, while the green curve shows the diffraction component given in Eq. \eqref{nue-spec-low}.
$m_{\nu_e}= 0.08$ eV, and $2\sigma_{\nu_e} = 12^\frac{2}{3}/m_\pi^2$ (${}^{12}C$ carbon target) were used in the numerical computation. Here, the detector is located  at $L=0$. The left figure corresponds to $cT=cT_D=1$, the right figure to $cT=cT_D=10$ m.
}
\label{fig:spec-nue}
\end{figure}

Eq. $(\ref{nue-spec-low})$ and the spectrum at asymptotic region, Eq. $(\ref{asy-rate})$ are  given in Fig. \ref{fig:spec-nue} for a suitable value of $\sigma_{\nu_e}$. At $cT=1$ m, the diffraction term $P^{(d)}$ is  considerably larger than the normal term in magnitude, and is about the same at $cT=10$ m. Moreover, the peak     
shifts to   lower energies by approximately $20$ MeV. This arises  from the fact that the dominant part of $P^{(d)}$ comes from the large momentum states and is derived from those  satisfying the condition in Eq. \eqref{conv-cond}. Currently, there are no precise data even for the electron in $x<1/2$, and it would be interesting to confirm this component. 
The effect is reduced if the muon is  a small wave packet. A low-energy negative muon in matter is trapped in an atom and forms a bound state with a small wave function; thus, the  effect is reduced.  

\subsubsection{Normal component}
From Eqs. $(\ref{asy-rate})$, and (A2), the energy spectrum  of the normal component of the muon decay at rest ($\mu$DAR) in the asymptotic region is written as
\begin{align}
 \frac{dP^0}{dx} = \frac{G_F^2m_\mu^5}{2(2\pi)^3}x^2(1-x)\tau_\mu(1-e^{-\frac{T_D}{\tau_\mu}}),\label{asy-spec-p}
\end{align}
where $T_D$ is the depth of the detector, which, for $\mu$DAR, determines the time width.
For the muon decay in flight($\mu$DIF), a decay region $cT$ restricts the time width where $\mu$ exists. 
Those with suitable values,  according to the experimental conditions, are used.
The normal component of the transition rate or probability  is independent of the size of the wave packets, from the completeness, but the energy spectrum depends on it and varies with the size. The spectrum for a plane wave given in Eq. \eqref{asy-spec} was confirmed in the upper energy region $x\geq 1/2$ of the electron spectrum.

\subsection{ Flavor mixing }
\subsubsection{Normal term}
With three neutrino mass eigenstates of $m_{\nu_i}$, $i=1-3$, and $U_{\alpha,i}$, 
the flavor neutrino fields $\nu_l(x)$ in Eq. \eqref{shrodinger-eq} are the linear combination of the fields of three $\nu_i(x)$ having mass $m_i$   as
\begin{align}
\nu_l(x) = \sum_{i}U_{l,i}\nu_i(x),\ l = e,\ \mu, \tau,
\end{align}
where the best-fit values of the mixing angles given in Ref. \cite{pdg}
\begin{align}
&\sin^2\left(\theta_{12}\right) = 0.304\pm 0.014\\
&\sin^2\left(\theta_{23}\right)=0.51\pm0.05\text{ (normal hierarchy)},\ \sin^2\left(\theta_{23}\right)=0.50\pm0.05\text{ (inverted)},\nonumber \\
&\sin^2\left(\theta_{13}\right)=(2.19\pm0.12)\times 10^{-2} \nonumber
\end{align} 
are used, and a CP violation phase $\delta_{CP}= 0$ is assumed.
The amplitudes for the anti-neutrino of flavor $\alpha$ to be detected in $\mu^+$ decay ($\mu^+\to \underline{\bar{\nu}_{\mu}} + e^+ + {\nu}_e$) and for the neutrino of flavor $\alpha'$ in $\mu^+$ decay ($\mu^+ \to \bar{\nu}_\mu + e^+ + \underline{\nu_e}$) are 
\begin{align}
& \mathcal{M}_{\alpha,\mu} = \sum_{i}U_{\alpha,i}\mathcal{M}(\mu^+,i)U_{\mu,i}^*,\\
&\mathcal{M}'_{\alpha',e} = \sum_{i}U_{\alpha',i}\mathcal{M}'(\mu^+,i)U_{e,i}^{*}. \nonumber
\end{align}
The probability is the square of the amplitude, 
\begin{eqnarray}
P^0_{\alpha, \beta}=|\mathcal{M}_{\alpha,\beta}|^2, \label{single-oscillation}
\end{eqnarray}
and is a function of $\sin^2 \frac{\Delta m_{\nu}^2 T}{4E_{\nu}}$ in the particle zone. For the two-flavor case,  that is proportional to  $4 (\cos \theta \sin \theta )^2 \sin^2 \frac{\Delta m_{\nu}^2 T}{4E_{\nu}}$.   
\subsubsection{Diffraction term}
 The probability amplitude in the wave zone is described by the  wavefunction of the correlation length $L_c$, and  depends on the absolute neutrino mass and   
 the flavor mixing matrix. The probability  
  for a flavor change from $\alpha$ to $\beta$ are expressed by   a new universal function 
$\tilde{g}_{\alpha,\beta}(\omega_{\nu_\alpha},T;\tau_\mu)$ as
\begin{align}
P_{\alpha \beta}^{(d)}  \propto \tilde{g}_{\alpha,\beta}(\omega_{\nu_\beta},T;\tau_\mu)
= \sum_{i,j}U_{\beta,i}U^*_{\alpha,i}U_{\beta,j}^*U_{\alpha,j}
\tilde{g}(\omega_i,\omega_j,T;\tau_\mu).\label{new-gtilde}
\end{align}
These reflect the enhanced probability  in short-distance regions, \cite{Ishikawa-tobita-neutrinomass}, accordingly the flavor change  can be much larger than that of the normal term.
They are    studied  in  Appendix \ref{app-gtilde}. Some specific features are the following:

(1) $\tilde{g}_{e,e}(\omega_{\nu_e})$ and $\tilde{g}_{\mu,e}(\omega_{\nu_e})$ are
 almost constant at $cT<100$ m, and decrease uniformly at larger $T$. The values are sensitive to the absolute  neutrino mass.

 (2) $\tilde{g}_{\mu,e}(\omega_{\nu_e})$ is significantly smaller than 
$\tilde{g}_{e,e}(\omega_{\nu_e})$, but not zero. That magnitude is proportional to $\tau_{\mu}(\omega_i+\omega_j)$ at a small $\tau \omega_i$ region. A new flavor changing effect arises 
in the short-distance region. 
\subsubsection{ Flavor change in the short-distance region}
The probability is the sum of those of Eqs.$(\ref{single-oscillation})$ and $(\ref{new-gtilde})$, and in the short-distance region, 
\begin{eqnarray}
P=C_0(   4 (\cos \theta \sin \theta )^2  ( \frac{\Delta m_{\nu}^2 T}{4E_{\nu}})^2+ \tilde{g}_{\alpha,\beta}(\omega_{\nu_\beta},T;\tau_\mu)),
\end{eqnarray}
for two flavor, where $C_0$ is a constant. The first term is proportional to the square of small quantity, but the second one is proportional to it.   Thus the probability is determined by the second term. 

The flavor change is determined by the diffraction in short-distance regions,  but that is by 
the flavor oscillation   in long-distance regions.  
\section{Implications to neutrino experiments}
Been determined by the neutrino wavefunction, the standard formula depends on single-body properties such as the mass-squared difference $\Delta m_{\nu_{21}}^2=m_{\nu_2}^2-m_{\nu_1}^2$ and mixing angles.  Now, $P^{(d)}$ is determined by
 the many-body wavefunction expressing the whole process of the long distance correlation of the length, $L_c$, which  is much longer than the typical detector size, and the final states from FQM. Accordingly,
  $P^{(d)}$    varies with   the  geometry and set up of the experiments.  $\Gamma T$ is equivalent in LSND and KARMEN, but $P^{(d)}$ is very different.     
\subsection{Boundary conditions  for muon decays}
In ground experiments, $\mu^+\ (\mu^-)$ is produced in $\pi^+\ (\pi^-)$ decay simultaneously with a ${\nu_\mu}\ (\bar{\nu}_\mu)$, and decays to $e^+\ (e^-)$, $\bar{\nu}_\mu\ (\nu_\mu)$,  and $\nu_e \ (\bar{\nu}_e)$. There are two typical types of experiments; one is the accelerator experiment that uses a neutrino beam from pion decay.
In this case, both decays of high-energy $\pi$ and $\mu$ are sources of neutrinos.  
The other is the experiment that observes the $\bar{\nu}_e$ or $\nu_\mu$ appearance in $\mu^+$ decay. In the latter, $\mu^+$ is extracted and used as a source of neutrino, and $\pi$ is not involved.

The   nuclei wavefunctions  in solid  have  ranges in space of the order 
of $10^{-15}$ m, while the electrons  bound in the atoms have ranges in
space of the order of $10^{-10}$ m.  The neutrino and anti-neutrino interact differently with the matter  \cite{Ishikawa-tobita-wavepacket-letter}. 
For  $\nu_e$ detected by the ${}^{12}C + \nu_e \to {}^{12}N_{g.s.}  + e^-$ process, that is the nucleus size
\begin{align}
 2\sigma_{\nu_e} = \frac{12^\frac{2}{3}}{m_\pi^2},
\end{align}
of ${}^{12}C$.  
For $\bar{\nu}_e$ detected by the inverse beta decay and delayed 
signal of neutron capture, the lightest nucleus $H$, i.e., the proton, and other
nuclei are the targets.  When $C_nH_{2n+2}$ is used for the scintillator,
 the size of the wave packet  calculated from the ratio of the proton between $C$ and $H$ is 
\begin{align}
 2\sigma_{\bar{\nu}_e} = \frac{3}{4}\frac{12^\frac{2}{3}}{m_\pi^2}  + \frac{1}{4}\left(\frac{m_ea_\infty}{m_p+m_e}\right)^2,
\end{align}
where $m_e$ and $m_p$ are the masses of the electron and proton, respectively, and $a_\infty$ is the used Bohr radius. For the detector, which consists of a mixture of several materials, the wave packet size becomes the value averaged over the abundance ratio of 
the materials.

The  incoming muon   is prepared in the apparatus. That  propagates in
matter or space with a  mean free path  
\begin{align}
 l_\mu = \frac{1}{n\sigma_\text{cross}},
\end{align}
where $n$ and  $\sigma_\text{ cross}$ are the density of the scatterers and  the
cross section, respectively. They are determined by  the  transition rate and 
average lifetime  of the velocity  $v$ 
\begin{align}
& \Gamma = n\sigma_\text{cross}v = \frac{v}{l_{\mu}},\ \tau_\text{int} = \frac{1}{\Gamma}.
\end{align}
This is summarized in Ref. \cite{pdg}.

A  pion is produced by a proton collision, hence,  its range in space
covered by the wave function is 
estimated from that of a proton.   A proton mean free path of 1 GeV/$c$ was estimated as $l_\text{proton} = 50 - 100$ cm, and, at a lower momentum of 2 MeV/$c$, $l_\text{proton} = 10$ cm.
Analyzing the decay processes, we found that the range in space covered
by the pion wave function at momentum 1 GeV/$c$, or larger, is $\delta
x_\pi \approx 40 $-$100$ cm, and  that by of the
muon wave function of the  momentum  around 1GeV/$c$  is $\delta   x_\mu \approx 40 
$-$100$ cm. Thus, $\sigma_{\mu} =\pi(0.4-1.0)^2
  \text{ m}^2$, and $\sigma_{\mu} =\infty$ is a good approximation. 

$ \mu^{+}$ and $\mu^{-}$ decay in a symmetric manner in vacuum   with the mean
lifetime at rest 
 $\tau_\mu = 2.2 \times 10^{-6} \text{ sec}$.
 However, due to the different charges they interact with atoms differently.
For the stopped $\mu^{+}$, the wavefunctions  in the periodic potential
of a solid  are extended waves with continuous energies. They  are plane
waves with phase shifts, and  $\mu^{+}$ at rest, $v=0$,  is described 
by a wavefunction covering 
a large  range in space.  
The stopped $\mu^{-}$ can form bound states of localized  wavefunctions
and  discrete  energies. Thus, the wave packet for  $\mu^{-}$ at rest 
has a short range. 
Consequently, $\mu^+$DAR is expressed by  plane 
waves, while  $\mu^-$DAR is expressed by  a wave
packet of the small range. 
Both $\mu^\pm$ DIF are produced from decays 
of $\pi^\pm$ and retain coherence of the same range of $\pi^\pm$. Therefore, it is a good approximation to treat $\mu^\pm$DIF as plane waves \cite{Ishikawa-Tobita-ANA}.
Hereafter, we focus only on the $\mu^+$ decay and study   the above two types of experiments with  $\sigma_{\mu} =\infty$.

\subsection{$\nu_e$ and $\bar{\nu}_\mu$ in $\mu^+$ decay}
\subsubsection{$\mu^+$ decay at rest ($\mu^{+}$DAR)} 
\begin{figure}[t]
\includegraphics[scale=.42,angle=-90]{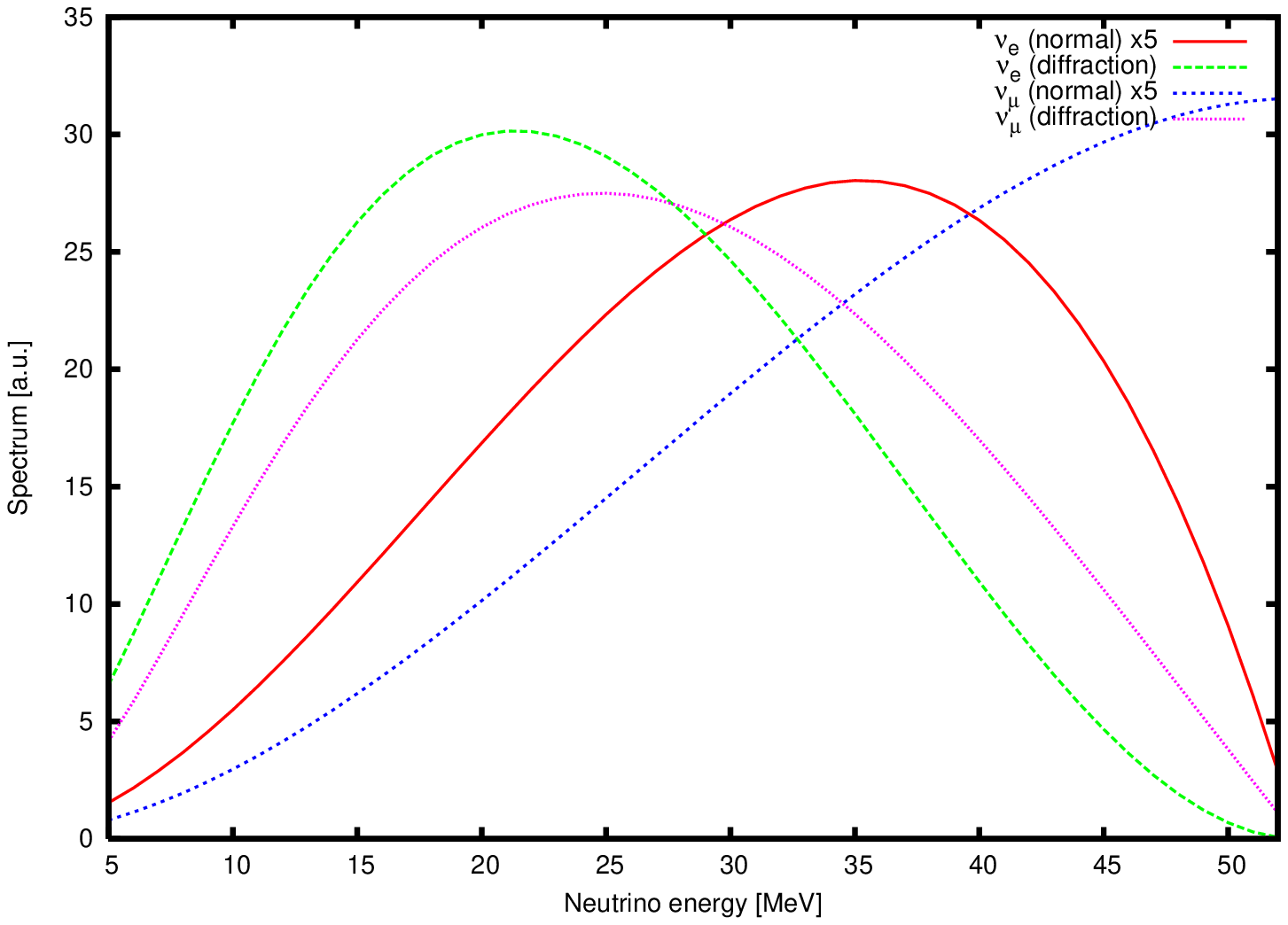}
\caption{(Color online) Spectra of normal and diffraction terms for $\nu_e$ and $\bar{\nu}_\mu$ in the $\mu^+$ decay at rest. The spectra of the normal terms for $\nu_e$ (red) and $\bar{\nu}_\mu$ (blue) are different from those of $\nu_e$ (green) and $\bar{\nu}_\mu$ (magenta). These properties can be used to eliminate background events
from $\mu^-$ decays, etc. $m_{\nu_h} = 0.08$ eV, $\sigma_\nu = 12^\frac{2}{3}/m_\pi^2$,   
$cT=cT_D=1.0$ m, and inverted hierarchy are assumed in the numerical calculation.}
\label{fig:spectra-dar}
\end{figure}
In $\mu^+$DAR,  the neutrino has the energy less than $m_{\mu}c^2$ and 
 can produce the electron but not the muon. $\nu_e$ is detectable 
but  $\bar{\nu}_\mu$  in charged current interactions is not. Nevertheless, the  $\bar{\nu}_\mu$ spectrum
is important   to distinguish background events of the flavor
oscillation phenomena.   
The   energy spectra of the normal and diffraction terms 
for $\bar{\nu}_\mu$ are 
\begin{align}
&\frac{dP^0}{dE_{\bar{\nu}_\mu}}=\frac{G_F^2m_\mu^2}{12\pi^3}E_{\bar{\nu}_\mu}^2\left(3-4\frac{E_{\bar{\nu}_\mu}}{m_\mu}\right)
\tau_\mu\left(1 - e^{-\frac{T_D}{\tau_\mu}}\right)\label{normal-spec-numubar-dar},\\
& \frac{dP^{(d)}}{dE_{\bar{\nu}_\mu}} = \frac{G_F^2m_\mu^2}{12\pi^3}
\frac{E_{\bar{\nu}_\mu}^2 m_\mu^2 \sigma_{\bar{\nu}_\mu}}{4\pi}
\left(1 - 2\frac{E_{\bar{\nu_\mu}}}{m_\mu}\right)
\left(5-6\frac{E_{\bar{\nu}_\mu}}{m_\mu}\right)\tau_\mu\tilde{g}(\omega_{\bar{\nu}_\mu},T;\tau_\mu)
\label{diff-spec-numubar-dar},
\end{align} 
and their  ratio is
\begin{align}
&R(E_{\bar{\nu}_\mu})= \frac{dP^{(d)}}{dE_{\bar{\nu}_\mu}}/ \frac{dP^{0}}{dE_{\bar{\nu}_\mu}}\nonumber \\
&=\frac{m_\mu^2\sigma_{\bar{\nu}_\mu}}{4\pi}\frac{\left(1-2\frac{E_{\bar{\nu}_\mu}}{m_\mu}
\right)\left(5-6\frac{E_{\bar{\nu}_\mu}}{m_\mu}\right)}{3-4\frac{E_{\bar{\nu}_\mu}}{m_\mu}}
\frac{\tilde{g}(\omega_{\bar{\nu}_\mu},T;\tau_\mu)}{1 -
 e^{-\frac{T_D}{\tau_\mu}}}.
\label{eq:ratio-numu-at-rest}
\end{align}

The  energy spectra of the diffraction and normal 
terms for $\nu_e$ are given in Eqs. \eqref{nue-spec-low}, and
\eqref{asy-spec-p} and their ratio is
\begin{align}
 R(E_{\nu_e}) = \frac{\sigma_{\nu_e}m_\mu^2\left(1-\frac{2E_{\nu_e}}{m_\mu}\right)}
 {4\pi\left(1- \exp[-T_D/\tau_\mu]\right)}
\tilde{g}_{e,e}(\omega_{\nu_e},T;\tau_\mu), \label{rate-mup-DAR}
\end{align}
Figure \ref{fig:spectra-dar} shows the spectra obtained from Eqs. \eqref{nue-spec-low}, \eqref{asy-spec-p}, \eqref{normal-spec-numubar-dar}, and \eqref{diff-spec-numubar-dar}. The distinctive spectra with  peaks in the lower energy regions and the unique property that the magnitude  varies with  $T$ and $\sigma_{\nu_e},$ facilitate their differentiation from background events. 

\begin{figure}[t]
\includegraphics[scale=.42,angle=-90]{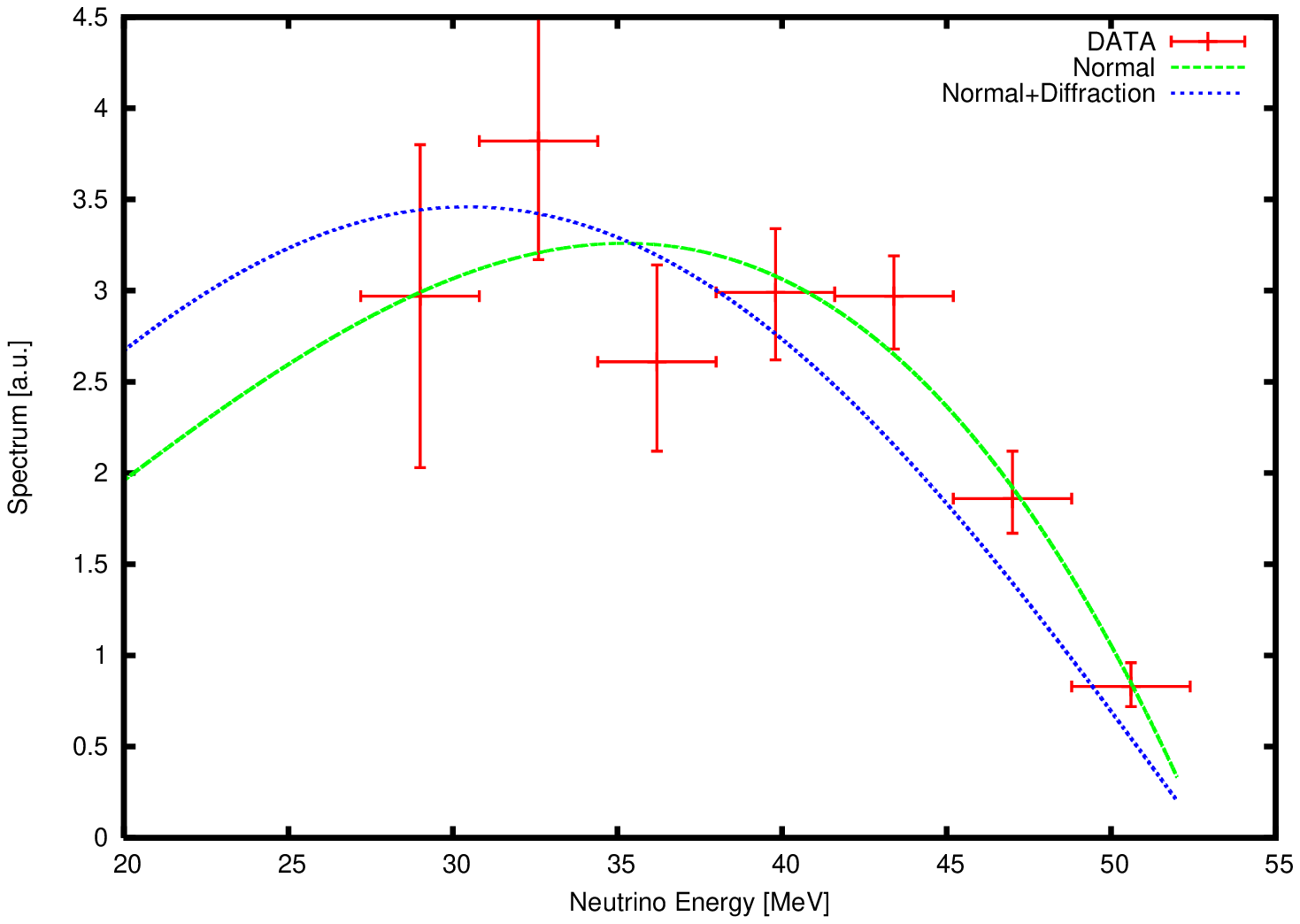}
\caption{(Color online) The spectra of normal and diffraction terms for $\nu_e$ are compared with KARMEN data.  The green curve indicates the $\nu_e$  spectrum with normal term, while the blue curve indicates the sum of normal and diffraction terms; the crosses indicate the KARMEN experiment. Suitable  parameters are chosen and relative magnitudes are shown.}
\label{fig:spectra-ne}
\end{figure}

According to Fig.\ref{fig:spectra-dar}, the ratios in Eqs. \eqref{eq:ratio-numu-at-rest} and \eqref{rate-mup-DAR}  are approximately equal to 5 at $cT=1$ m with  $m_{\nu_h}=0.08$ eV of the inverted hierarchy.
This value is quite large compared with that of the $\pi$ decay \cite{Ishikawa-tobita-neutrinomass}: if it is possible to identify $\nu_e$ from $\mu^+$DAR and collect 
sufficient statistics, it may be feasible to observe  the excess of the $\nu_e$ flux
and measure the absolute neutrino mass. The spectrum 
 from  KARMEN experiment is compared with the normal and diffraction terms   in Fig. \ref{fig:spectra-ne}.   The
 statistics are insufficient, and both theories are not in-consistent with  the experimental results.  
\subsubsection{
$\mu^{+}$ decay in flight ($\mu^+$DIF)}
The energy spectra of $\nu_e$ in $\mu^+$DIF are
\begin{align}
 \frac{d^2P^{0}_{\nu_e}}{dE_{{\nu}_e}d\cos\theta}  
= &\frac{2G_F^2E_{\nu_e}^2}{(2\pi)^3E_\mu}{(E_\mu - p_\mu\cos\theta)(m_\mu^2-2p_\mu\cdot p_{\nu_e})}
\gamma\tau_\mu \left(1 - \exp\left[-\frac{T}{\gamma\tau_\mu}\right]\right),\label{eq:nue-normal-spectrum}\\
   \frac{d^2P^{(d)}_{{\nu}_e}}{dE_{\nu_e}d\cos\theta}=&
\frac{G_F^2E_{\nu_e}^2}{(2\pi)^4E_\mu}(E_\mu-p_\mu\cos\theta)(m_\mu^2 - 2p_\mu\cdot p_{\nu_e})^2
\sigma_{\nu_e}\gamma\tau_\mu\tilde{g}_{e,e}(\omega_{\nu_e},T;\gamma\tau_\mu)\label{eq:nue-diffraction-spectrum},
\end{align}
where $\theta$ is an angle between $\vec{p}_\mu$ and $\vec{p}_{\bar{\nu}_\mu}$; and $\gamma = E_\mu/m_\mu$;
\begin{figure}[t]
 \includegraphics[scale=.42,angle=-90]{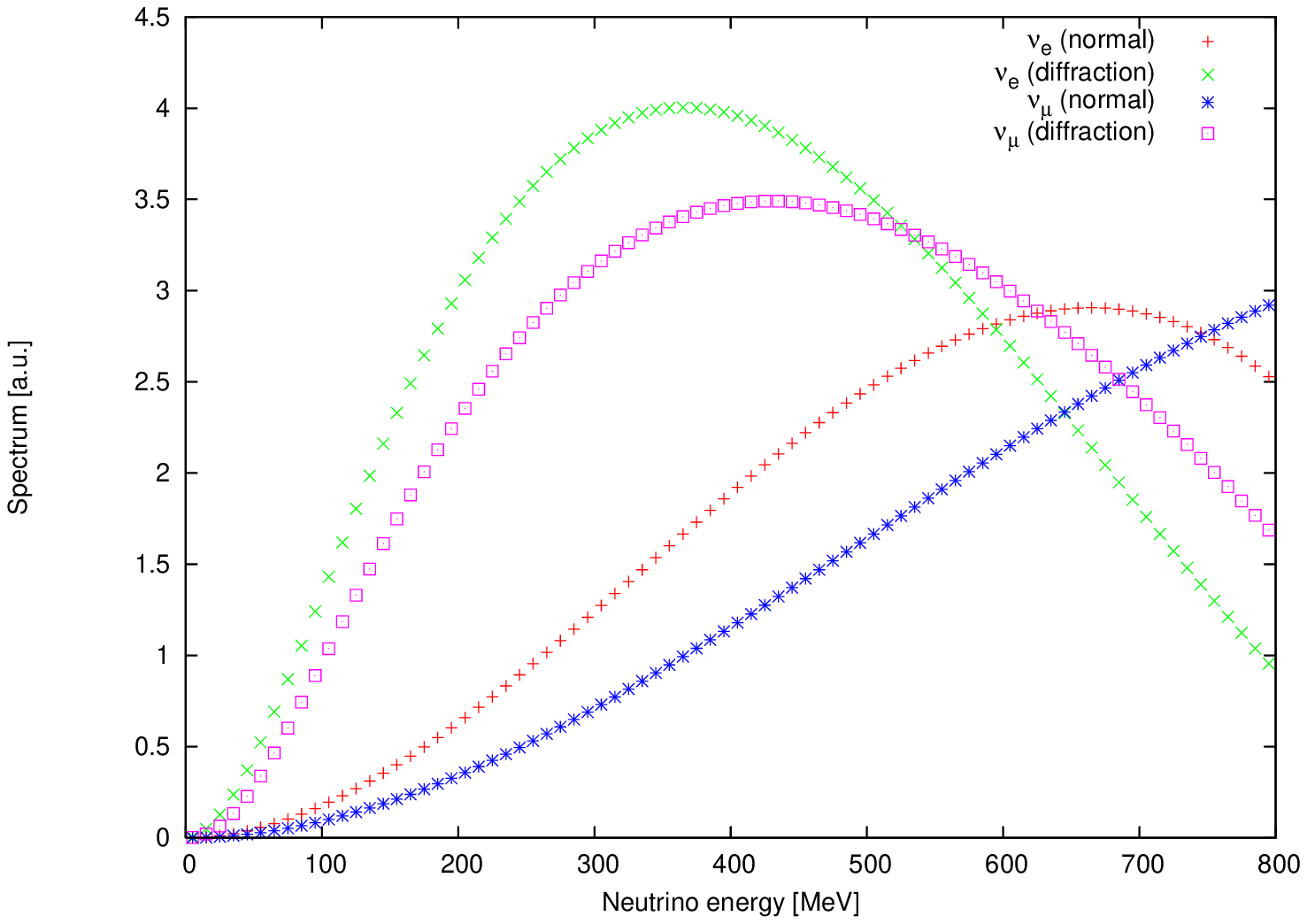}
 \caption{(Color online) Spectra of the normal and diffraction terms for $\nu_e$ and $\bar{\nu}_\mu$ in $\mu^+$DIF. The spectra of the normal terms for $\nu_e$ (red) and $\bar{\nu}_\mu$ (blue) are different from those of $\nu_e$ (green) and $\bar{\nu}_\mu$ (magenta). These properties can be used to eliminate background events from $\mu^-$ decays, etc. $m_{\nu_h} = 0.08$ eV, $\sigma_\nu = 12^\frac{2}{3}/m_\pi^2$, $cT=200$ m, $\cos\theta=1$, and $E_\mu=1$ GeV were used, and inverted hierarchy was assumed in the numerical calculation.}
\label{fig:spectrum-1GeV}
\end{figure}
The energy spectra of $\bar{\nu}_\mu$  are 
\begin{align}
& \frac{d^2P^{0}_{\bar{\nu}_\mu}}{dE_{\bar{\nu}_\mu}d\cos\theta} 
= \frac{G_F^2E_{\bar{\nu}_\mu}^2}{24\pi^3E_\mu}\left[
(E_\mu-p_\mu\cos\theta)(3m_\mu^2-4E_{\bar{\nu}_\mu}(E_\mu-p_\mu\cos\theta))\right]
\nonumber\\&\times
{\gamma\tau_\mu}\left(1 - \exp\left[-\frac{T}{\gamma\tau_\mu}\right]\right),\label{eq:numu-normal-spectrum}\\
&   \frac{d^2P^{(d)}_{\bar{\nu}_\mu}}{dE_{\bar{\nu}_\mu}d\cos\theta}=
\frac{G_F^2}{24\pi^3E_\mu}
\frac{E_{\bar{\nu}_\mu}^2}{4\pi}
(E_\mu-p_\mu\cos\theta)
(m_\mu^2-2E_{\bar{\nu}_\mu}(E_\mu-p_\mu\cos\theta))\nonumber\\
&\times(5m_\mu^2 - 6E_{\bar{\nu}_\mu}(E_\mu-p_\mu\cos\theta))
\gamma\tau_\mu\sigma_{\bar{\nu}_\mu}\tilde{g}_{\mu,\mu}(\omega_{\bar{\nu}_\mu},T;\gamma\tau_\mu).
\label{eq:numu-diffraction-spectrum}
\end{align}

The ratios between the normal and diffraction terms for $\nu_e$ and $\bar{\nu}_\mu$ are
\begin{align}
 R(E_{\nu_e},\cos\theta) = &\frac{\sigma_{\nu_e}}{4\pi}(m_\mu^2-2E_{\nu_e}(E_\mu-p_\mu\cos\theta))
\frac{\tilde{g}_{e,e}(\omega_{\nu_e},T;\gamma\tau_\mu)}
{1 - \exp[-T/\gamma\tau_\mu]}
,\label{eq:nue-ratio-dif}\\
 R(E_{\bar{\nu}_\mu},\cos\theta) =
&\frac{\sigma_{\bar{\nu}_\mu}}{4\pi}\frac{(m_\mu^2-2E_{\bar{\nu}_\mu}(E_\mu-p_\mu\cos\theta))
(5m_\mu^2 - 6E_{\bar{\nu}_\mu}(E_\mu-p_\mu\cos\theta))}
{3m_\mu^2 - 4E_{\bar{\nu}_\mu}(E_\mu-p_\mu\cos\theta)}
\frac{\tilde{g}_{\mu,\mu}(\omega_{\bar{\nu}_\mu},T,\gamma\tau_\mu)}
{(1-\exp[-T/\gamma\tau_\mu])}\label{eq:numu-ratio-dif}.
\end{align}

The spectra obtained from Eqs. \eqref{eq:nue-normal-spectrum}--\eqref{eq:numu-diffraction-spectrum} are shown in Fig. \ref{fig:spectrum-1GeV}. They indicate that 
$R(E_{\bar{\nu}_\mu})$ and $R(E_{{\nu}_e})$ for on-axis $\bar{\nu}_\mu$ and ${\nu}_e$ are approximately equal to one with $E_\mu = 1$ GeV, $m_{\nu_h}=0.08$ eV with inverted hierarchy,  and $\sigma_{\nu}$ of  nuclear size at $cT=200$ m. 
As this effect is clear and unique, this  may be observed in high-energy neutrino experiments, even at small neutrino flux. In the next section, these findings will be compared with the existing experimental results.

\subsubsection{Excess of electron neutrino  in accelerator  experiments}
The accelerator experiments use a neutrino beam  produced by $\pi$ decays. In $\pi^+$ decay, $\nu_e$ is produced by the following processes
\begin{align}
 \pi^+ \to &\mu^+ + \nu_\mu,\\
&\mu^+ \to e^+ + \bar{\nu}_\mu + \underline{\nu_e},\label{mode:mu-to-nue}\\
       \to &e^+ + \underline{\nu_e},\label{mode:pi-to-nue}
\end{align}
and has  two sources.  $P^{(d)}$ in the   $\nu_e$ mode is not suppressed, because  the helicity-suppression works only to $\Gamma T$. That 
 has been compared with the existing data of MiniBooNE in \cite{Ishikawa-tobita-neutrinomass,Miniboone-nue},  and all corrections, including those from the  muon decay,  are compared here. Lifetime of $\mu^+$ and  $\pi^+$  are also included. 
\begin{figure}[t]
 \includegraphics[scale=.42,angle=-90]{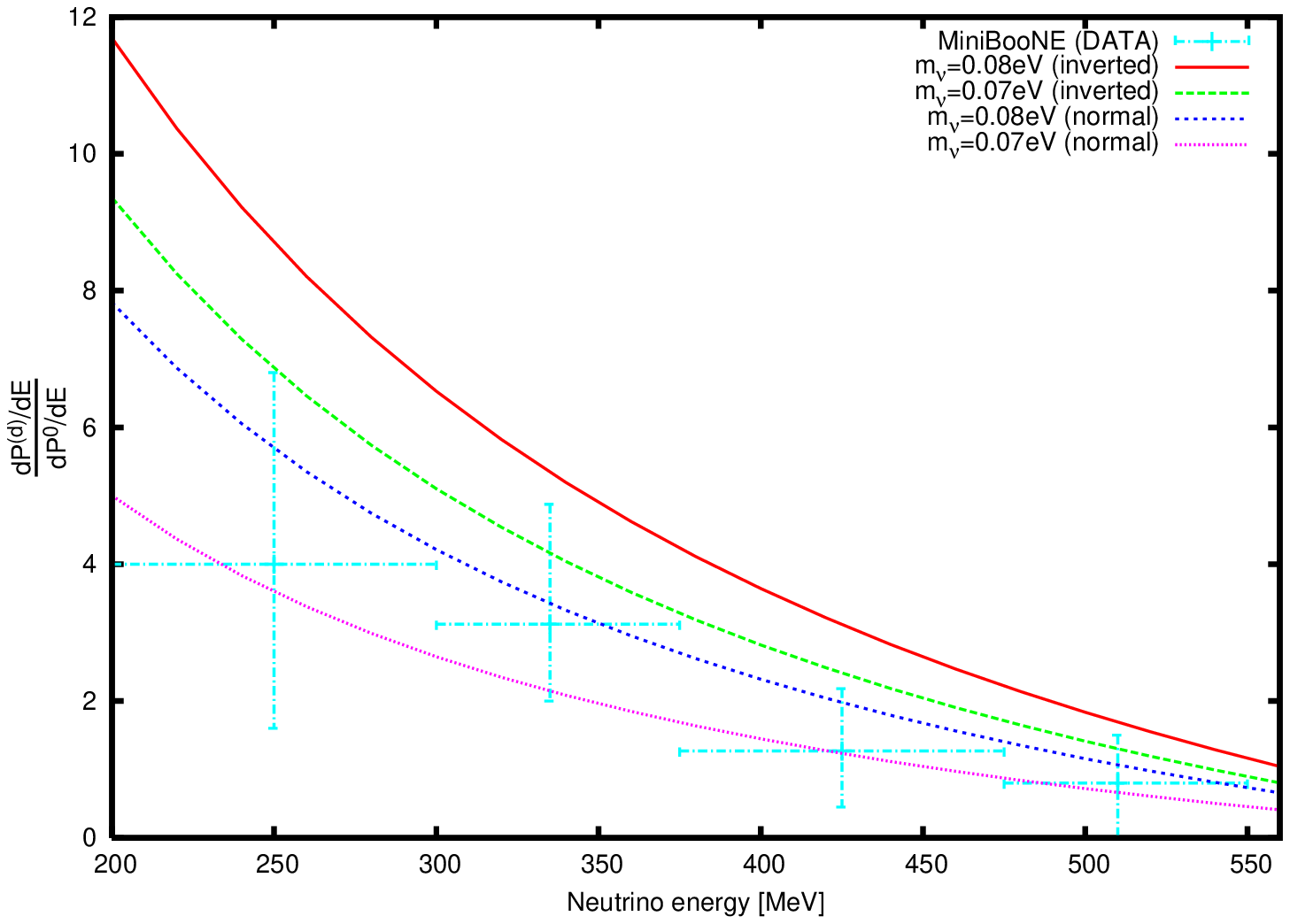}
\caption{(Color online) $P^{(d)}_{\nu_e}/P_{\nu_e}^{(0)}(\mu)$ is  compared with the MiniBooNE data (light-blue) including statistic and systematic errors \cite{Miniboone-nue}. In the numerical calculation, $m_{\nu_h} = 0.07 \text{ eV (green: inverted, magenta: normal)}$, $m_{\nu_h}= 0.08\text{ eV (red: inverted, blue: normal)}$,
$E_\mu=670$ MeV, and $E_\pi=1.15$ GeV are used.}
\label{fig:miniboone-nue}
\end{figure}

We compare the result $\Gamma T+P^{(d)}$ with  the MiniBooNE.  
$P^{(d)}$ of   $\nu_e$ in Eq. \eqref{mode:pi-to-nue}, from  Ref. \cite{Ishikawa-tobita-neutrinomass}, and   that of $\nu_e$ from $\mu^+$ of 
Eq. \eqref{mode:mu-to-nue}, is given in Fig. \ref{fig:miniboone-nue}. Figure \ref{fig:miniboone-nue} indicates that  our numerical results for the ratio between normal and diffraction modes are in agreement with the data.
Furthermore, this shows that $P^{(d)}$  in the $\mu^+$ decay dominates in $\nu_e$ events of MiniBooNE experiments, and is sensitive to the absolute mass value of the neutrino and the mass hierarchy. Our results are consistent within experimental uncertainties and Monte Carlo simulations with absolute neutrino mass values of $m_{\nu_h} = 0.07-0.08$ eV. This is consistent with our previous results \cite{Ishikawa-tobita-neutrinomass}.
\begin{figure}[t]
 \includegraphics[scale=.32,angle=-90]{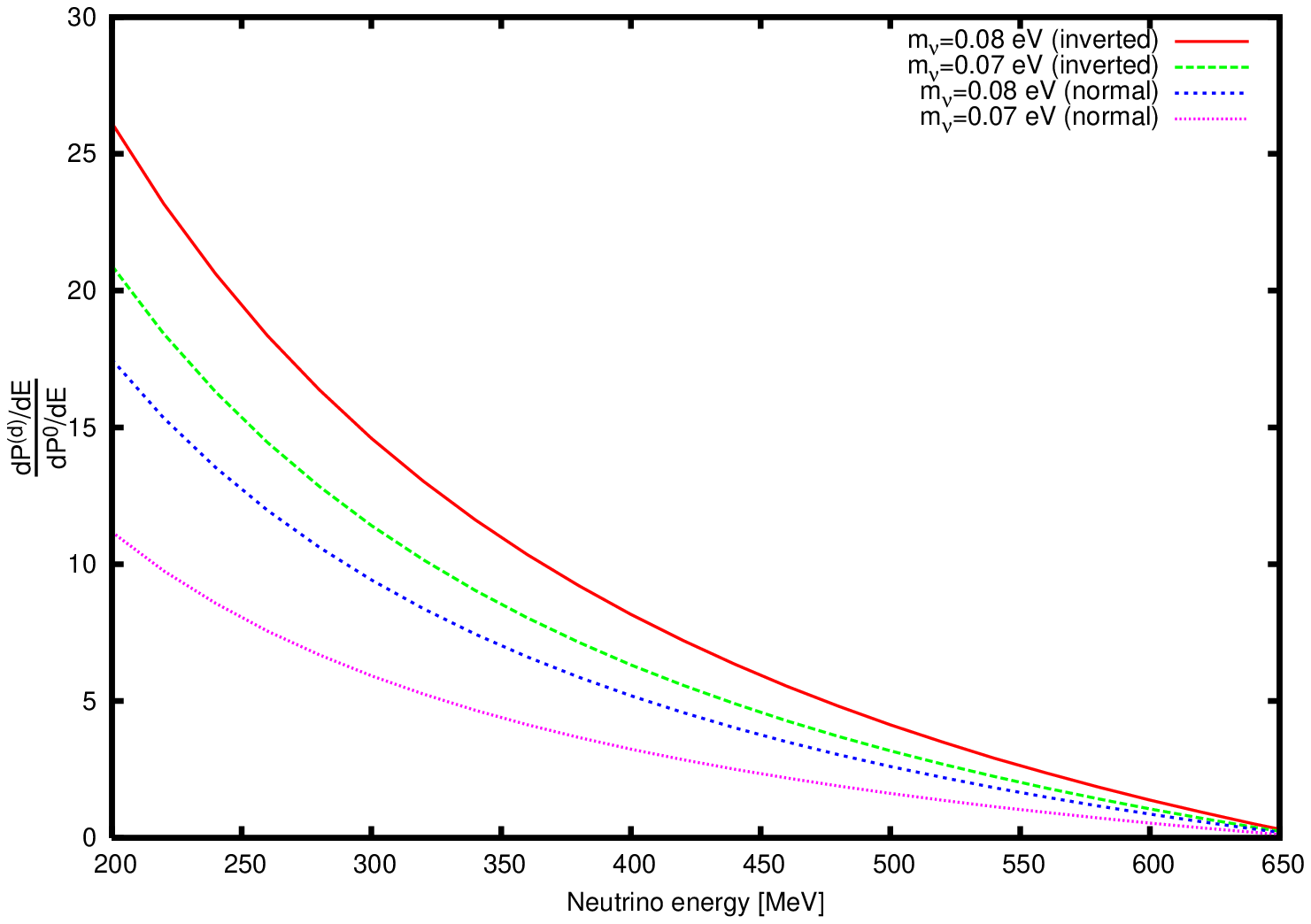}
 \includegraphics[scale=.32,angle=-90]{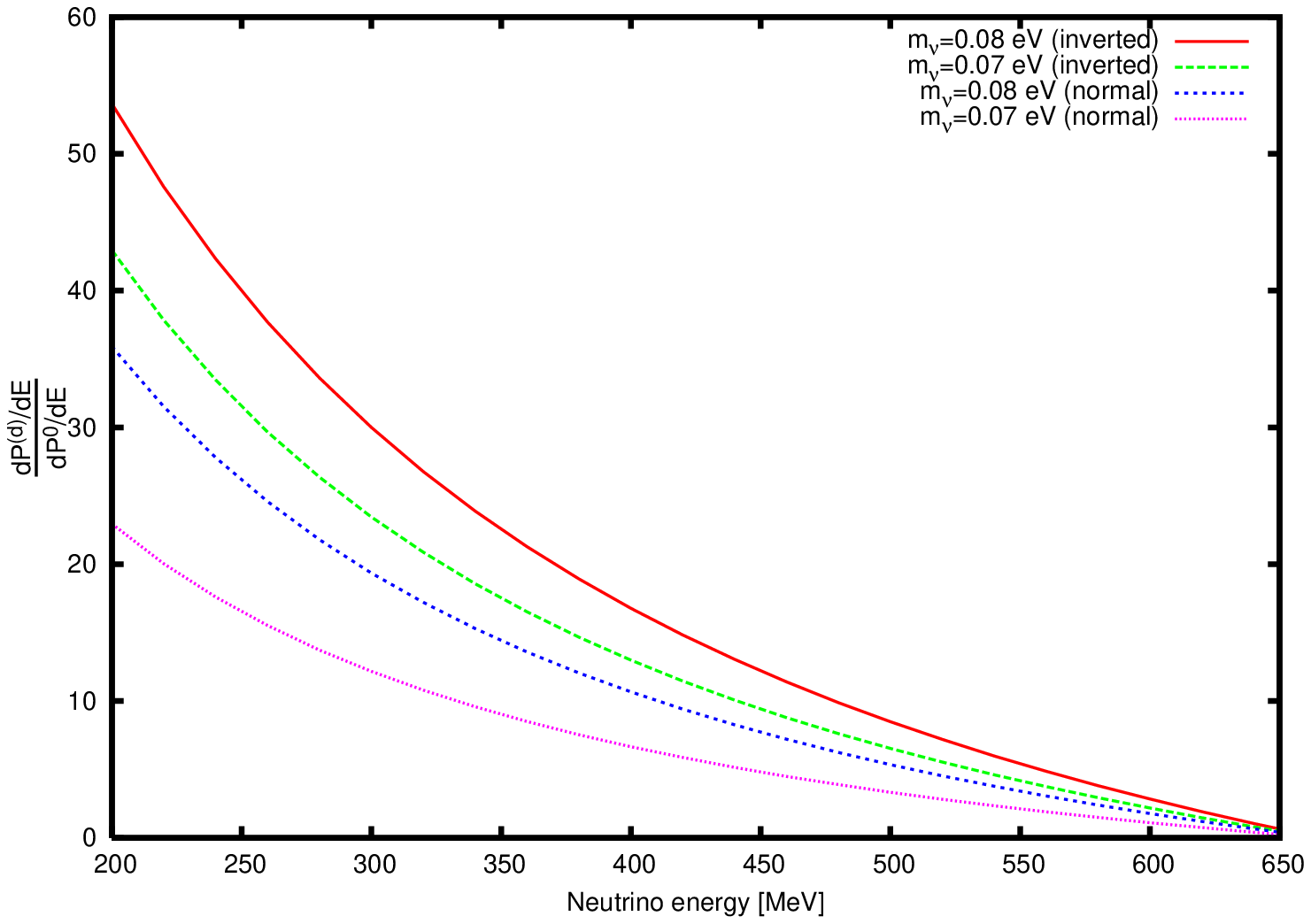}
\caption{(Color online) Energy dependences of the ratio $P^{(d)}_{\nu_e}/P_{\nu_e}^{0}$ 
from our theory for MicroBooNE \cite{Microboone-prop} setup. The left figure corresponds to $cT=50$ m; the right figure to $cT=25$ m. In the numerical calculation, $m_{\nu_h} = 0.07 \text{ eV (green: inverted, magenta: normal)}$,$m_{\nu_h}= 0.08$ eV (red: inverted, green: normal), $E_\mu=670$ MeV, and $E_\pi=1.15$ GeV were used. The target nucleus is ${}^{40}$Ar. }
\label{fig:microboone-nue}
\end{figure}
Figure \ref{fig:microboone-nue} shows the energy dependence of the ratio $P^{(d)}_{\nu_e}/P_{\nu_e}^{(0)}$ of our theory in a MicroBooNE experiment
\cite{Microboone-prop}. In this experiment, the neutrino beam is the
same as that in the  MiniBooNE experiment. The MicroBooNE detector is
smaller than that of MiniBooNE, but the target nucleus is ${}^{40}$Ar,
whose range in space covered is substantially larger than that of ${}^{12}$C, and the finite-size correction becomes also large. By using two different lengths of the decay region and sufficient statistics, not only  the mass hierarchy, but also the absolute neutrino mass  can be determined.
\subsection{Neutrino flavor changes through diffraction }
$P^{(d)}$ in the system of flavor mixing shows  a unique behavior distinct from the standard formula.  In $\mu^+$DAR, Eq.($\ref{mu+decay} $), $\bar{\nu}_e$ 
 produced  by the flavour mixing is sizable numbers and is detected at  
detectors with $L\sim 10-100$ m, if  a range in space covered by the 
wavefunctions is large even with a small $\Delta m^2$. 
In $\mu^+$DIF, both ${\nu}_\mu$ and $\bar{\nu}_\mu$   have   excesses.

\begin{center}
\begin{figure}[t]
 \includegraphics[scale=.45]{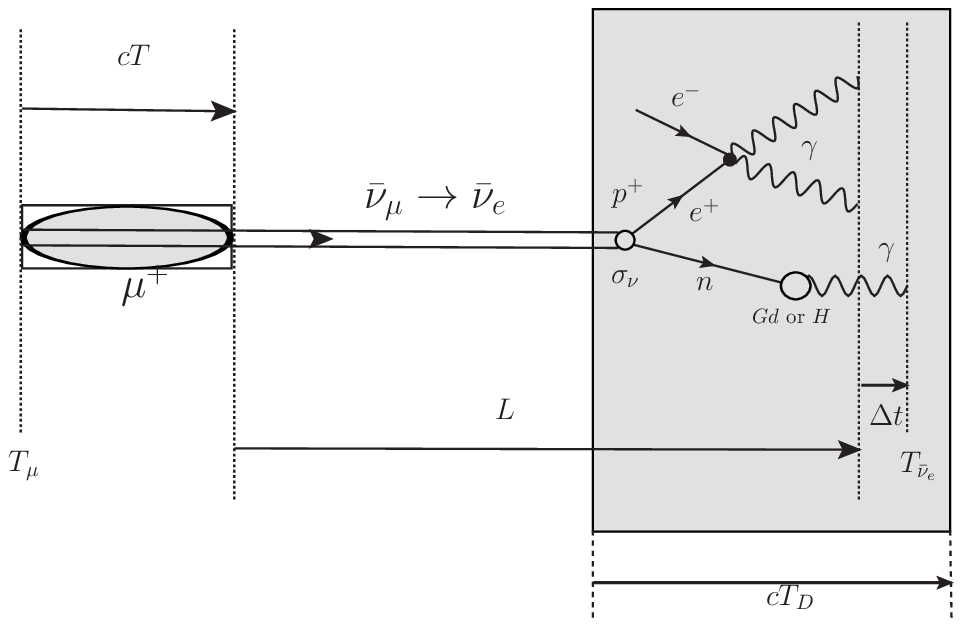}
 \caption{Space-time geometry  of the detection of $\bar{\nu}_e$ through  inverse beta process  with the delayed coincidence in $\mu^+$DAR. $T$ is the time width where $\mu^+$ and its decay products can overlap. $L$ is the length between the decay region and the detector used in the flavor oscillation formula. $\Delta t$ is the time difference between the photon signals of the positron and neutron captures, used for event selection in KARMEN experiments. 
}
\label{fig:anti-beta}
\end{figure}
\end{center}
\subsubsection{LSND and KARMEN ($\mu^+$DAR)}
 Detector geometries  of  $\mu^+$DAR in LSND  \cite{LSND-nuebar} and KARMEN \cite{KARMEN-nuebar} are similar   but not identical. 
That  of the
$\mu^+$DAR experiment is shown in Fig. \ref{fig:anti-beta}. $\mu^{+}$ at rest
decays in matter and  $\bar{\nu}_e$, which is  from $P^{(d)}$,  propagates and is detected 
at downstream   through  inverse beta process  with the delayed
coincidence.  $T$ is a time interval in which $\mu^+$ and its decay 
products co-exist and  overlap. $L$ is the spatial length 
between  the decay region and the detector, and  $\Delta t$ is the 
time interval between the photon signals of the positron and 
neutron captures, used for event selection in KARMEN experiments. 
$\Delta t$ is not taken into account in LSND. In the anti-neutrino events,  photons from the positron annihilation arrive to the detector first, and the photon from the neutron capture arrive later, for the neutrino in the particle zone. The neutrino  has the short-correlation length, and the delay time is sharply distributed in the  order few $\mu $ seconds. Now, the neutrino in the wave zone  represents  the rapid transition in the short-time interval, and spreads  over wide  kinetic-energy region.   That has the correlation length $L_c$. The neutron has the same properties , and interacts strongly with nucleus due to strong interaction through $P^{(d)}$ without thermalize, and emits a photon in the short-time interval.   Hence  photons  from two processes arrive to detector almost simultaneously.  By  the timing  cut, the neutrino events by $P^{(d)}$  may be rejected.

The diffraction term has an energy spectrum 
\begin{align}
&\frac{dP^{(d)}}{dE_{\bar{\nu}_e}} = \frac{G_F^2m_\mu^2\tau_\mu}{12\pi^3}E_{\bar{\nu}_e}^2\frac{m_\mu^2\sigma_{\bar{\nu}_e}}
{8\pi}
\left(1-2\frac{E_{\bar{\nu}_e}}{m_\mu}\right)\left(5-6\frac{E_{\bar{\nu}_e}}{m_\mu}\right)
\tilde{g}_{\mu,e}(\omega_{\bar{\nu}_e},T;\tau_\mu), \label{nuebar-spectrum-DAR}
\end{align}
where $\tilde g_{\mu,e}  $ is given in Appendix C. 
\begin{figure}[t]
\begin{minipage}[h]{8cm}
\includegraphics[scale=.35,angle=0]{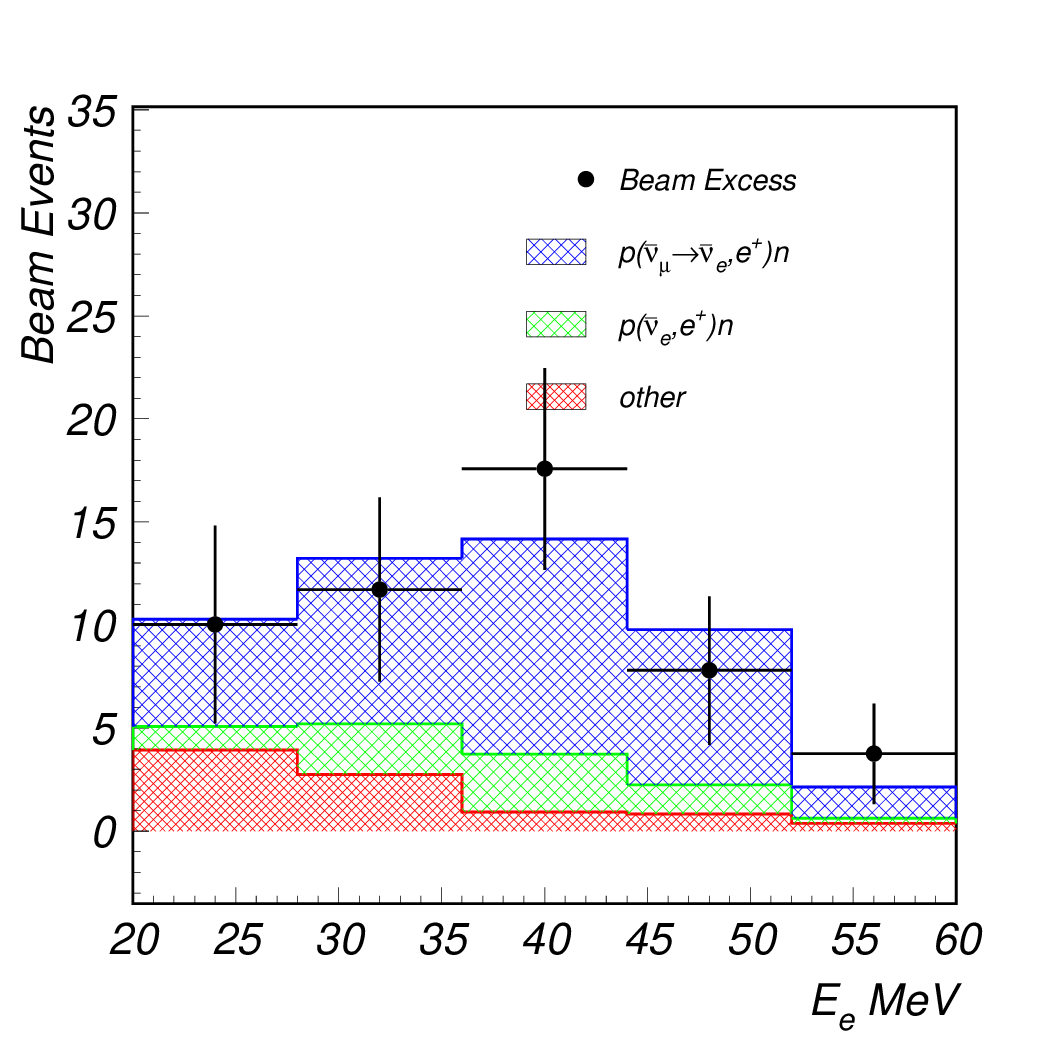}
\end{minipage}
\begin{minipage}[h]{8cm}
\includegraphics[scale=.30,angle=-90]{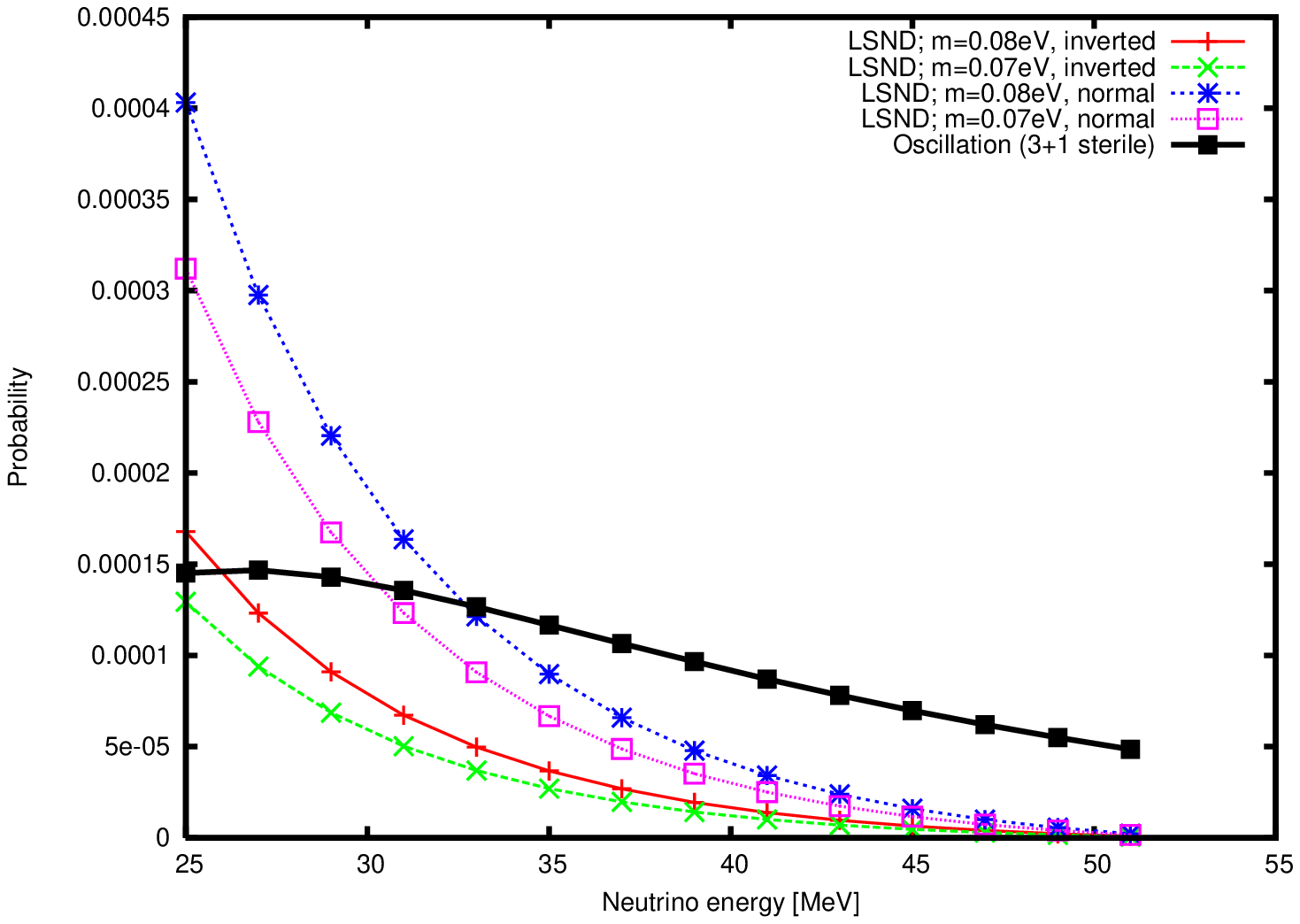}
\end{minipage}
\caption{(Color online) Spectra of LSND \cite{LSND-data} (top) is compared with those of the neutrino oscillation due to the sterile neutrino (black line ) and to the 
diffraction terms (colored line) in the bottom. 
$T_\mu=0$, $cT=0.8$ m ($T\sim 2.5$ ns), 
$cT_D=8.3$ m, $L=29.8$ m, $\sigma_{\bar{\nu}_e}^\text{LSND}$
 of $C_{2n}H_{2n+2}$,  $\Delta m^2_{LSND} = 1.2$ eV${}^2$, and
$\sin^22\theta_{LSND} = 0.003$ are used. The red and green curves show the
 inverted hierarchy of $m_{\nu_h} = 0.08$ eV and  $m_{\nu_h} = 0.07$ eV; the blue and magenta curves show
 the normal hierarchy of $m_{\nu_h}=0.08$ eV and  $m_{\nu_h}=0.07$ eV. 
 }
\label{fig:lsnd-spectra}
\end{figure}
The spectrum for three flavor are given  in Fig.\ref{fig:lsnd-spectra}, and are compared with the spectrum from  LSND experiment.
The event selection using the $\Delta t$  is not made, but LSND  adopted the
likelihood ratio instead of $\Delta t$.
  The experimental parameters  are
summarized in Table \ref{table:LSND-KARMEN}. 
  The flavor change through $P^{(d)}$ in the experimental condition of KARMEN, $5\mu\text{s}<\Delta t<300\ \mu\text{s}$ is given  in Fig.11, and  almost disappears.
Thus a small difference in $\Delta t$ causes  an
essential difference. This is understandable from the fact that  the photons from two processes arrive to the detector almost simultaneously. 
The flavor change  through $P^{(d)}$ in the experimental conditions of LSND and that of KARMEN are compared further.
\begin{figure}[t]
 \includegraphics[scale=.42,angle=-90]{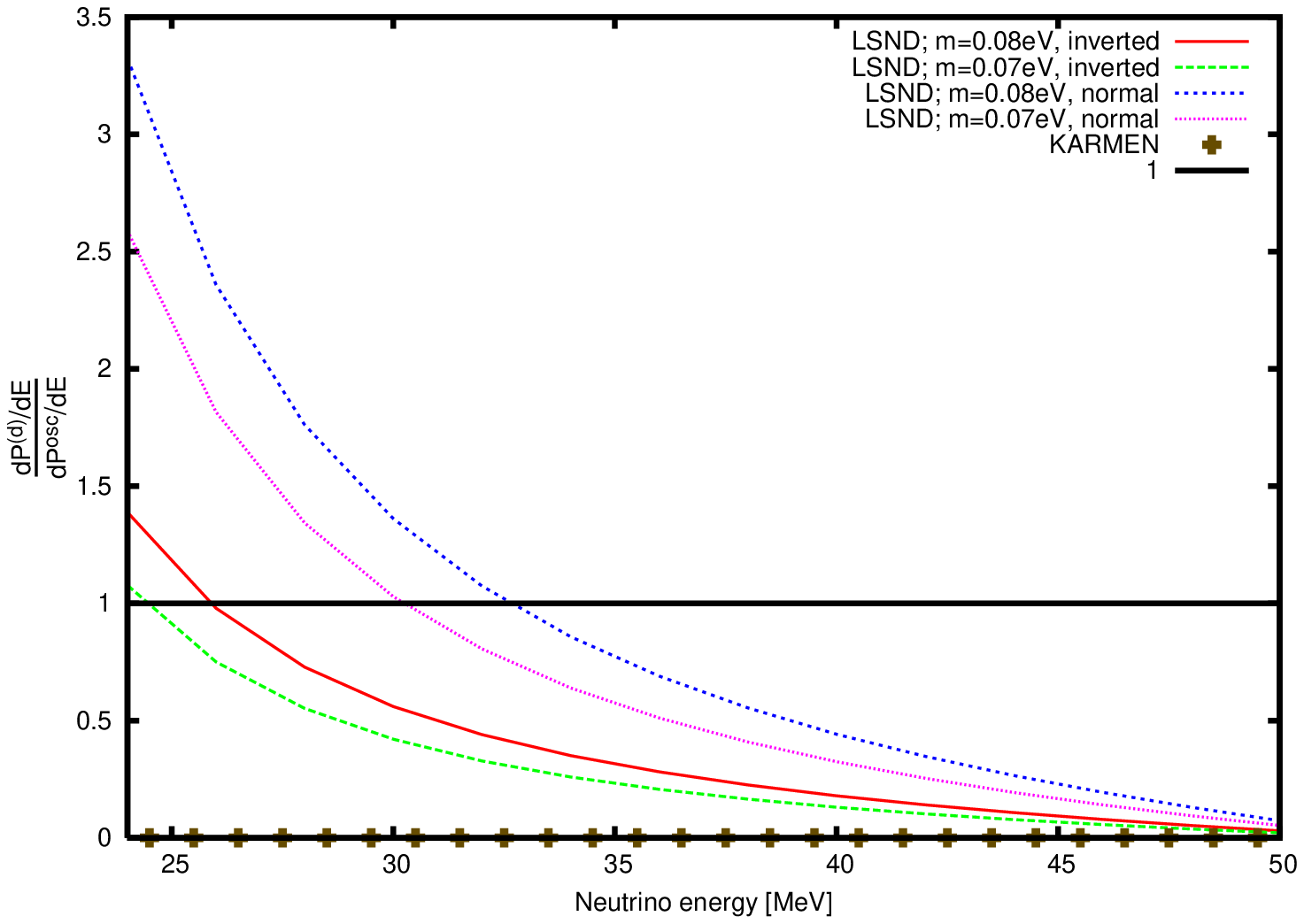}
 \caption{(Color online) The ratios $\frac{dP^{(d)}/dE}{dP^\text{osc}/dE}$
are shown. For LSND, $T_\mu=0$, $cT=0.8$ m ($T\sim 2.5$ ns), 
$cT_D=8.3$ m, $L=29.8$ m, $\sigma_{\bar{\nu}_e}^\text{LSND}$
 of $C_{2n}H_{2n+2}$,  $\Delta m^2_{LSND} = 1.2$ eV${}^2$, and
 $\sin^22\theta_{LSND} = 0.003$ are used. The red curve shows the
 inverted hierarchy of $m_{\nu_h} = 0.08$ eV; the green curve shows
 the inverted hierarchy of $m_{\nu_h} = 0.07$ eV; the blue curve shows
 the normal hierarchy of $m_{\nu_h}=0.08$ eV; the magenta curve shows
 the normal of $m_{\nu_h}=0.07$ eV. For KARMEN, $T_\mu=0.3\ \mu$s,
 $\Delta t=5\ \mu$s, $cT=0.3$ m $(T \sim 1.0\text{ ns})$, $cT_D=3.5$ m,
 $m_{\nu_h}=0.08$ eV of inverted hierarchy,  $L=17.7$ m, angle between
 proton beam and detector $\theta=100^\circ$,
 $\sigma_{\bar{\nu}_e}^\text{KARMEN}$,
 $\Delta m^2_{LSND} = 1.2$ eV${}^2$ and $\sin^22\theta_{LSND} = 0.003$
 are used. A geometry for $\mu^+$DAR is shown in
Fig. \ref{fig:anti-beta} and a relation between $T$ and $\theta$ is
 given in Appendix \ref{app:angle-dep}. }
\label{fig:LSND-KARMEN}
\end{figure}
\begin{center}
\begin{table}[h]
\begin{tabular}{|l|c|c|}\hline
 \ &LSND &KARMEN\\\hline
Size of beam stop (D$\times$W$\times$H)& 1 m$\times$0.2 m$\times$0.2 m  
& 0.5 m$\times$ 0.25 m$\times$0.25 m\\\hline
 $L$ & 29.8 m & 17.7 m\\\hline
$\Delta t$ & No& 5 $\mu$s$<\Delta t < 300\ \mu$s\\\hline
Scintillator & $CH_2$& $C_nH_{2n+2}(75\%) + C_9H_{12}(25\%)$\\\hline
$E_{\bar{\nu}_e}$  & 36(20)--60 MeV& 16--50 MeV\\\hline
Primary positron time window &No, $T_\mu=0$ & 0.6 $\mu$s $<T_\mu< 10\ \mu$s\\\hline
$T_D$, Depth of detector & 8.3 m& 3.5 m\\\hline
Detector angle & $10^\circ$ & $100^\circ$\\\hline
$\bar{\nu}_e$ event excess & 87.9$\pm$22.4$\pm$6.0&No excess\\\hline
Best fit $\Delta m^2$ and $\sin^2\theta$ & $\Delta m^2 = 1.2$ eV${}^2$, $\sin^2\theta=0.003$ &None\\\hline
\end{tabular}
\caption{Parameters and results of LSND and KARMEN.}\label{table:LSND-KARMEN}
\end{table}
\end{center}
For a comparison of the flavor change  through $P^{(d)}$ in the experimental conditions of LSND and  of KARMEN ,  the standard oscillation formula of two neutrino of the mass-squared difference $\Delta m^2$ derived from  $\Gamma$  
\begin{align}
& \frac{dP^\text{osc}}{dE_{\bar{\nu}_e}} = \frac{G_F^2m_\mu^2\tau_\mu}{12\pi^3}E_{\bar{\nu}_e}^2
\left(3-4\frac{E_{\bar{\nu}_e}}{m_\mu}\right)
\left(e^{-\frac{T_\mu}{\tau_\mu}}-e^{-\frac{(T_\mu+T_D)}{\tau_\mu}}\right)
\sin^22\theta\sin\left(1.27\frac{\Delta m^2}{E_\nu}L\right),
\end{align}
which is less sensitive to the experimental conditions, is used as a reference. 
Their ratios  $\frac{dP^{d}/dE}{dP^\text{osc}/dE}$, following
the parameters shown in Table \ref{table:LSND-KARMEN} are computed.   
As  Fig. \ref{fig:anti-beta} shows, the detector is located
away from the muon decay area, and  the parent and daughters 
overlap in a finite range covered by the wave functions   
$T=T_{\bar{\nu}_e} - T_\mu - L/c$. There are two cases:

1. No cut is required on the time difference $\Delta t$  between the photons
from positron and those from neutron:

In this case,   the overlapping region  is determined by the 
interval between the initial and final instant of times,  and the range
covered by  the beam stop, and is wide.  Hence, the diffraction term is included 
into the event of $\bar{\nu}_e$. 

2. Cut on $\Delta t$ is made. 

In this case,  the
overlapping region is reduced  by $\Delta t$ as 
$T = T_{\bar{\nu}_e} -T_\mu - L/c - \Delta t$. Only for
$T>0$, the waves  overlap, and $P^{(d)} \neq 0$ . In other case, the waves do not overlap and $P^{(d)} =0$. 

 LSND did not use the cut and included all the events. Consequently,
 the diffraction was  included. 
Now,   KARMEN selected the events of satisfying $\Delta t> 5\ \mu$s, 
and excluded the events of $\Delta t< 5\ \mu$s, which corresponds to  the length $\Delta
L=1.5 \times 10^3$ meters from our estimation. 
   This
condition does not affects the events due to $\Gamma$, because the correlation length    
 is much shorter.

\begin{figure}[t]
 \includegraphics[scale=.42,angle=-90]{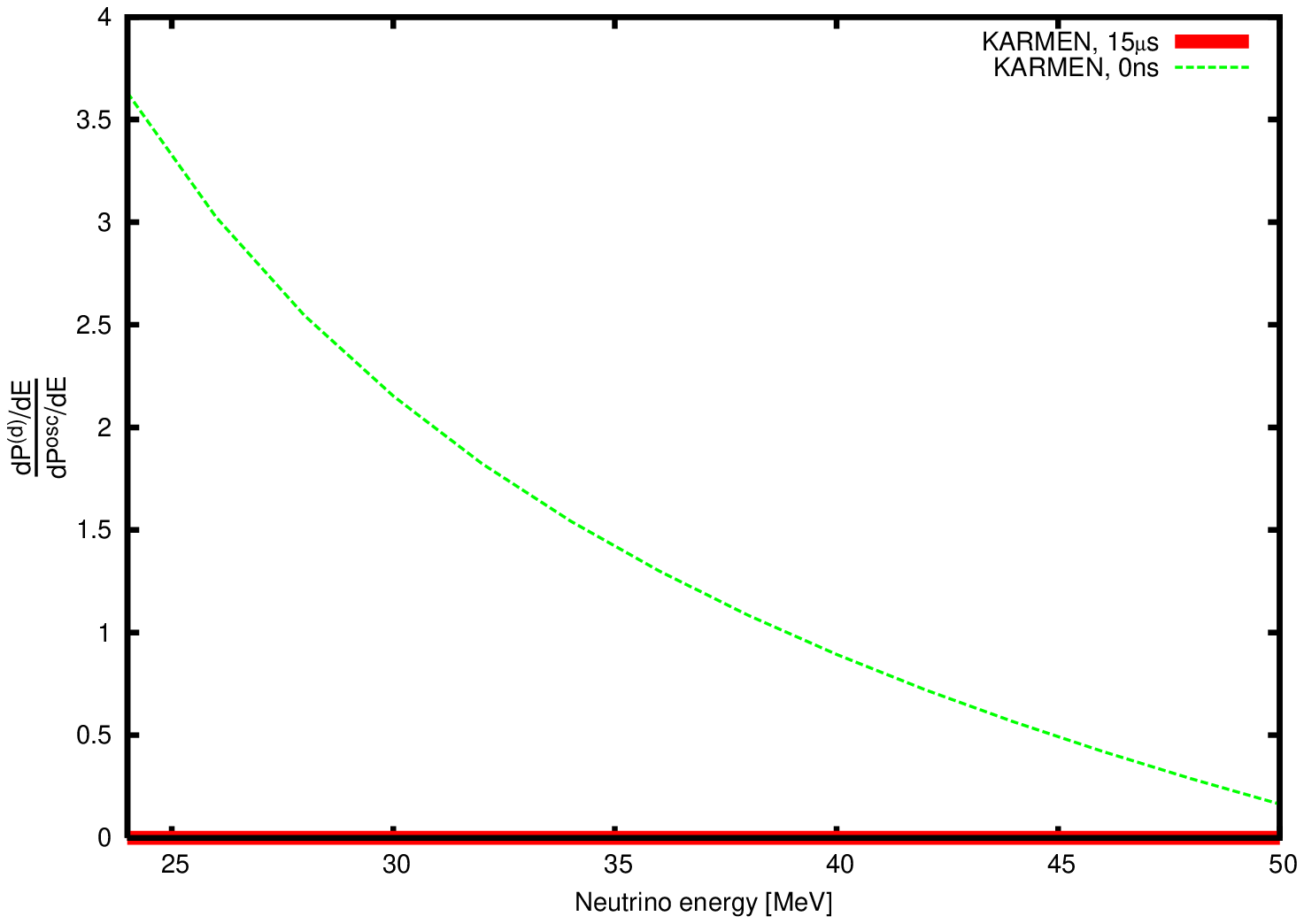}
\caption{(Color online) The $\Delta t$ dependence of the ratio for KARMEN. $\Delta t=0$ ns (green)  and $\Delta t = 15\mu$s (red). We used $m_{\nu_h}=0.08$ eV of inverted hierarchy; all other parameters are the same as those in Fig. \ref{fig:LSND-KARMEN}.}
\label{fig:deltat-dep-KARMEN}
\end{figure}

The theoretical value  of $P^{(d)}$ for LSND and KARMEN are shown  in
Fig. \ref{fig:LSND-KARMEN}. $P^{(d)}$ in LSND configuration  is large,   while 
that for KARMEN configuration vanishes.  If $\bar{\nu}_e$ excess is verified from events 
 of $\Delta t<5 \mu$s in the KARMEN experiment, that  proves the present theory.
A precise  dependence on $\Delta t$ of the ratio is given in Fig. \ref{fig:deltat-dep-KARMEN}. The magnitude of $P^\text{(d)}$ is 
approximately ten times larger than that of LSND because $L$ of KARMEN
is shorter than that of LSND.

Thus the difference between LSND and KARMEN results from the methods of the event selection, and an indication may appear as a sharp peak in the small $\Delta t$ region in Ref. \cite{KARMEN-nuebar}.
Accordingly, the experimental results of both LSND and
KARMEN can be explained from $P^{(d)}$, and are consistent with the previous result of LSND $\pi$DIF \cite{Ishikawa-tobita-neutrinomass}. With a precise energy spectrum and suitable selection criteria,  it would be  possible to confirm $P^{(d)}$ by the excess of $\bar{\nu}_e$ at near detector. 

\subsubsection{Future experiments of $\mu^+$DAR and $\mu^+$DIF}
There are two possible experiments to confirm $P^{(d)}$ from the  excess of the neutrino flux.   One is a $\bar{\nu}_e$ appearance experiment with $\mu^+$DAR \cite{J-PARC-sterile}; the other is a $\nu_\mu$ appearance and $\bar{\nu}_\mu$ disappearance with $\mu^+$DIF \cite{nuSTORM}. Here, we give predictions for future experiments regarding both cases.

\noindent 1. $\mu^+$DAR

\begin{figure}[t]
 \includegraphics[scale=.42,angle=-90]{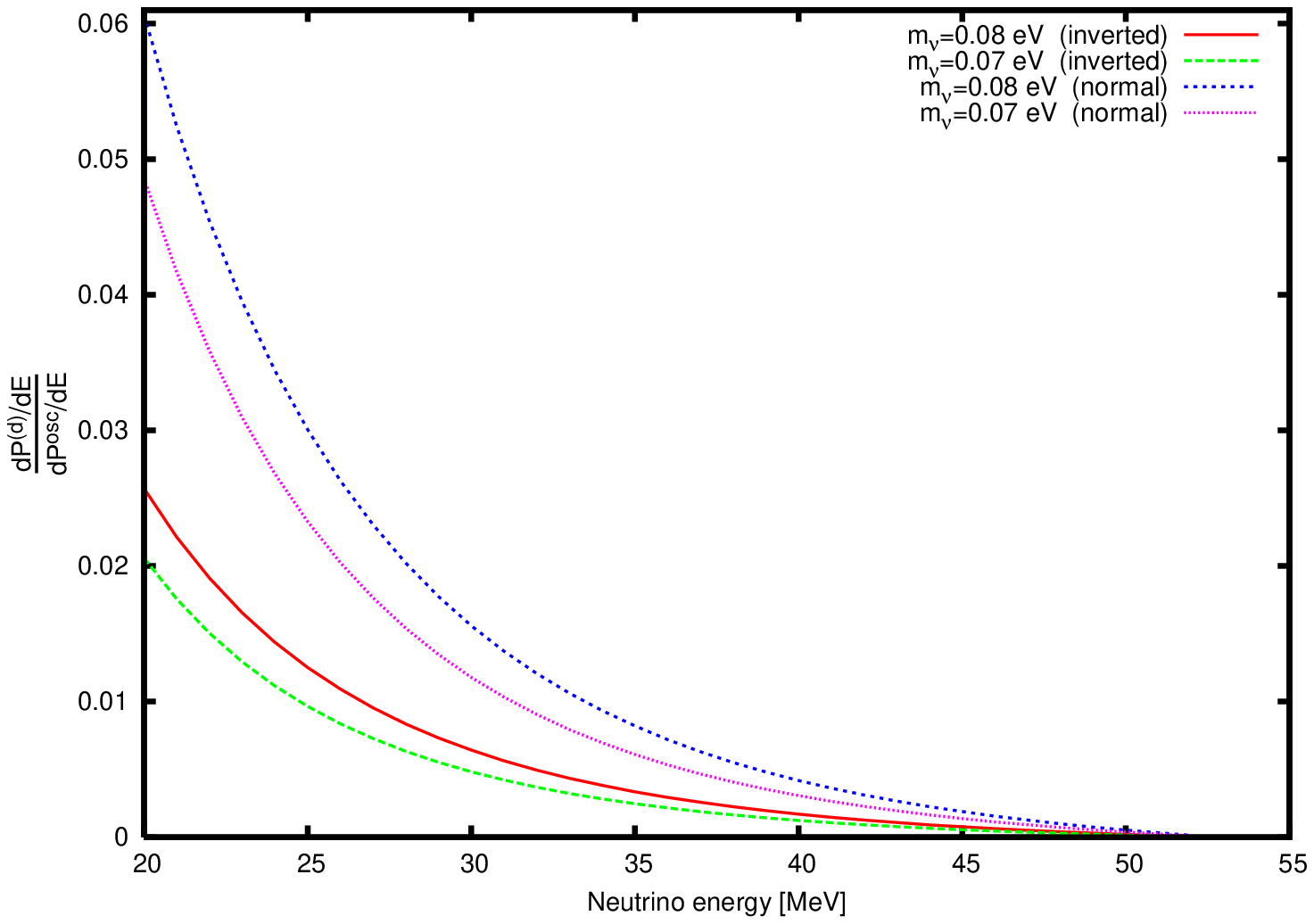}
 \caption{(Color online) The ratios $\frac{dP_{\bar{\nu}_e}^{(d)}/dE}{dP_{\bar{\nu}_\mu}^{0}/dE}$, Eq. \eqref{eq:jparc-ratio-normal}, 
for $\mu^+$DAR are shown, where
$T_\mu=1$ $\mu$s, $cT=1.0$ m, $L=17.0$m, and $\sigma_{\bar{\nu}_e}=7.3$ of $C_{2n}H_{2n+2}$ and  $cT_D=3.4$ m are used \cite{J-PARC-sterile}. The red curve shows the inverted hierarchy of $m_{\nu_h} = 0.08$ eV, the green curve shows the inverted hierarchy of $m_{\nu_h} = 0.07$ eV, the blue curve shows the normal hierarchy of $m_{\nu_h}=0.08$ eV, and the magenta curve shows the normal hierarchy of $m_{\nu_h}=0.07$ eV.
}
\label{fig:J-PARC}
\end{figure}

In $\mu^+$DAR experiments, the $\bar{\nu}_e$ spectrum is given by Eq. \eqref{nuebar-spectrum-DAR}. The ratio between $\bar{\nu}_e$ and $\bar{\nu}_\mu$ spectra for 
events free from   the double coincidence condition of $\Delta t$ is written as
\begin{align}
 R_{\bar{\nu}_e}(E_{\nu}) = \frac{m_\mu^2\sigma_{\bar{\nu}_e}}{8\pi}\frac{\left(1 - 2\frac{E_\nu}{m_\mu}\right)\left(
5-6\frac{E_\nu}{m_\mu}\right)}{\left(3 - 4\frac{E_\nu}{m_\mu}\right)(e^{-\frac{T_\mu}{\tau_\mu}}
 - e^{-\frac{T_\mu+T_D}{\tau_\mu}})}
\tilde{g}_{\mu,e}(\omega_{\bar{\nu}_e},T;\tau_\mu).\label{eq:jparc-ratio-normal}
\end{align}
This is the  value under  ideal conditions. The value  for an experimental setup \cite{J-PARC-sterile} is  shown in Fig. \ref{fig:J-PARC}, where $cT$, the size of the beam stop, is 1 m, 
$T_\mu=1$ $\mu$s, $cT_D=3.4$ m, and $\sigma_{\bar{\nu}_e}$ is of $C_{2n}H_{2n+2}$ in the liquid scintillator. The ratio from the flavor oscillation with one sterile neutrino is, $\frac{dP^d/dE}{dP^{\text{osc}}/dE}$,
\begin{align}
 R^\text{osc}_{\bar{\nu}_e}(E_{\nu}) = \frac{m_\mu^2\sigma_{\bar{\nu}_e}}{8\pi}
\frac{\left(1 - 2\frac{E_\nu}{m_\mu}\right)\left(
5-6\frac{E_\nu}{m_\mu}\right)\tilde{g}_{\mu,e}(\omega_{\bar{\nu}_e},T;\tau_\mu)}
{\left(3 - 4\frac{E_\nu}{m_\mu}\right)(e^{-\frac{T_\mu}{\tau_\mu}}
 - e^{-\frac{T_\mu+T_D}{\tau_\mu}})\sin^22\theta_{\mu e}\sin^2\left(1.27\frac{\Delta m_{41}^2}{E_\nu}L\right)}
. \label{eq:jparc-ratio-flavor}
\end{align}

\begin{figure}[t]
 \includegraphics[scale=.42,angle=-90]{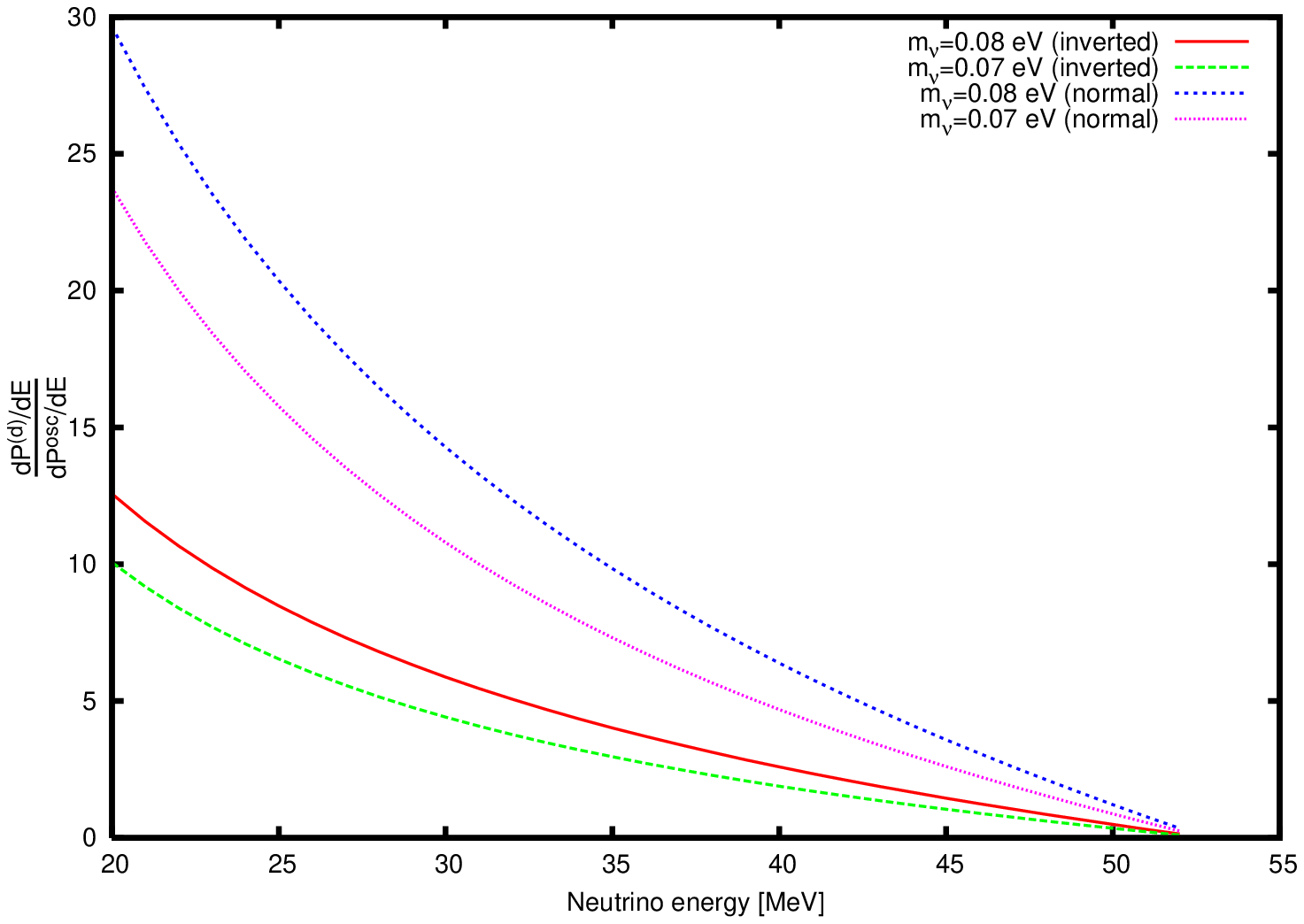}
 \caption{(Color online) The ratios $\frac{dP_{\bar{\nu}_e}^{(d)}/dE}{dP^{osc}_{\bar{\nu}_e}/dE}$, Eq. \eqref{eq:jparc-ratio-flavor}, 
for $\mu^+$DAR are shown, where
$T_\mu=1$ $\ \mu$s, $cT=1.0$ m, $L=17.0$ m, $cT_D=3.4$ m, and $\sigma_{\bar{\nu}_e}=7.3$ of $C_{2n}H_{2n+2}$
and  are used \cite{J-PARC-sterile}. The red curve shows the ratio of the inverted hierarchy with $m_{\nu_h} = 0.08$ eV, the green curve shows that of the inverted hierarchy with $m_{\nu_h} = 0.07$ eV, the blue curve shows the ratio of the normal hierarchy with $m_{\nu_h}=0.08$ eV, and the magenta curve shows the ratio of the normal hierarchy with $m_{\nu_h}=0.07$ eV.
}
\label{fig:J-PARC-OSCI}
\end{figure}

Figure \ref{fig:J-PARC-OSCI} shows the ratio in Eq. \eqref{eq:jparc-ratio-flavor}, where the experimental parameters are the same as those in Fig. \eqref{fig:J-PARC} and the parameters of the sterile neutrino are \cite{sterile-review2013}
\begin{align}
&\Delta m^2_{41} = 0.9 \text{ eV}^2,\ U_{e 4} = 0.15, \  U_{\mu 4} = 0.17,\label{eq:delta-uemu-sterile}\\
& \sin^2\theta_{\mu e} = 4\left|U_{\mu 4}U_{e 4}\right|^2,\ 
\sin^2\theta_{\mu\mu} = 4\left|U_{\mu 4}\right|^2\left(1 - \left|U_{\mu 4}\right|^2\right).\label{eq:sin-mumu-mue}
\end{align}
These values are also used in the next $\mu^+$DIF case. According to Figs. \ref{fig:J-PARC} and \ref{fig:J-PARC-OSCI},  the magnitude of the $\bar{\nu}_e$ appearance through $P^{(d)}$  can be almost the same as, or larger than, that of the flavor oscillation with the sterile neutrinos. Furthermore, the effect is sensitive to the absolute neutrino mass and the mass hierarchy of the neutrino.

\noindent 2. $\mu^+$ DIF

\begin{figure}[t]
 \includegraphics[scale=.32,angle=-90]{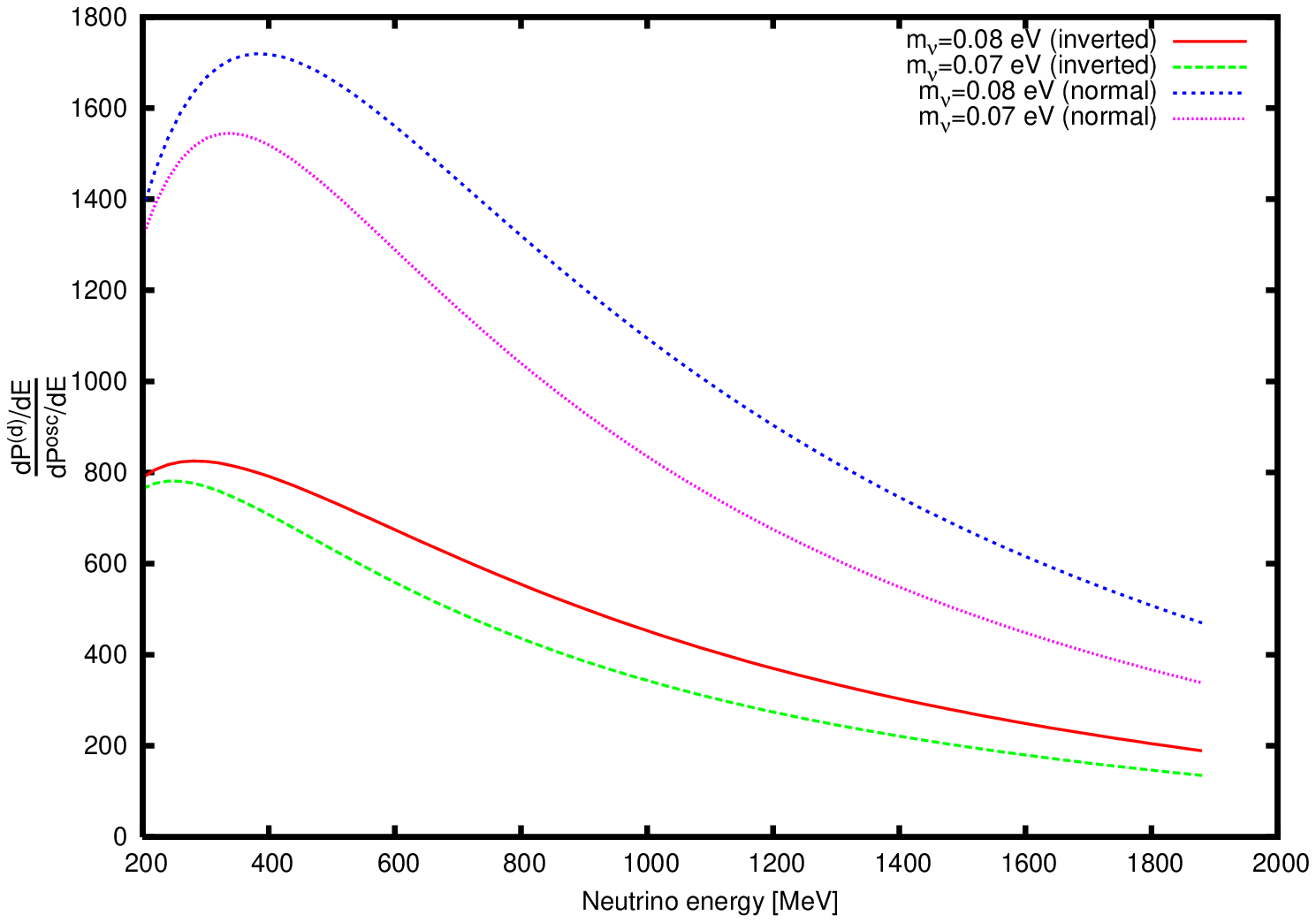}
\includegraphics[scale=.32,angle=-90]{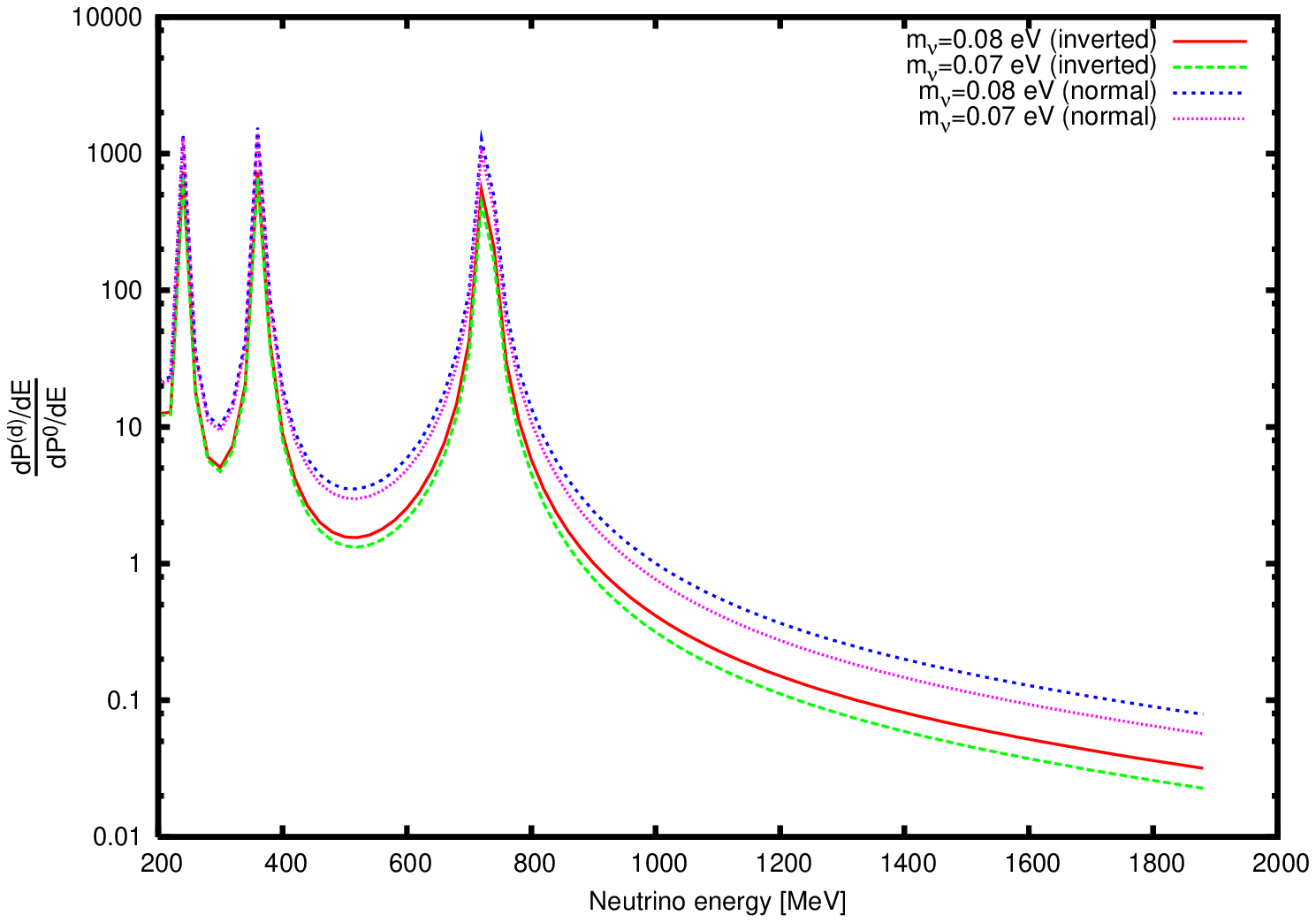}
 \caption{(Color online) Ratios of $\nu_\mu$ appearance $\frac{dP_{{\nu}_\mu}^{(d)}/dE_\nu}{dP^{osc}_{{\nu}_\mu}/dE_\nu}$, 
Eq. \eqref{eq:jparc-ratio-flavor}, 
for $\mu^+$DIF, where
$T_\mu=0$ $\mu$s, $cT=226.0$ m, $L=20.0$ m (left) and $L=2000.0$ m (right), and $\sigma_{{\nu}_\mu}=$ of ${}^{56}$Fe
and $\cos\theta=1$ are used \cite{nuSTORM}. The red curve shows the ratio of the inverted hierarchy with $m_{\nu_h} = 0.08$ eV, the green curve shows the ratio of the inverted hierarchy with $m_{\nu_h} = 0.07$ eV, the blue curve shows the ratio of the normal hierarchy with $m_{\nu_h}=0.08$ eV, and the magenta curve shows the ratio of the normal hierarchy with $m_{\nu_h}=0.07$ eV. In the right figure, there are three sharp peaks because the denominator oscillates in energy and becomes very small.
}
\label{fig:nustorm-app}
\end{figure}

\begin{figure}[t]
 \includegraphics[scale=.32,angle=-90]{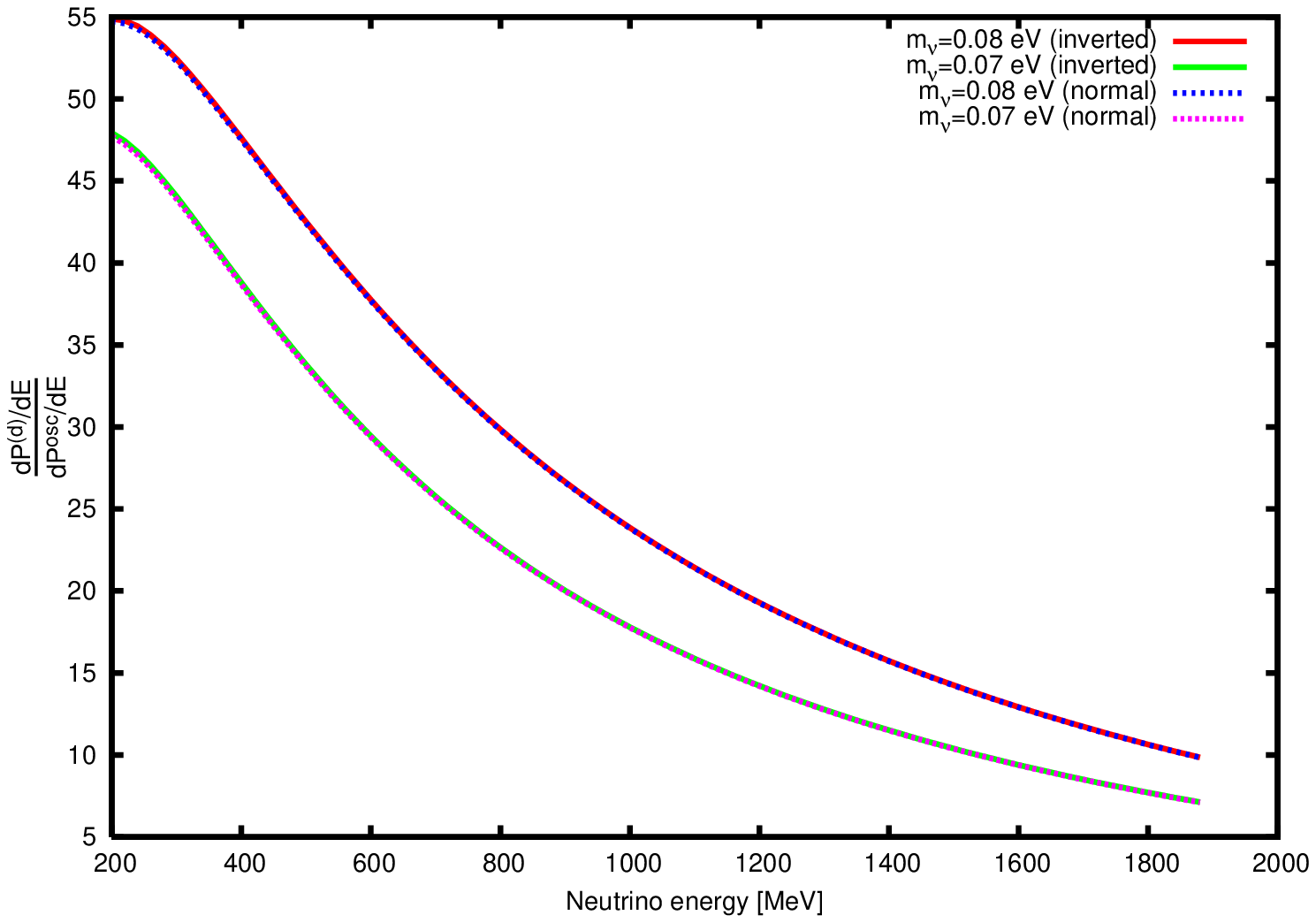}
 \includegraphics[scale=.32,angle=-90]{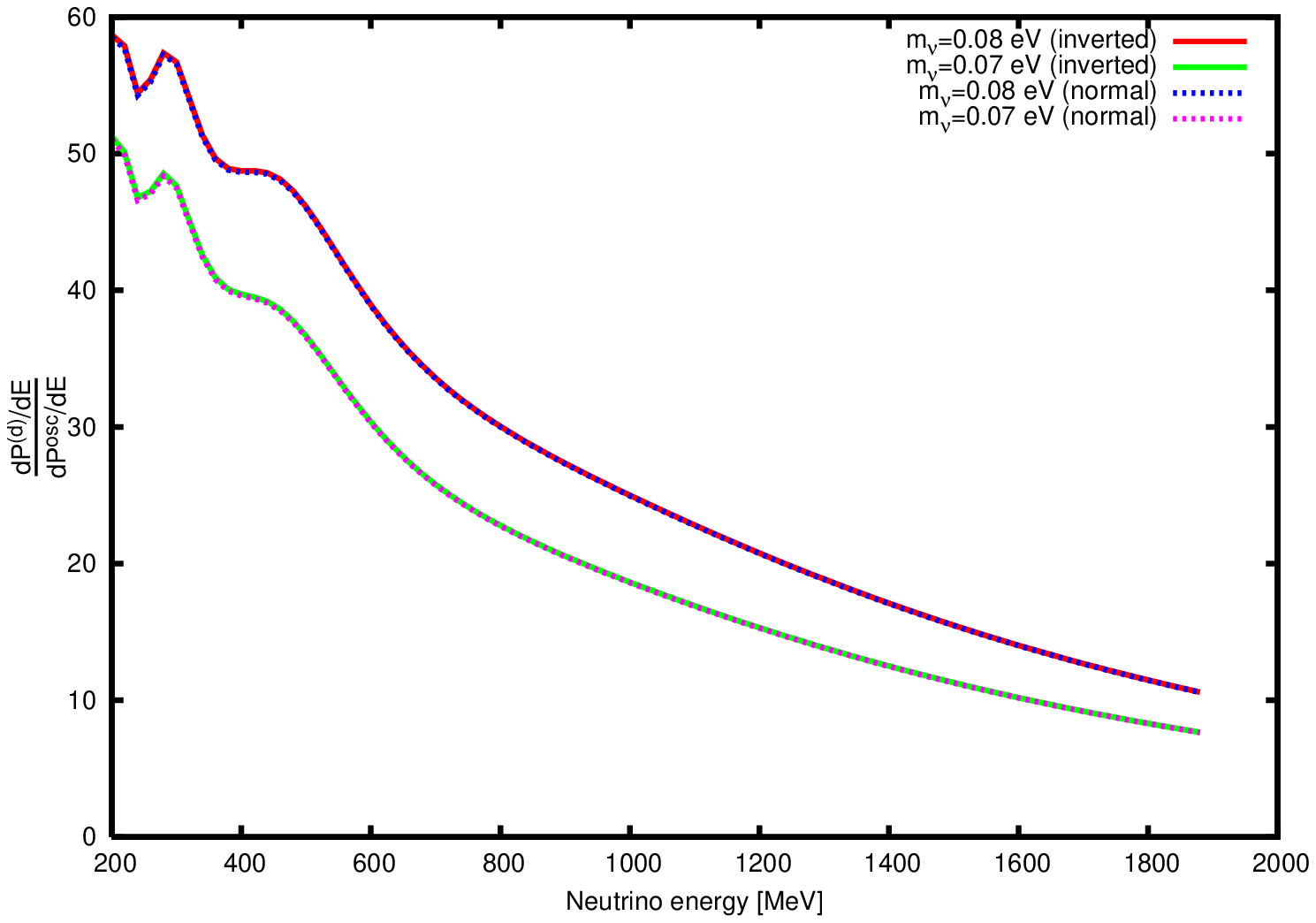}
 \caption{(Color online) Ratios of excess and disappearance of $\bar{\nu}_\mu$,
$\frac{dP_{\bar{\nu}_\mu}^{(d)}/dE}{dP^{osc}_{\bar{\nu}_\mu}/dE}$, 
Eq. \eqref{eq:jparc-ratio-flavor}, for $\mu^+$DIF, where $T_\mu=0$ $\mu$s, $cT=226.0$ m, $L=20$ m (Left) and $L=2000$ m (Right), and $\sigma_{\bar{\nu}_\mu}=$ of ${}^{56}$Fe and $\cos\theta=1$ are used \cite{nuSTORM}. The red curve shows the ratio of the inverted hierarchy with $m_{\nu_h} = 0.08$ eV, the green curve shows the ratio of the inverted hierarchy with $m_{\nu_h} = 0.07$ eV,  the blue curve shows the ratio of the normal hierarchy with $m_{\nu_h}=0.08$ eV, and the magenta curve shows the ratio of the normal hierarchy with $m_{\nu_h}=0.07$ eV.
}
\label{fig:nustorm-dis}
\end{figure}

In the $\mu^+$DIF experiment, an appearance of $\nu_\mu$ ($\nu_e\to\nu_\mu$) from $\nu_e$ and a disappearance of $\bar{\nu}_\mu$ ($\bar{\nu}_\mu\to\bar{\nu}_e$) will be searched, as the flavor oscillations with sterile neutrinos are considered.
The finite-size corrections provide the appearance of $\nu_\mu$ and $\bar{\nu}_\mu$, but their magnitudes and spectra are very different from those of the sterile neutrinos.
Using Eqs. \eqref{eq:nue-normal-spectrum} -- \eqref{eq:numu-diffraction-spectrum}, 
the ratios $\frac{P^{(d)}_{\nu_\mu}(E_\nu)}{P^{osc}_{\nu_\mu}(E_\nu)}$
 and $\frac{P^{(d)}_{\bar{\nu}_\mu}(E_\nu)}{P^{osc}_{\bar{\nu}_\mu}(E_\nu)}$
 are written as
\begin{align}
 &R_{\nu_\mu}({E_{\nu}},\cos\theta) = \frac{\sigma_{\nu_\mu}(m_\mu^2 - 2E_{\nu}(E_\mu-p_\mu\cos\theta))
\tilde{g}_{e,\mu}(\omega_{\nu},T;\gamma\tau_\mu)}
{2\pi\left(
\exp\left[-\frac{T_\mu}{\gamma\tau_\mu}\right] - \exp\left[
-\frac{T_\mu+T}{\gamma\tau_\mu}\right]\right)\sin^22\theta_{\mu e}\sin^2\left(1.27\frac{\Delta m^2_{41}}{E_\nu}L\right)},\label{eq:nue-ratio-dif-osc}\\
& R_{\bar{\nu}_\mu}({E_{\nu}},\cos\theta) = 
\frac{\sigma_{\bar{\nu}_\mu}(m_\mu^2 - 2E_{\nu}(E_\mu-p_\mu\cos\theta))
\tilde{g}_{e,\mu}(\omega_{\nu},T;\gamma\tau_\mu)}{2\pi\left(
\exp\left[-\frac{T_\mu}{\gamma\tau_\mu}\right] - \exp\left[
-\frac{T_\mu+T}{\gamma\tau_\mu}\right]\right)
\left(1 - \sin^22\theta_{\mu\mu}
\sin^2\left(1.27\frac{\Delta m^2_{41}}{E_\nu}L\right)\right)}
,\label{eq:numu-ratio-dif-osc}
\end{align}
and that of $\bar{\nu}_\mu$ is given by Eq. \eqref{eq:numu-ratio-dif}.
Eqs. \eqref{eq:nue-ratio-dif} and \eqref{eq:numu-ratio-dif} are given in Figs. \ref{fig:nustorm-app} and \ref{fig:nustorm-dis}.
The target nucleus is ${}^{56}$Fe and the parameters given in Eqs. \eqref{eq:delta-uemu-sterile} and \eqref{eq:sin-mumu-mue} are used.
The ratio of $\nu_\mu$ appearance at $L=20$ m is very large. This is because $T$ dependences are different in normal and diffraction terms. The normal term is very small at $T\ll\gamma\tau_\mu$ and $P^0$ increases with $T$, but the diffraction term $P^{(d)}$ is constant in $T$.
In addition, the oscillation length with $\Delta m_{41}^2 =0.9$ eV${}^2$ at $E_\nu=1$ GeV is 800--900 m, and the effect of the flavor oscillation is not significant at $L=20$ m. 
Thus, the $10^3$ larger magnitude of the $\nu_\mu$ appearance through the finite-size correction at $L=20$ m is a natural consequence of the nature of the diffraction term.
Then, the relative magnitude between them becomes very large, of the order of $10^3$, at $cT=226$ m $\ll \gamma\tau_\mu \text{ at } E_\mu=3$ GeV. 
  
The ratio of the $\nu_\mu$ appearance  at $L=2000$ m has three peaks because the numerator $dP^{(d)}_{{\nu}_\mu}/dE_\nu$ varies uniformly in energy and 
the denominator $dP^{osc}_{{\nu}_\mu}/dE_\nu$ oscillates in energy, becoming very small at certain energies.

\subsection{Comparison of the flavor-changing probability with the flavor oscillation    }

 $\tilde g_{\alpha \beta}(\omega_{\nu},T,\tau_{\mu})$ for $\alpha \neq \beta$ is approximately proportional to $\omega_{\nu}T$, whereas that of  the single-particle formula is proportional to $(\omega_{\nu_i}-\omega_{\nu_j})^2T^2$. Accordingly  LSND data is explained using the single-particle  formula of   $\Delta m^2 \approx 1$  eV${}^2$, or  using $P^{(d)}$ of much smaller masses of the standard 
three  neutrinos.  From $P^{(d)}$,  excesses   of  
$\bar{\nu}_\mu$ and  $\nu_\mu$  arise in  $\mu^+$DIF, but that is not the case from  the $\Gamma$.    The two mechanisms can be clearly 
distinguished.

\subsection{Modified survival probability due to  $ P^{(d)}$}
For the neutrino, $\tau_{\mu} \ll T_0$, and the life-time effect was  dominant in $t <\tau_{\mu}$.  For the electron,  $\omega_e$ is substantially  larger and $T_0(e) \ll \tau_{\mu}$,  and the effect due  to  $P^{(d)}_e$ appears before $\tau_{\mu}$. The survival probability  has a contribution from   $P^{(d)}_e$ in addition to  $\Gamma T$; the final value is given by
\begin{align}
P^{\text{new}}(T)=\frac{1}{1+P^{(d)}_e} e^{-\frac{T}{\tau_{\mu}}}. 
\end{align}
$P^{(d)}_{e}$  becomes sizable for $\sigma_e\to\infty$, and  the correction due to $P^{(d)}_{e}$ becomes then important.
\section{Summary}

It was found that the neutrinos in the wave zone participate in the muon decays.  The transition probability at $T$  is modified from the golden rule
 to $\Gamma T+P^{(d)}$. $P^{(d)}$ has the form Eq.$(\ref{diffraction-form})$, and shows the rapid transition at  small $T$.    These neutrinos reveal 
 exciting behaviors within the standard model of three neutrino flavor.
They are able to resolve  the discrepancies among   the neutrino experiments in short-distance region, and to provide the information on the neutrino absolute mass.  
 $P^{(d)}$ is due to
  the non-stationary states  of the overlapping  initial and final waves, which extend to  the macroscopic length $L_c$. $L_c$ is inversely proportional to the neutrino mass-squared, and is much longer than the detector size.   These  states in the wave zone have  the continuous spectrum of the kinetic energy which includes     $\Delta E  \to \infty $ in Eq.$(\ref{weight})$.  
Accordingly $P^{(d)}$ which rises steeply in  small $T$  becomes  constant in large $T$.  
   $P^{(d)}$ for the neutrino $P^{(d)}_{\nu}$  has the same expression with 
that of the electron  $P^{(d)}_e $, but $P^{(d)}_{\nu} \gg P^{(d)}_e$ because $\frac{m_\nu}{m_e} \ll 1$.    For the electron,  the spectrum at $T$ is expressed by the golden rule,  and the probability that the electron
 is detected is in agreement with the probability that the muon has decayed.  However  for the neutrino, that is expressed by the  
sum of the terms from the golden rule and its correction, and depends on the initial state and the final state. 

Since $L_c$ is much longer than the detector size,  the probability amplitude defined by the FQM between the non-stationary states and the outgoing states  is necessary sensitive to the latter wavefunctions. The final states   depend on  the cut on the time difference $\Delta t$ of the detection method. For the case of no cut, these  are  extended in large area in the configuration space  and overlap with the initial state.   For the case of the cut of a small $\Delta t$, a small portion of the wavefunction is excluded, and
 for that of a large $\Delta t$, a large portion of the wavefunction is excluded.     Consequently $P^{(d)}$ is large for the case of no cut,
 but small for the case of large  $\Delta t $. 
 The analysis on  the experiments is altered by $P^{(d)}$. 
  The LSND, MiniBooNE, and KARMEN  experiments have
  different geometries and used different $\Delta t$. Their $P^{(d)}$ are thus  different.  Surprisingly, the experiments seemingly inconsistent
  with each other if the probability by the Fermi's golden rule, $\Gamma$, is used
 become consistent when   $\Gamma T +P^{(d)}$ is used.    In the geometry of  long-distance  oscillation experiments,   $P^{(d)}$ is negligible 
  \cite{Ishikawa-Tobita-ANA}, and  the neutrino oscillations  are described by the standard formula. Therefore, the neutrino 
  oscillation experiments  are in  agreement  with each other and with the theory within the three flavors, when 
 the probability $\Gamma T +P^{(d)}$ is  used.     
By measuring  $P^{(d)}$ accurately,  the absolute 
neutrino masses will be determined.  The 
effects studied  here are not modified by   higher-order corrections. 

 The rigorous transition probability at $T$ 
 is derived from FQM  and describes transitions  in other 
systems of  light particles as well.   $P^{(d)}$ of intriguing properties appears in many transitions involving light particles, and  gives sizable effects. 
 They    will be presented in separate publications.

\begin{acknowledgments}
This work was partially supported by a Grant-in-Aid for Scientific Research (Grant No. 24340043).
Authors thank Dr. Kobayashi,   Dr. Maruyama, Dr. Suekane, Dr. Nakaya  for useful discussions on the neutrino experiments, Dr. Asai, Dr. Kobayashi, Dr. Mori, Dr.Minowa, Dr. Yamada, Dr. Sloan, Dr.Arai, and Dr.Ojima,  
for useful discussions on quantum interferences.

\end{acknowledgments}
\appendix
\section{Normal term}\label{App-normal}
In $T \ll \tau_\mu$ the muon lifetime can be ignored, but in $ T\approx \tau_\mu$, that cannot. In both cases,  we have
\begin{align}
 \left|I^\text{normal}(\delta p)\right|^2 &=\left(\frac{2\pi\sigma_\mu\sigma_{\nu_e}}{\sigma_\mu+\sigma_{\nu_e}}\right)^3
\exp\left[
-\frac{\sigma_\mu\sigma_{\nu_e}}{(\sigma_\mu+\sigma_{\nu_e})}\delta\vec{p}^{\,2}
-\frac{\left(\tilde{\vec{X}}_\mu-\tilde{\vec{X}}_{\nu_e}\right)^2}{(\sigma_\mu+\sigma_{\nu_e})}\right]\nonumber\\
&\times\int dt_1dt_2\,e^{-\frac{t_1+t_2}{\tau_\mu}+i\left(\delta p^0-\vec{v}_0\cdot\delta \vec{p}\right)(t_1-t_2)}\nonumber\\
&\times\exp\left[\frac{\left(\vec{v}_{\nu_e}-\vec{v}_\mu\right)\cdot\left(\tilde{\vec{X}}_{\nu_e}-\tilde{\vec{X}}_\mu
\right)}{\sigma_\mu+\sigma_{\nu_e}}(t_1+t_2)
-\frac{\left(\vec{v}_{\nu_e}-\vec{v}_\mu\right)^2}{2(\sigma_\mu+\sigma_{\nu_e})}(t_1^2+t_2^2)\right]
\label{sq-I-normal-tau}.
\end{align}
Integrating over $\vec{X}_{\nu_e}$,
\begin{align}
 \int d\vec{X}_{\nu_e} \left|I^\text{normal}(\delta p)\right|^2 
&=16\frac{\pi^5\left(\sigma_\mu\sigma_{\nu_e}\right)^\frac{3}{2}}
{(\sigma_\mu+\sigma_{\nu_e})|\vec{v}_{\nu_e}-\vec{v}_\mu|}
\exp\left[-\frac{\sigma_\mu+\sigma_{\nu_e}}{\left(\vec{v}_{\nu_e}-\vec{v}_{\mu}\right)^2}\left(
\delta p^0-\vec{v}_0\cdot\delta \vec{p}\,\right)^2\right]\nonumber\\
&\times\exp\left[-\frac{\sigma_\mu\sigma_{\nu_e}}{\sigma_\mu+\sigma_{\nu_e}}\delta\vec{p}^{\,2}\right]
\frac{\tau_\mu}{2}\left(1-e^{-\frac{2T}{\tau_\mu}}\right),\label{app-Xint-sqr-I-noraml}
\end{align}
for the wave packets of
$
 \sqrt{\frac{\sigma_\mu+\sigma_{\nu_e}}{\left(\vec{v}_\mu-\vec{v}_{\nu_e}\right)^2}}\ll\tau_\mu,T$, and outside  the probabilities from the normal term are computed numerically. 
 For $T \ll \tau_{\mu}$, $\frac{\tau_\mu}{2} (1-e^{-\frac{2T}{\tau_\mu}})=T$.
\section{Light-cone singularity}
The light-cone singularities  are explained  in many textbooks
and in Ref. \cite{Ishikawa-Tobita-PTEP,Ishikawa-Tobita-ANA}; the new formulae used in this paper  are briefly summarized.



(1) $D^+(\delta t,\delta\vec{x};m)$ and $D^+(\delta t,\delta\vec{x};i m)$ are
 single particle correlation functions of the real mass and imaginary
 mass.

(2) The correlation function $D^+(\delta t,\delta \vec{x};p,m)$ of an
external momentum p is written  as
$D^+(\delta t,\delta \vec{x};p,m) = e^{-ip\cdot\delta x}D^+(\delta
 t,\delta\vec{x},m),\ p^2 = m_0^2\label{1-particle1} $
or $D^+(\delta t,\delta\vec{x};p,m) = D_\infty^+(\delta t,\delta \vec{x};p,m) + 
D^+_\text{finite}(\delta t,\delta \vec{x};p,m)$.
The latter is written further 
\begin{align}
D^+_\infty(\delta t,\delta \vec{x};p,m)  =&D_m\left(\xi\right)D_\infty^+(\delta t,\delta\vec{x};i\tilde{m}),\\
\tilde{m} =& \sqrt{m_0^2 - m^2},\ \xi = -2ip\cdot\frac{\partial}{\partial \delta x},\ 
D_m\left(\xi\right) = \sum_{l=0}\frac{\xi^l}{l!}\left(\frac{\partial}{\partial \tilde{m}^2}\right)^l,\nonumber\\
D_m\left(\xi\right)D_\infty^+(\delta t,\delta\vec{x};i\tilde{m})& = \frac{\epsilon(\delta t)}{4\pi}\delta(\lambda)
+ D_m\left(\xi\right)\Biggl[
-\frac{\tilde{m}}{8\pi\sqrt{-\lambda}}\theta(-\lambda)
\Bigl\{
N_1\left(\tilde{m}\sqrt{-\lambda}\right)\nonumber\\
&-i\epsilon(\delta t)J_1\left(\tilde{m}\sqrt{-\lambda}\right)
\Bigr\} + \theta(\lambda)\frac{\tilde{m}}{4\pi^2\sqrt{\lambda}}K_1\left(\tilde{m}\sqrt{\lambda}\right)
\Biggr].
\end{align}
The light-cone singularity of $D^+_\infty(\delta t,\delta\vec{x};p,m)$ is independent of $p$ and $m$, and exists  in the region expressed by Eq. \eqref{conv-cond} in the text. 

The integral over the region $-p_0\leq r_0 \leq 0$ is written as,
\begin{align}
 D_\text{finite}^+(\delta t,\delta \vec{x};p,m) = \frac{1}{i(2\pi)^3}\int \frac{d\vec{q}}{E(\vec{q}\,)}
e^{i(q - p)\cdot\delta x}\theta(p_0 - E(\vec{q}\,)).
\end{align}
For $\tilde{m}^2 < 0$, $D^+(\delta t,\delta\vec{x};p,m)=0$.

(3) $D^+(\delta t,\delta\vec{x};p,m)_{\alpha_1,\alpha_2,\cdots}$ of many-body states are
 expressed with the mass spectrum, $\rho(m^2)$, as
\begin{align}
& \int dm^2\rho(m^2)D^+(\delta t,\delta \vec{x};p,m),\\
& \rho(m^2) = \int d(\text{phase space})\delta\left(m^2 -
 \left(\sum_lp_l\right)^2\right). \nonumber
\end{align}
For two particles  
\begin{align}
 \Delta_{e,\nu_\mu}(\delta x) &= \frac{1}{(2\pi)^6}\int_{-\infty}^\infty \frac{d\vec{p}_{e}d\vec{p}_{\nu_\mu}}
{E_eE_{\nu_\mu}}(p_\mu\cdot p_{\nu_e})(p_e\cdot p_{\nu_\mu})e^{i(p_{e} +
 p_{\nu_\mu} - p_\mu)\cdot \delta x},\\
&  
=\frac{p_\mu\cdot p_{\nu_e}}{2(2\pi)^2}\int dm^2\left(m^2 - 2m_e^2 +
 \frac{m_e^4}{m^2}\right)iD^+(\delta t,\delta\vec{x};p_\mu,m). \nonumber \\
\Delta_{\nu_\mu,\nu_e}(\delta x) &= \frac{1}{(2\pi)^6}\int_{-\infty}^\infty\frac{d\vec{p}_{\nu_\mu}d\vec{p}_{\nu_e}}
{E_{\nu_\mu}E_{\nu_e}}(p_\mu\cdot p_{\nu_e})(p_e\cdot p_{\nu_\mu})
e^{i(p_{\nu_\mu} + p_{\nu_e} - p_\mu)\cdot \delta x}, \\
 =&\frac{i}{12(2\pi)^2}\int dm^2 \left(m^2(p_\mu\cdot p_e) + 2 p_\mu\cdot
\left(p_\mu -i\frac{\partial}{\partial\delta x}\right)p_e\cdot\left(p_\mu-i\frac{\partial}{\partial\delta x}\right)
\right)\nonumber\\
&\times D^{+}(\delta t,\delta \vec{x};p_\mu,m).\label{correlation-function-app-nu}
\end{align}
\section{Universal function $\tilde{g}(\omega,T;\tau_\mu)$}\label{app-gtilde}
\subsection{ Asymptotic behaviors of integrals }
The constant term of the integral over times plays a crucial role.   
First we study  an integral of a function  $f(t)$ of the property,
\begin{eqnarray}
{\mathcal I}(T)=\int_0^T dt_1 \int_0^T dt_2 f(t_1-t_2) = C T+I_0 ,  \label{short-range}
\end{eqnarray}
for a large $T$ and evaluate the constants $C$ and  $I_0$. For  a short-range 
function  $f(t)$ that decreases rapidly and is
 negligible in $ |t| \geq \epsilon$, the integral for   $T \gg \epsilon$ ,
\begin{eqnarray}
& &{\mathcal I}(T)=\int_{0+\epsilon}^{T-\epsilon} dt_1 \int_0^T dt_2 f(t_1-t_2) 
+\epsilon( \int_0^T dt_1 (f(T-t_1)+f(t_1-T)) =C(T- \epsilon),   \\
& &C= \int_{-\infty}^{\infty} d t f(t), I_0=-\epsilon C. \nonumber
\end{eqnarray}
Thus $I_0$ is  negligible for a microscopic $\epsilon$.

 $I_0$  is significant only for a macroscopic $\epsilon$ or  
for a long-range $f(t)$.  One example is
\begin{align}
T g(\omega T)= \int_0^T dt_1\int_0^T dt_2\frac{\sin(\omega_\nu( t_1-t_2))}{t_1-t_2} =T {\pi}+  T\tilde g(\omega_{\nu} T),  \label{lifetime-g_0}
\end{align}
where $T \tilde g(\omega_{\nu} T)$ was given in \cite{Ishikawa-Tobita-PTEP} and in Eq.$(\ref{asymptotic-gtilde1})$, and is not negligible.   

From Eq.$(\ref{total-probability1})$, the probability is expressed with  the integral over the coordinates $x_1$, $x_2$, and $\vec{X}_{\nu_e}$  
\begin{align}
& I(T)=\int d\vec{X}_{\nu_e}\int d^4x_1d^4x_2e^{ip_{\nu_e}\cdot\delta x} f(\delta x)\prod_{i=1,2}
w(x_i,X_\mu;\sigma_\mu)w(x_i,X_{\nu_e};\sigma_{\nu_e}) \nonumber\\
&=\left(\frac{\pi\sigma_\mu\sigma_{\nu_e}}{\sigma_\mu+\sigma_{\nu_e}}\right)^\frac{3}{2}
\int d\vec{X}_{\nu_e}e^{-\frac{\left(\tilde{\vec{X}}_\mu-\tilde{\vec{X}}_{\nu_e}\right)_T^2}{\sigma_\mu+\sigma_{\nu_e}}}
 \int dt_1dt_2d\delta\vec{x}\,e^{ip_{\nu_e}\cdot\delta x}
e^{-\frac{1}{4\sigma_\mu}\left(\delta \vec{x}-\vec{v}_\mu\delta t\right)^2
-\frac{1}{4\sigma_{\nu_e}}\left(\delta\vec{x}-\vec{v}_{\nu_e}\delta t\right)^2}\nonumber\\
&\times
e^{-\frac{t_1+t_2}{\tau_\mu}}\exp\left[-\frac{\left(\vec{v}_\mu-\vec{v}_{\nu_e}\right)^2}
{\sigma_\mu+\sigma_{\nu_e}}
\left(
\frac{t_1+t_2}{2}-\tilde{T}_L
\right)^2 \right]f(\delta x),\label{arb-spacetime-int}\\
&\tilde{T}_L = 
\frac{\left(\vec{v}_\mu-\vec{v}_{\nu_e}\right)
\cdot\left(\tilde{\vec{X}}_\mu-\tilde{\vec{X}}_{\nu_e}\right)}
{\left(\vec{v}_{\mu}-\vec{v}_{\nu_e}\right)^2},\nonumber
\end{align}
where   $f(\delta x)$ is derived from the correlation function. This is 
 reduced to an integral of the form of Eq.$(\ref{short-range})$  using a Gaussian approximation for the integration in $\vec{X}_{\nu_e}$, 
\begin{align}
I(T)=\left({\pi^2\sigma_\mu\sigma_{\nu_e}}\right)^\frac{3}{2}
 \int_0^T dt_1\int_0^T dt_2d\delta\vec{x}\,e^{ip_{\nu_e}\cdot\delta x}
e^{-\frac{1}{4\sigma_\mu}\left(\delta \vec{x}-\vec{v}_\mu\delta t\right)^2
-\frac{1}{4\sigma_{\nu_e}}\left(\delta\vec{x}-\vec{v}_{\nu_e}\delta t\right)^2}
e^{-\frac{t_1+t_2}{2\tau_\mu}}f(\delta x).\label{app-Xint}
\end{align}
Eq.$(\ref{app-Xint})$ for  $f(x)=D^{+}(\delta t,\delta {\vec x};p_{\mu},m)^{(1)}$ or $f(x)=D^{+}(\delta t,\delta {\vec x};p_{\mu},m)^{(3)}$ is
 short-range in $t_1-t_2$ and $I(T)$ is proportional to $T$. Now  for   $f(\delta x)= i\frac{\epsilon(\delta t)}{4\pi}\delta(\lambda)$, 
\begin{align}
& I(T) \simeq \int_0^T dt_1 \int_0^T  dt_2  \left({\pi^2\sigma_\mu\sigma_{\nu_e}}\right)^\frac{3}{2} \frac{i}{2}\sigma_{\nu_e}e^{-\frac{(\vec{v}_\mu-\vec{v}_{\nu_e})^2}{4\sigma_\mu}\delta t^2}
e^{-\frac{(1-|\vec{v}_{\nu_e}|)^2}{4\sigma_{\nu_e}}\delta t^2}\frac{e^{i\omega_{\nu_e}\delta t}}{\delta t}
e^{-\frac{t_1+t_2}{2\tau_\mu}},\label{app-Xint-delta}
\end{align}
 has both components and $I(T)=CT+I_0$, where $\omega_{\nu_e} = \frac{m_{\nu_e}^2}{2E_{\nu_e}}$, and $\sigma_{\nu_e}|\vec{p}_{\nu_e}|\ll T$ is used. Owing to the small mass of the neutrino, $e^{-\frac{(1-|\vec{v}_{\nu_e}|)^2}{4\sigma_{\nu_e}}} \approx 1$. For the case$ \frac{(\vec{v}_\mu-\vec{v}_{\nu_e})^2}{4\sigma_\mu} T^2 \ll 1$,  that is reduced to  the  integral
\begin{align}
 \tau_{\mu} g(\omega_{\nu},T;\tau_\mu) 
=-\tau_{\mu} \int_0^T dt\frac{\sin(\omega_\nu t)}{t}\left(
e^{-\frac{t}{2\tau_\mu}}
-e^{-\frac{1}{2\tau_\mu}\left(2T-{t}\right)}
\right),\label{lifetime-g}
\end{align}
which is an extended version of the   function Eq.$(\ref{lifetime-g_0})$. For $\tau_{\mu} =\infty$, the constant term in $T {g}(\omega_{\nu},T;\tau_\mu)$ is obtained from $T \tilde g(\omega_{\nu} T)$.  
For finite $\tau_{\mu}$, that varies  depending on the relative magnitudes of $\tau_{\mu}$ and $T_0={1 \over
\omega_{\nu}}$. The constant term of ${g}(\omega_{\nu},T;\tau_\mu)$ is expressed  differently. 

1. $\tau_{\mu} \gg  T_0 $. In this case, $g(\omega_{\nu},t;\tau_\mu)$ at $  t <T_0 $ is equivalent to $g(\omega_{\nu},t;\infty)$. Its short-range component is  combined with the
 short-range terms from  $D^{+}(\delta t,\delta {\vec x};p_{\mu},m)^{(1)}$ and $D^{+}(\delta t,\delta {\vec x};p_{\mu},m)^{3)}$. The rest  $\tilde{g}^{(1)}(\omega_\nu,T;\tau_\mu) = {g}(\omega_\nu,T;\tau_\mu) - g(\omega_\nu,\infty;\infty)$,  determines the constant term. 
 
2. $T_0 \gg \tau_{\mu}  $. In this case, in $g(\omega_{\nu},t;\tau_\mu)$ at $ t <  \tau_{\mu}
$, the short-range component is expressed by  $g(\omega_{\nu},t=\infty ;\tau_\mu)$, and  is
combined with  the other  short-range terms. The rest 
$\tilde{g}^{(2)}(\omega_\nu,T;\tau_\mu) = {g}(\omega_\nu,T;\tau_\mu) - g(\omega_\nu,\infty;\tau_\mu)$
determines the constant term.  The
neutrino masses around $0.08$ eV correspond to the second case.
Hereafter, we study the second case. 
 \subsection{General form of $\tilde{g}(\omega_\nu,T;\tau_\mu)$}
\subsubsection{Without mixing}
The universal function $\tilde  g(\omega_{\nu},T;\tau_\mu) $ is written from  Eq.$(\ref{lifetime-g})$  as 
\begin{align}
& \tilde{g}(\omega_\nu,T;\tau_\mu) = \arctan(2\omega_\nu\tau_\mu)-
\int_0^T dt\frac{\sin(\omega_\nu t)}{t}\left(
e^{-\frac{t}{2\tau_\mu}}
-e^{-\frac{1}{2\tau_\mu}\left(2T-{t}\right)}
\right).\label{lifetime-gtilde} \\
&g(\omega_\nu,\infty;\tau_\mu) = -\arctan({2\omega_\nu\tau_\mu}). \nonumber
\end{align}
\begin{figure}[t]
 \includegraphics[scale=.42,angle=-90]{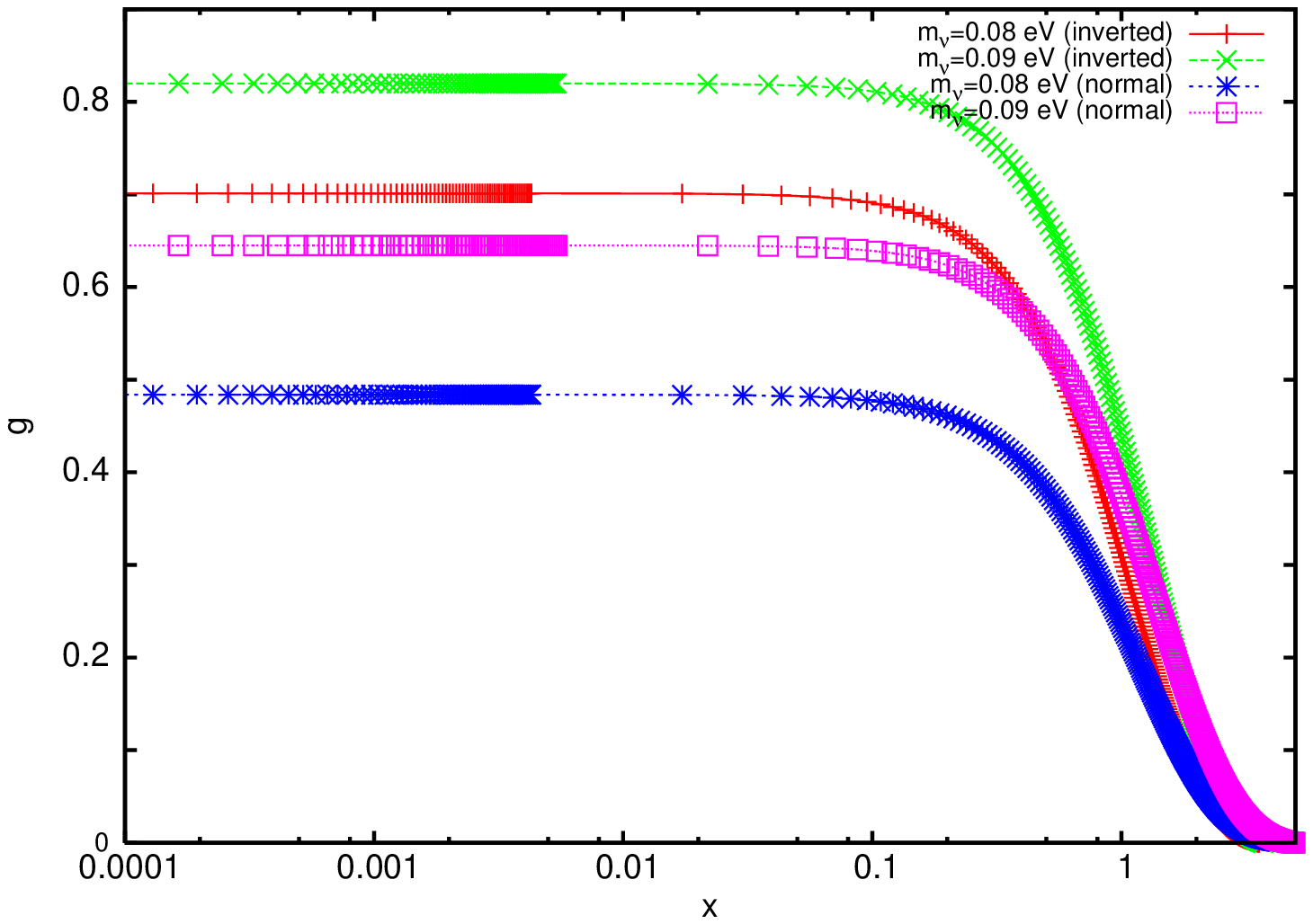}
\caption{(Color online) $\tilde{g}_{e,e}(\omega_{\nu_e},T;\tau_\mu)$
 including the flavor mixing, Eq. \eqref{eq:tildeg-alpha-beta}. The horizontal axis is $x=\omega_{\nu_h}T$, where $\omega_{\nu_h}=\frac{m_{\nu_h}^2}{2E_{\nu_e}}$, $m_{\nu_h}$ is the mass of the heaviest neutrino. The different colors indicate different mass values of $m_{\nu_h}=0.09$ eV (green: inverted hierarchy, magenta: normal hierarchy) and $m_{\nu_h}=0.08$ eV (red: inverted hierarchy, blue: normal hierarchy).
}\label{fig:tildeg-ee}
\end{figure}
\subsubsection{Mixing case}
\begin{figure}[t]
 \includegraphics[scale=.42,angle=-90]{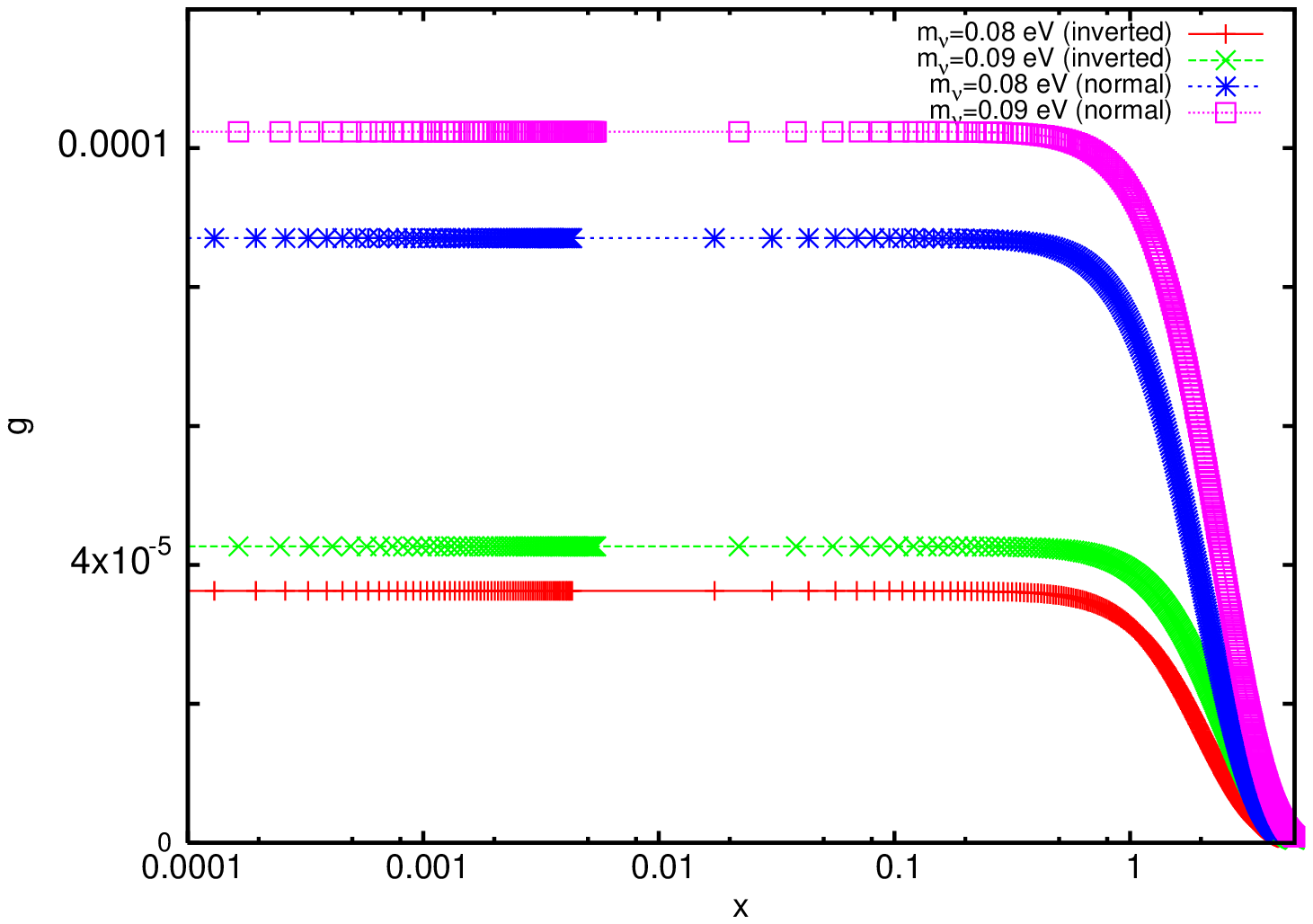}
\caption{(Color online) 
$\tilde{g}_{\mu,e}(\omega_{\nu_e},T;\tau_\mu)$  including flavor mixing, Eq. \eqref{eq:tildeg-alpha-beta}. The horizontal axis is $x=\omega_{\nu_h}T$, where $\omega_{\nu_h}=\frac{m_{\nu_h}^2}{2E_{\nu_e}}$, $m_{\nu_h}$ is the mass of the heaviest neutrino. Different colors represent different mass values of $m_{\nu_h}=0.09$ eV 
(green: inverted hierarchy, magenta: normal hierarchy) and $m_{\nu_h}=0.08$ eV (red: inverted hierarchy, blue: normal hierarchy).
}\label{fig:tildeg-mue}
\end{figure}
$P^{(d)}$  in the system of  the mixing is expressed with the integral for 
 the mass eigenstates of $m_i$ and $m_j$, 
\begin{align}
\mathcal{C}{g}(\omega_{i},\omega_{j},T;\tau_\mu) =
&-\frac{\tau_\mu}{1+\tau_\mu^2(\omega_i-\omega_j)^2}\int_0^T dt
\frac{\sin\left((\omega_i+\omega_j)\frac{t}{2}\right)}{{t}}\nonumber\\
&\hspace{-2cm}\times\Biggl[
e^{-\frac{t}{2\tau_\mu}}
\left\{
\cos\left((\omega_i-\omega_j)\frac{t}{2}\right) -\tau_\mu(\omega_i-\omega_j)\sin\left((\omega_i-\omega_j)\frac{t}{2}\right)
\right\}\nonumber\\
&\hspace{-2cm}-e^{-\frac{1}{2\tau_\mu}\left(2T-{t}\right)}
\left\{
\cos\left((\omega_i-\omega_j)\left(T-\frac{t}{2}\right)\right)
-\tau_\mu(\omega_i-\omega_j)\sin\left((\omega_i-\omega_j)\left(T-\frac{t}{2}\right)\right)
\right\}
\Biggr], \\
\mathcal{C} g(\omega_i,\omega_j,\infty;\tau_\mu) = &\frac{\tau_\mu}{1+\tau_\mu^2(\omega_i-\omega_j)^2}\Biggl[
-\frac{1}{2}(\arctan(2\omega_i\tau_\mu)+\arctan(2\omega_j\tau_\mu))\nonumber\\
&+\frac{\tau_\mu(\omega_i-\omega_j)}{4}\left[
-\log(1+4\omega_j^2\tau_\mu^2) + \log(1+4\omega_i^2\tau_\mu^2)\right]\Biggr].
\end{align}
This is combined with the mixing matrix and the universal function is written as
\begin{align}
 \tilde{g}_{\alpha,\beta}(\omega_{\nu_\beta},T;\tau_\mu) = 
\sum_{i,j}U_{\beta,i}U^*_{\alpha,i}U^*_{\beta,j}U_{\alpha,j}
(g(\omega_i,\omega_j,T;\tau_\mu)-g(\omega_i,\omega_j,\infty;\tau_\mu)),\label{eq:tildeg-alpha-beta}
\end{align}
and shown in Figs. \ref{fig:tildeg-ee} and \ref{fig:tildeg-mue} as a function of $x=\omega_{\nu} T$. 
According to those figures, $\tilde{g}_{\alpha,\beta}(\omega_{\nu_e},T;\tau_\mu)$
is sensitive to the absolute neutrino mass and mass hierarchy, and  causes
the observable effects discussed in the text. The value of $x$ is  $0.0006 \text{ or } 0.0007$  
for $cT=1$ m, $m_{\nu_2}=0.08$ eV, and  $m_{\nu_2}=0.09$ eV, and
is $0.08$ for $cT=100$ m, $m_{\nu_2}=0.08$ eV, and $m_{\nu_2}=0.09$ eV; $\tilde{g}_{e,e}(\omega_{\nu_e})$ and $\tilde{g}_{\mu,e}(\omega_{\nu_e})$ are
 almost constant at $cT<100$ m, but the magnitude is  sensitive to the absolute neutrino mass.
Thus, there is a wide window where the absolute neutrino mass can be measured by using the finite-size corrections. Furthermore, $\tilde{g}_{\mu,e}(\omega_{\nu_e})$ is considerably smaller than $\tilde{g}_{e,e}(\omega_{\nu_e})$, but does not vanish, and causes a new flavor changing effect on the neutrino. 

\subsection{Large life-time and $T$}
For $\omega T\gg 1$, the universal function Eq. \eqref{lifetime-gtilde} behaves as
\begin{align}
& \tilde{g}(\omega,T;\tau) \propto \frac{1}{\omega T},~~\omega\tau \approx 1~\label{asymptotic-gtilde}\\
& \tilde{g}(\omega,T;\tau) \sim \frac{2}{\omega T},~~\tau\to\infty.\label{asymptotic-gtilde1}
\end{align}
\subsection{Angle dependence of overlapping region $T$}\label{app:angle-dep}
For $\mu^+$DAR, the region where parent and daughters overlap is sensitive to the geometry of the experiments, and the diffraction term depends on the angle between the beam axis and the detector, even if the decay is spherically symmetric. 
\begin{figure}[t]
 \includegraphics[scale=.35]{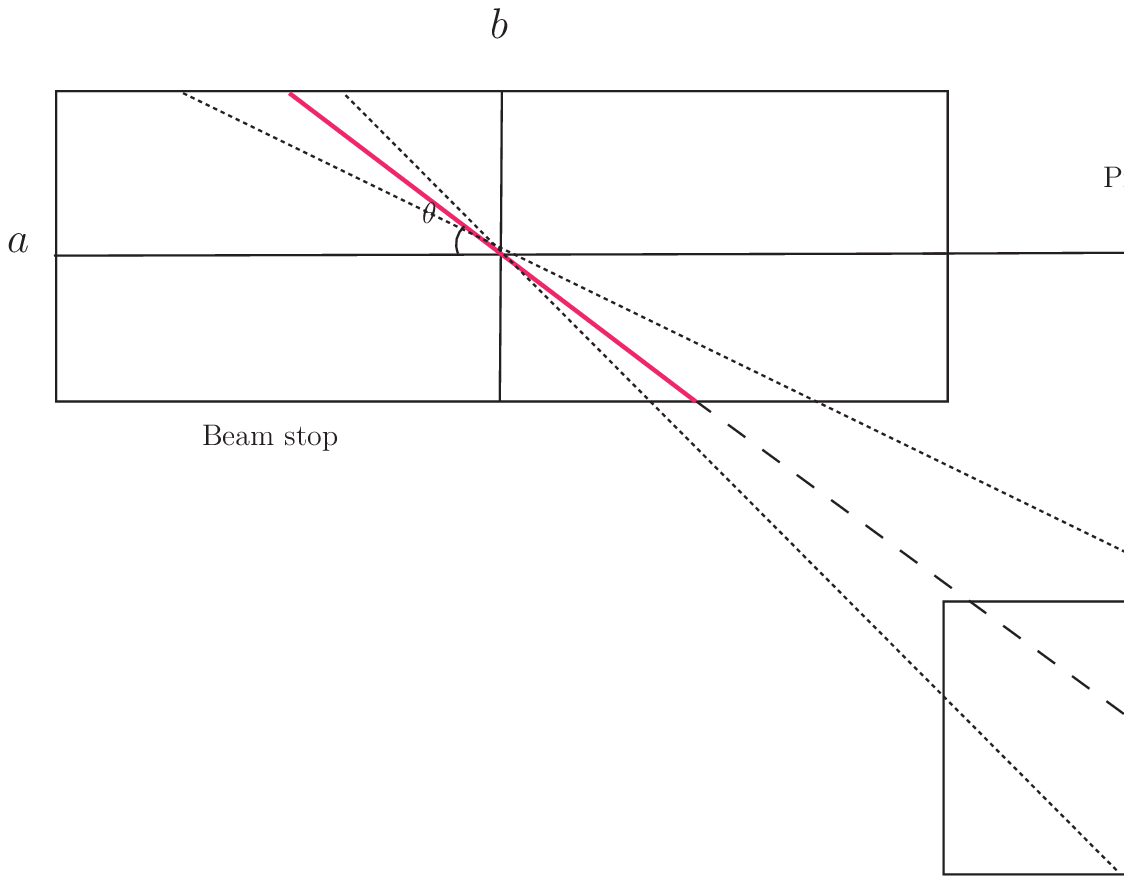}
\caption{(Color online) Angle dependence of the overlapping region (red line) for $\mu^+$DAR. It changes with the position and size of the detector.}
\label{fig:T-region-DAR}
\end{figure}
Following Fig. \ref{fig:T-region-DAR}, the angle dependence region denoted as $T$ in the text is written as
\begin{align}
T = \begin{cases}
     \frac{2a}{\cos\left(\frac{\pi}{2} - \theta\right)}\ \text{for } 0\leq \cos\theta<\frac{b}{\sqrt{a^2+b^2}}\\
\frac{2b}{\cos\theta}\text{ for } \frac{b}{\sqrt{a^2+b^2}}\leq \cos\theta,
    \end{cases}\label{eq:angle-dep-experiment}
\end{align}
where $\theta$ is the angle between the beam axis and the detector, and $a$ and $b$ are the height and length of the detector, respectively. The probability of the event detected at the detector is averaged over the angle within the detector.

\subsection{$\sigma_\mu$ dependence}
$g$ for a finite $\sigma_\mu$, which  corresponds to $\mu^-$ DAR forming
a bound state, is
\begin{align}
  &g(\omega_{\nu},T;\tau_\mu)= \int_{-T}^0dt_-\int_{-\frac{1}{2}t_-}^{T+\frac{1}{2}t_-}dt_+\frac{|\vec{v}_{\nu_e}|t_-\sin(\omega_{\nu_e}t_-)}
{\vec{v}_{\nu_e}^{\,2}t_-^2 + 4|\vec{p}_{\nu_e}|^2\sigma_{\nu_e}^2}
e^{-\frac{\left(\vec{v}_\mu-\vec{v}_{\nu_e}\right)^2}
{\sigma_\mu+\sigma_{\nu_e}}
\left(
t_+-\tilde{T}_L
\right)^2}e^{-\frac{2t_+}{\tau_\mu}}
\nonumber\\
&\hspace{1.2cm}+\int^{T}_0dt_-
\int_{\frac{1}{2}t_-}^{T-\frac{1}{2}t_-}dt_+
\frac{|\vec{v}_{\nu_e}|t_-\sin(\omega_{\nu_e}t_-)}
{\vec{v}_{\nu_e}^{\,2}t_-^2 + 4|\vec{p}_{\nu_e}|^2\sigma_{\nu_e}^2}
e^{-\frac{\left(\vec{v}_\mu-\vec{v}_{\nu_e}\right)^2}
{\sigma_\mu+\sigma_{\nu_e}}
\left(
t_+-\tilde{T}_L
\right)^2}e^{-\frac{2t_+}{\tau_\mu}}
,\nonumber\\
&t_+=\frac{t_1+t_2}{2},\ t_- = t_1-t_2
.\nonumber
\end{align}
For $\mu^+$DAR, $\mu^+$ expands within the beam stop, while $\sigma_\mu$ almost does not change with $T$. Therefore, the $\sigma_\mu$ dependence can be included in the $T$-dependence.
For $\mu^\pm$DIF, $\sigma_\mu$ is determined by the coherence lengths of
the parent particles, and is estimated as 0.1--1 m
\cite{Ishikawa-Tobita-ANA}. Then it is good to approximate the muon 
by  a plane wave.

\end{document}